\documentclass[final]{jfm}

\usepackage{graphicx}
\usepackage{multirow}%
\usepackage{amsmath}
\usepackage{newtxtext}
\usepackage{overpic}
\usepackage{booktabs}%
\usepackage{newtxmath}
\usepackage{natbib}
\usepackage{bm}
\usepackage{hyperref}
\hypersetup{
    colorlinks = true,
    urlcolor   = blue,
    linkcolor = blue,
    citecolor  = blue,
}

\newcommand{\RomanNumeralCaps}[1]
\linenumbers

\graphicspath{{Figures/}}

\title{Numerical simulations of attachment-line boundary layer in hypersonic flow, Part II: the features of three-dimensional turbulent boundary layer}

\author{Youcheng, Xi\aff{1}
  \corresp{\email{xiyc@mail.tsinghua.edu.cn}}, 
  Bowen, Yan\aff{2}, Guangwen Yang\aff{2}
 \and Song, Fu\aff{1}}

\affiliation{\aff{1}School of Aerospace Engineering, Tsinghua University, 100084 Beijing, China
\aff{2}Institute of High Performance Computing, Department of Computer Science and Technology, Tsinghua University, 100084 Beijing, China}

\begin{document}
\maketitle

\begin{abstract}
In this study, we investigate the characteristics of three-dimensional turbulent boundary layers influenced by transverse flow and pressure gradients. Our findings reveal that even without assuming an infinite sweep, a fully developed turbulent boundary layer over the present swept blunt body maintains spanwise homogeneity, consistent with infinite sweep assumptions. We critically examine the law-of-the and temperature-velocity relationships, typically applied two-dimensional turbulent boundary layers, in three-dimensional contexts. Results show that with transverse velocity and pressure gradient, streamwise velocity adheres to classical velocity transformation relationships and the predictive accuracy of classical temperature-velocity relationships diminishes because of pressure gradient. We show that near-wall streak structures persist and correspond with energetic structures in the outer region, though three-dimensional effects redistribute energy to align more with the external flow direction. Analysis of shear Reynolds stress and mean flow shear directions reveals in near-wall regions with low transverse flow velocity, but significant deviations at higher transverse velocities. Introduction of transverse pressure gradients together with the transverse velocities alter the velocity profile and mean flow shear directions, with shear Reynolds stress experiencing similar changes but with a lag increasing with transverse. Consistent directional alignment in outer regions suggests a partitioned relationship between shear Reynolds stress and mean flow shear: nonlinear in the inner region and approximately linear in the outer region.
\end{abstract}


{\bf MSC Codes }  {\it(Optional)} Please enter your MSC Codes here

\section{Introduction}\label{sec1}
Boundary layer flows in practical engineering problems are generally turbulent and three-dimensional. However, to date, most theoretical, experimental, and numerical simulation studies on turbulent boundary layers have been confined to two-dimensional boundary layers. These studies have provided fundamental characteristics of turbulent boundary layers and have guided the development of turbulence models, such as Reynolds-Averaged Navier-Stokes (RANS) models.

Although there have been some studies on three-dimensional turbulent boundary layers, such as the experimental study of a three-dimensional supersonic turbulent boundary layer induced by a curved fin\citep{Konrad1998}, the three-dimensional shock-wave/turbulent boundary-layer interactions\citep{Gaitonde2023}, the three-dimensional turbulent boundary layer over a model lifting body\citep{Dong2024}, these investigations are mostly case-specific, and the relationships between three-dimensional and two-dimensional turbulent boundary layers remain unclear. 
Recently, we conducted simulations of the attachment-line boundary layer of a hypersonic leading edge. 
We realized that this configuration serves as a natural model to bridge two-dimensional and three-dimensional turbulent boundary layers: In the attachment-line region, the flow approximates a two-dimensional turbulent boundary layer, whereas as the flow develops downstream along the chordwise direction, away from the attachment-line region, its three-dimensional characteristics gradually become more pronounced. 
This progression evolves the flow from a two-dimensional to a three-dimensional turbulent boundary layer. 
Therefore, the swept blunt body model can be considered a standardized model for studying three-dimensional turbulent boundary layers. 

This paper is the second part of our study on the numerical simulations of attachment-line boundary layer in hypersonic flow. In the first part, the roughness-induced subcritical transitions of attachment-line boundary layer for a swept blunt body have been studied. Two roughness elements of different heights are examined. For the lower-height roughness element, additional unsteady perturbations are required to trigger a transition in the wake, suggesting that the flow field behind the roughness element acts as a disturbance amplifier for upstream perturbations. 
Conversely, a higher roughness element can independently induce the transition. A low-frequency absolute instability is detected 
behind the roughness, leading to the formation of streaks.
The secondary instabilities of these streaks are identified as the direct cause of the final transition. Following the transition of the boundary layer to a fully developed turbulent state, we continued our computations further downstream. The analysis of this fully developed turbulent boundary layer in the downstream region the core of this study.

In this paper, we describe the general characteristics of three-dimensional turbulent boundary layers in this configuration, focusing on the properties of the mean flow and low-order turbulence statistics. This paper is organized as follows. The methodology and models of this study are also presented for the sake of completeness of the paper in section \ref{sec2}. The major results of the three-dimensional turbulent boundary are shown in section \ref{sec3} and the conclusions are given in section \ref{sec5}.
%Further discussions of the features are present in section \ref{sec4} and the conclusions are given in section \ref{sec5}.

\section{Methodology}\label{sec2}
The governing equations for all simulations in this work are the dimensionless compressible Navier–Stokes(NS) equations for a Newtonian fluid, which can be written as:
\begin{equation}
\frac{\partial Q}{\partial t}+\frac{\partial F_j}{\partial x_j} + \frac{\partial F_{j}^{v}}{\partial x_j}=0,
\end{equation}
\begin{equation}
Q = \left[
  \rho,{\rho {u_1}},{\rho {u_2}},{\rho {u_3}},{{E_t}} 
\right]^{T},
\end{equation}
\begin{equation}
{F_j} = \left[ {\begin{array}{c}
  {\rho {u_j}} \\ 
  {\rho {u_1}{u_j} + p{\delta _{1j}}} \\ 
  {\rho {u_2}{u_j} + p{\delta _{2j}}} \\ 
  {\rho {u_3}{u_j} + p{\delta _{3j}}} \\ 
  {\left( {{E_t} + p} \right){u_j}} 
\end{array}} \right],{F_{j}^{v}} = \left[ {\begin{array}{c}
  0 \\ 
  {{\tau _{1j}}} \\ 
  {{\tau _{2j}}} \\ 
  {{\tau _{3j}}} \\ 
  {{\tau _{jk}}{u_k} - {q_j}} 
\end{array}} \right].
\end{equation}
Throughout this work the coordinates $x_i, (i =1, 2, 3)$ are referred to as $x, y, z$, respectively, with corresponding velocity components $u_1 = u, u_2 = v, u_3 = w$. $F_j$ and $F_j^{v}$ stand for the inviscid and viscous flux.
The total energy \(E_t\) and the viscous stress \(\tau_{ij}\) are given as, respectively, 
\begin{equation}
\begin{aligned}
E_{t}&=\rho\left(\frac{T}{\gamma(\gamma-1) M^{2}_{\infty}}+\frac{u_{k} u_{k}}{2}\right), \\
\tau_{i j}&=\frac{\mu}{Re_{\infty}}\left(\frac{\partial u_{i}}{\partial x_{j}}+\frac{\partial u_{j}}{\partial x_{i}}-\frac{2}{3} \delta_{i j} \frac{\partial u_{k}}{\partial x_{k}}\right).
\end{aligned}
\end{equation}
The pressure \(p\) and heat flux \(q_i\) are obtained from:
\begin{equation}   \label{eq2_6}
p=\frac{\rho T}{\gamma M_{\infty}^{2}}, \quad q_{i}=-\frac{\mu}{(\gamma-1) M_{\infty}^{2} Re_{\infty} Pr} \frac{\partial T}{\partial x_{i}}.
\end{equation}
The viscosity is calculated using the Sutherland law
\begin{equation}   \label{eq2_7}
\mu = T^{3/2}\frac{T_{\infty} + C}{T\cdot T_{\infty} + C},
\end{equation}
with \(C = 110.4K\).
The free-stream Reynolds number $Re_{\infty}$, Mach number $M_{\infty}$ and Prandtl number $Pr$ are defined as
\begin{equation}
Re_{\infty} = \frac{\rho_{\infty}^* U_{\infty}^* l_0^*}{\mu_{\infty}^*}, \quad
M_{\infty} = \frac{U_{\infty}^*}{\sqrt{\gamma R_g^* T_{\infty}^*}}, \quad
Pr = 0.72,
\end{equation}
where $\rho_{\infty}^*$, $U_{\infty}^*$, $T_{\infty}^*$ and $\mu_{\infty}^*$  
stand for the freestream density, velocity, temperature and viscosity, respectively. $R_g^* = 287 \text{J}/(\text{K}\cdot \text{Kg})$ represents the specific gas constant and $\gamma = 1.4$ stands for the ratio of specific heat. The length scale $l_0^*$ is chosen as $1$ millimeter in this research. The $*$ denotes dimensional flow parameters in this section.

\begin{figure}
  \centering
  \begin{overpic}[width=\textwidth]{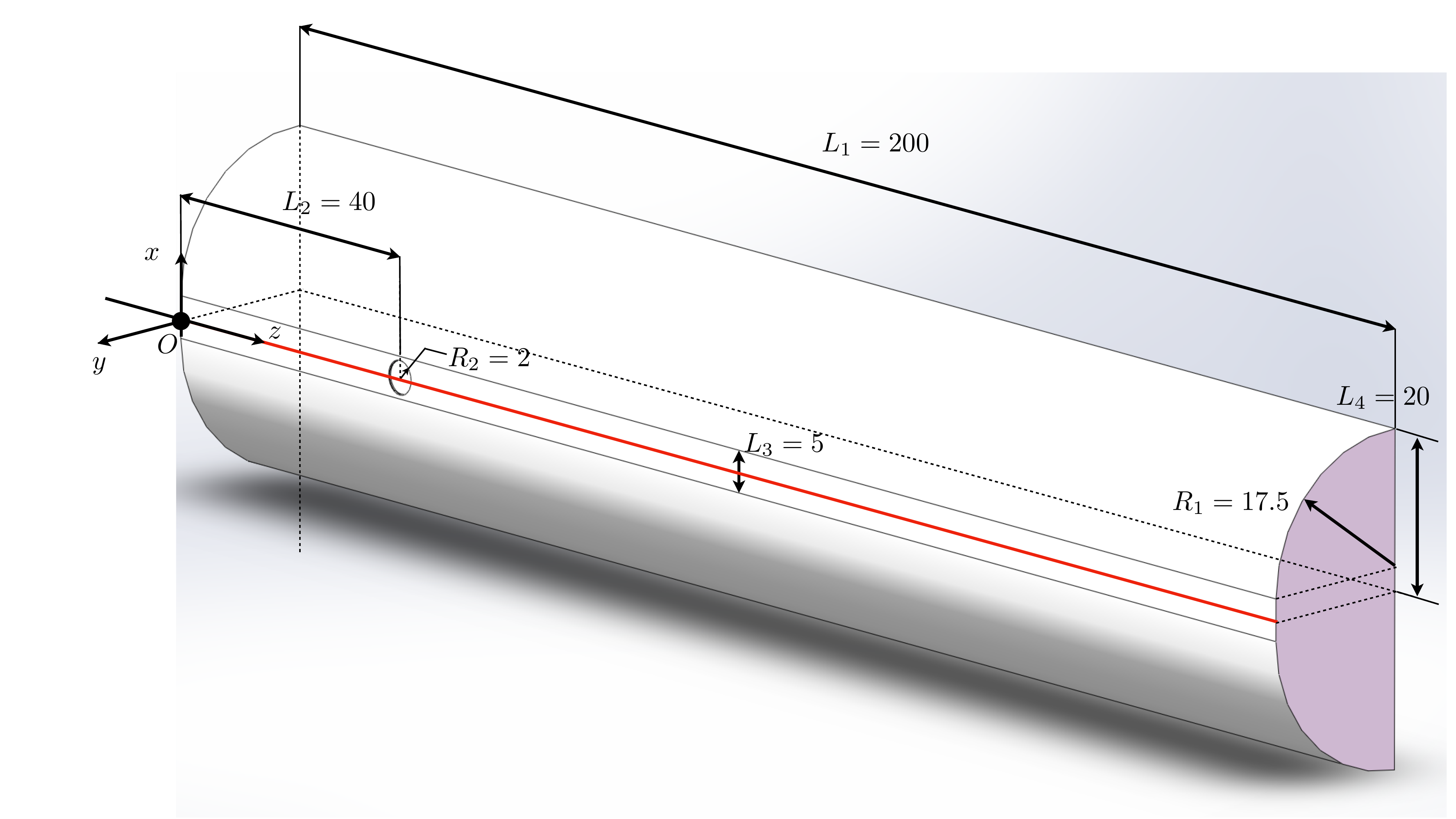}
  \end{overpic}
  \caption{Schematics of the swept blunt leading edge used for numerical simulations.}
  \label{Schematics}
\end{figure}

The computational model is designed based on the recent experimental test and the schematics of the model are shown in figure \ref{Schematics}. 
The computational model can be likened to a sandwich-like configuration, where the top and bottom layers consist of semicircles with a radius of $R_1 = 17.5\text{mm}$, and the intermediate layer is a flat plate with a width of $5\text{mm}$. Together, these three layers form the complete swept blunt body configuration. The roughness elements is located at $z = L_2 = 40\text{mm}$, at the center of the leading plate. The radius of the roughness is $R_2 = 2\text{mm}$. The length of the whole model is designed as $L_1 = 200\text{mm}$. Based on experiments, the surface temperature is set to $T_w^* = 370\text{K}$. The blunt leading edge is placed in a hypersonic flow of $M_{\infty} = 6.0$, the swept angle is $\Lambda = 45^o$ and the temperature of the incoming flow is set to $56.58K$ according to experimental conditions. 
Grids that used in the present simulations are presented in table \ref{BasicGridWallUnits}, based on wall units. 
Based on that grid size, as there are no additional stress and heat flux terms in the present simulations, resolutions for an implicit large-eddy simulations (ILES) can be recognized for the turbulent region. More details about the numerical methods, the strategies, the computational grids and boundary conditions used for performing simulations have already been presented in the first part of this series of study. The H0200 cases in the first part of this series of study is used for further analysis.

It is important to note that our calculations are based on results obtained from ILES, rather than DNS, some specific quantitative  trends or small scale features may exhibit deviations. However, the first-order statistical quantities and mean-flow results are considered to be reliable.
Based on our experience, the grid requirements for DNS of general supersonic three-dimensional boundary layers are as follows: The grid spacing in the two directions parallel to the wall, based on wall units, should ideally be isotropic and satisfy the criteria $\Delta \xi \approx \Delta \zeta \leqslant 4$; The grid spacing in the wall-normal direction should meet the specified requirement $\Delta \eta \leqslant 0.5$. These lead to the condition that effective mesh spacing $\Delta \approx \left(\Delta \xi \times \Delta \zeta \times \Delta\eta\right)^{1/3}$ is always smaller than two times the local Kolmogorov length scale. Also, there should be at least $100-200$ points within the boundary layer. These conditions ensure the grid convergence of high-order turbulence statistics (with errors less than 0.1\%). Given the conditions of the present configuration, we need approximately 8 times more grids than we currently have to resolve all the scales, which exceeds our existing computational capabilities. 

\begin{table}
\centering
\begin{tabular}{c}
\begin{overpic}[width=0.55\textwidth]{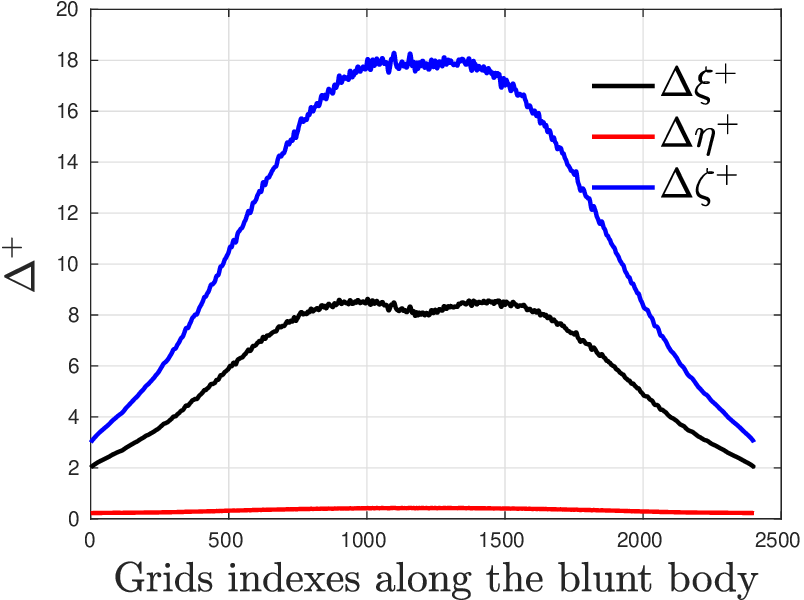}
\end{overpic} \\
\begin{tabular}{ccccccc}
\toprule
 $\xi$ & $\eta$ & $\zeta$ & Total points & $\Delta \xi_{max}^+$ & $\Delta \eta_{max}^+$ & $\Delta \zeta_{max}^+$ \\
 \midrule
 2401 & 601 & 4401 & $6.35\times10^9$ & 8.5 & 0.4 & 18.2 \\ 
\bottomrule
\end{tabular}
\end{tabular}
\caption{Grid points and maximum grid sizes in wall units at the turbulent boundary region.}
\label{BasicGridWallUnits}
\end{table}%

To analyze the characteristics of the turbulent boundary layer flow, we extracted a fully developed turbulent section of the previously computed three-dimensional transitional boundary layer along the spanwise direction for detailed analysis.
To facilitate the analysis investigating the mean field characteristics, both Reynolds-averaged and Favre-averaged mean quantities are employed, following the approach of \citet{Huang1995}. The Reynolds-averaged mean of an arbitrary variable $f$ is denoted by $\overline{f}$, while the Favre-averaged mean is denoted by $\tilde{f}=\overline{\rho f}/\overline{\rho}$. 
In addition, the fluctuations around the Reynolds and Favre averages are represented by single and double primes, respectively. That is, $f^{\prime} = f - \overline{f}$ and $f^{\prime\prime} = f - \tilde{f}$.
Given the absence of homogeneous directions in the configurations under consideration, achieving ideally averaged states necessitates a considerable amount of time. The statistical analysis of the flows was conducted over a span of 1800 time units, encompassing approximately 900,000 steps. This duration is roughly sixfold the time required for a flow to evolve from the inlet to the spanwise flow outlet. To ascertain the independence of flow statistics from the duration of statistical analysis, we performed an additional test. This test involved comparing flow statistics derived from two distinct averaging intervals, with one interval containing 50\% more statistical steps than the other, to ensure that any discrepancies of the flow statistics for major variables between the two intervals were small enough, not exceeding 0.1\%.

Some common variables are defined along the whole three-dimensional surface. Here, as the usual boundary layers in previously, we define the velocity $\overline{u}^{+}$, based on inner scale as
\begin{equation}
\left.
\begin{aligned}
 \overline{u}^+ &= \frac{|\overline{u}_p|}{\overline{u}_\tau}, \quad
|\overline{u}_p| = \sqrt{\overline{u}_{\xi}^2 + \overline{w}^2}, 
\quad h^+ = \frac{\overline{\rho}_w \overline{u}_{\tau} h}{\overline{\mu}_w}, \\
\overline{u}_{\tau} &= \sqrt{\frac{\overline{\tau}_w}{\overline{\rho}_w}}, \quad
\overline{\tau}_w = \frac{\overline{\mu}}{Re}\left. \frac{\partial \overline{u}_p}{\partial h} \right|_{h=0},
h^* = \frac{\overline{\rho} \sqrt{\overline{\tau_w}/\rho} h}{\overline{\mu}}
\end{aligned}
\right\},
\end{equation}  
where, $\overline{u}_{p}$ is the velocity parallel to the surface and $\overline{u}_{\xi}$ is also the velocity parallel to the surface but confined to the $x-y$ plane. The skin-friction coefficient $C_f$ and surface heat-flux ${\theta}_{tw}$ for this kind of flow are defined as
\begin{equation}
C_f = \frac{2 \overline{\mu}_w^*}{\rho_{\infty}^* U_{\infty}^{*2}} = \frac{2 \overline{\mu}_w}{Re}\frac{\partial \overline{u}_{p}}{\partial h}
= 2 \overline{\tau}_w, {\theta}_{tw} = - \left|\kappa \nabla \overline{T} \cdot \boldsymbol{n}\right|.
\end{equation}
The derivatives of surface normals, denoted as ${\partial}/{\partial h}$, for arbitrary variables $f_{\psi}$, are determined through a two-step process. Initially, the gradients of the variables $f_{\psi}$ are computed utilizing the identical scheme adopted for the calculation of viscous fluxes during the simulations. Subsequently, the derivatives of the surface normals ${\partial f_{\psi}}/{\partial h}$ are obtained by projecting the calculated gradients $\nabla f_{\psi}$ onto the surface normal vectors $\bm{n}$. The thickness of boundary layer $\delta_{99}$ is mainly defined based on spanwise velocity as the wall-normal distance from the wall to the location where ${\overline{w}} = 0.99{\overline{w}_e}$.

\section{Results}\label{sec3}

\subsection{Validation of the numerical results}
\begin{figure}
  \centering
  \begin{overpic}[width=0.5\textwidth]{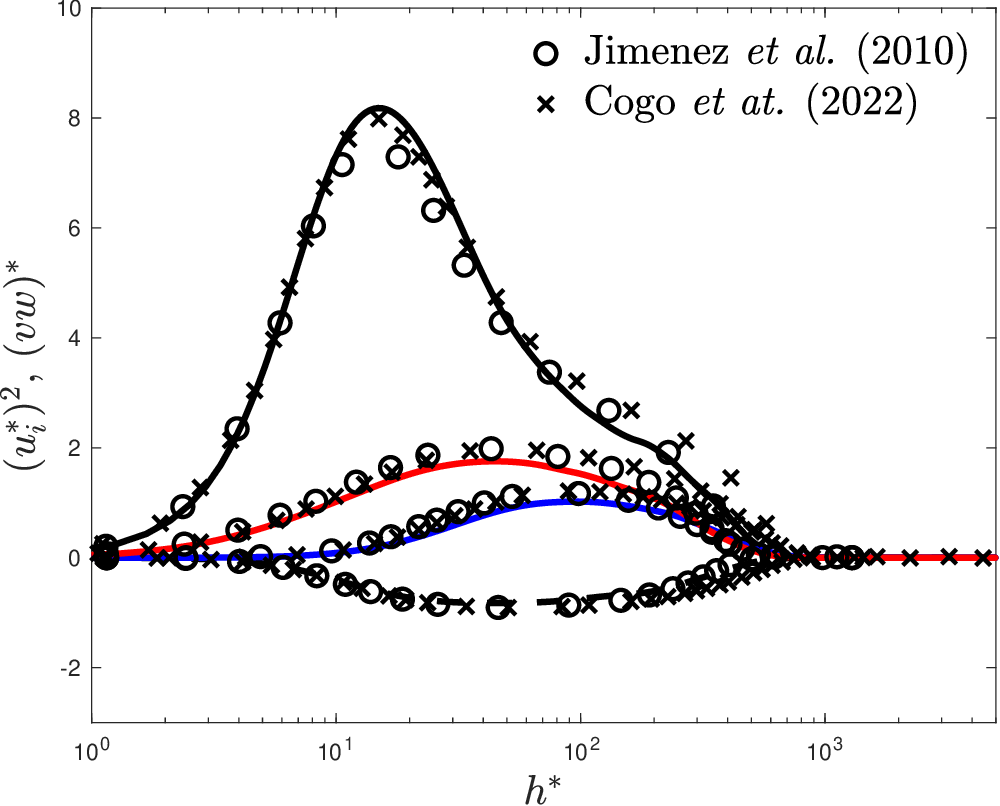}
  \end{overpic}
  \caption{Comparison of the scaled Reynolds stress obtained from the current simulation with the reference values at the spanwise location $z=145.5$.}
  \label{Check_Reynolds_Number}
\end{figure}

Before conducting the related analysis, it is crucial to first verify the reliability of the data obtained from the current simulation. This is particularly challenging in compressible flows. 
Unlike incompressible flows, compressible flows lack sufficient experimental data support, mainly because accurate measurements in the supersonic regime are very difficult\citep{Smits2005,Gatski2013}. 
For general high Mach number three-dimensional compressible boundary layers, this is undoubtedly even more challenging. Fortunately, in our current designed case, we can somewhat refer back to two-dimensional boundary layers, at least along the attachment line, where the flow reverts to a statistically two-dimensional boundary layer. 
Consequently, validation of the present data is conducted by comparing the characteristics of the boundary layer along the attachment line. 
For compressible two-dimensional boundary layers, Morkovin's hypothesis\citep{Morkovin1962} is generally considered valid, which posits that compressibility effects do not alter the time and length scales of turbulence. 
By considering the density variations in the turbulent boundary layer region, compressible turbulence can be analogized to incompressible turbulence through local parameters. Then, the scaled Reynolds stress can be written as
\begin{equation}
\left.
\begin{aligned}
\left(u_i^*\right)^2&=\frac{\overline{\rho}}{\overline{\rho}_w}\frac{\widetilde{u_i^{\prime\prime 2}}}{\overline{u}_\tau^2},\quad i=1,2,3 \\
\left(u_iu_j\right)^*&=\frac{\overline{\rho}}{\overline{\rho}_w}\frac{\widetilde{u_i^{\prime\prime}u_j^{\prime\prime}}}{\overline{u}_\tau^2},\quad i\neq j
\end{aligned}
\right\},
\end{equation}
which should collapse the data to the incompressible ones. Here, $\overline{u}_{\tau}$ is the friction velocity and subscript $w$ represents the variables at the wall surface. The results of a incompressible turbulent boundary layer at $Re_{\tau} = 445$ simulated by
\citet{Jimenez2010} and a compressible turbulent boundary layer at $Re_{\tau} = 453$ simulated by \citet{Cogo2022} are used as references. 
The comparison for major components of Reynolds stress is shown in figure \ref{Check_Reynolds_Number}. All calculated profiles are presented as a function of the semilocal scaling $h^*$, which is based on $\tau_w$ and local properties, proposed by \citet{Huang1995}. 
In the figure, the black solid line represents the spanwise-transformed Reynolds stress component $(w^*)^2$. The red and blue solid lines indicate the Reynolds stress components in the $x-$direction $(u^*)^2$ and the wall-normal $y-$direction $(v^*)^2$, respectively. The black dashed line represents the shear Reynolds stress component $(wv)^*$. As can be viewed, the computed and statistically analyzed Reynolds stresses at the exact attachment line generally align with the results obtained from direct numerical simulations for both incompressible and compressible flat-plate boundary layers. The peak values and shapes of the Reynolds stresses in the streamwise direction are consistent with the corresponding reference values. This agreement validates the resolution of our current computational setup. However, the Reynolds stresses in the wall-normal and chordwise directions are smaller than the reference values. This discrepancy is likely due to the influence of the leading-edge shock wave and the chordwise pressure gradient in the conditions of this study. At the leading edge, the height of the shock wave from the wall is approximately 13mm, which is about sixteen times the turbulent boundary layer thickness at that location. This suppresses the development of fluctuations in this direction, resulting in lower corresponding Reynolds stresses $(v^*)^2$. Along the chordwise direction at the leading edge, the condition $\partial/\partial x = 0$ is satisfied only at the exact attachment line($x=0$), hence the Reynolds stresses $(u^*)^2$ in this direction are also somewhat suppressed.

\subsection{General features of the turbulent boundary layer}

\subsubsection{Mean field}
\begin{figure}
  \centering
  \begin{overpic}[width=\textwidth]{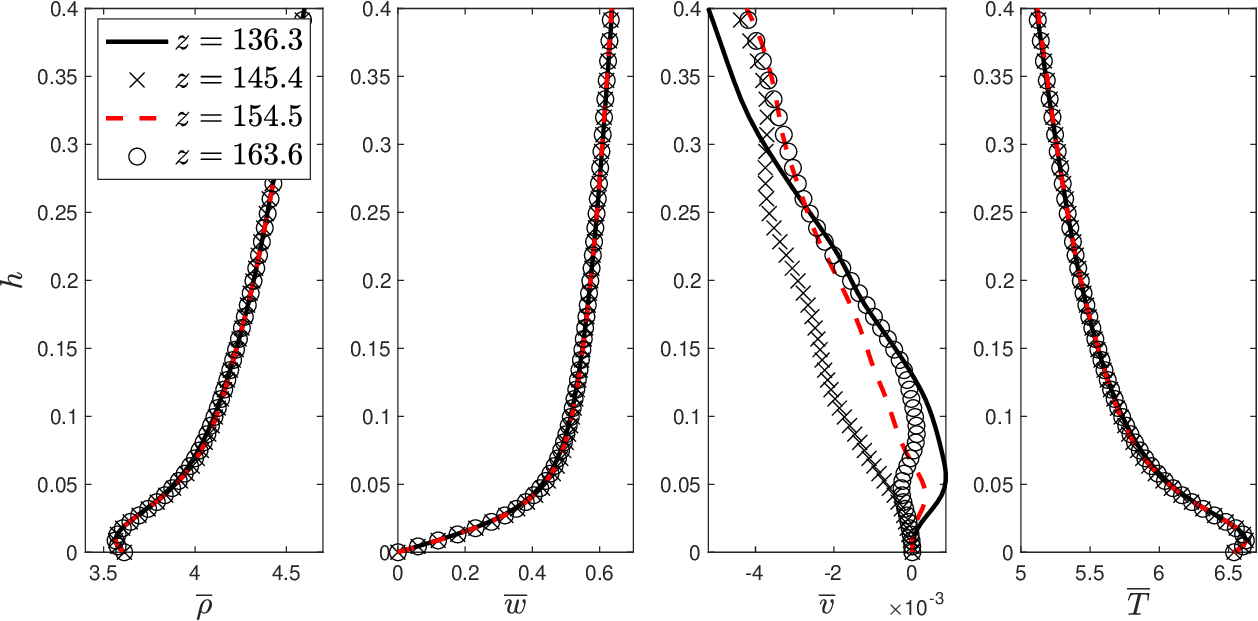}
  \end{overpic}
  \caption{The wall-normal profiles of mean-flow variables at the attachment-line plane.}
  \label{Check_Mean_Flow_AT}
\end{figure}

\begin{figure}
  \centering
  \begin{overpic}[width=\textwidth]{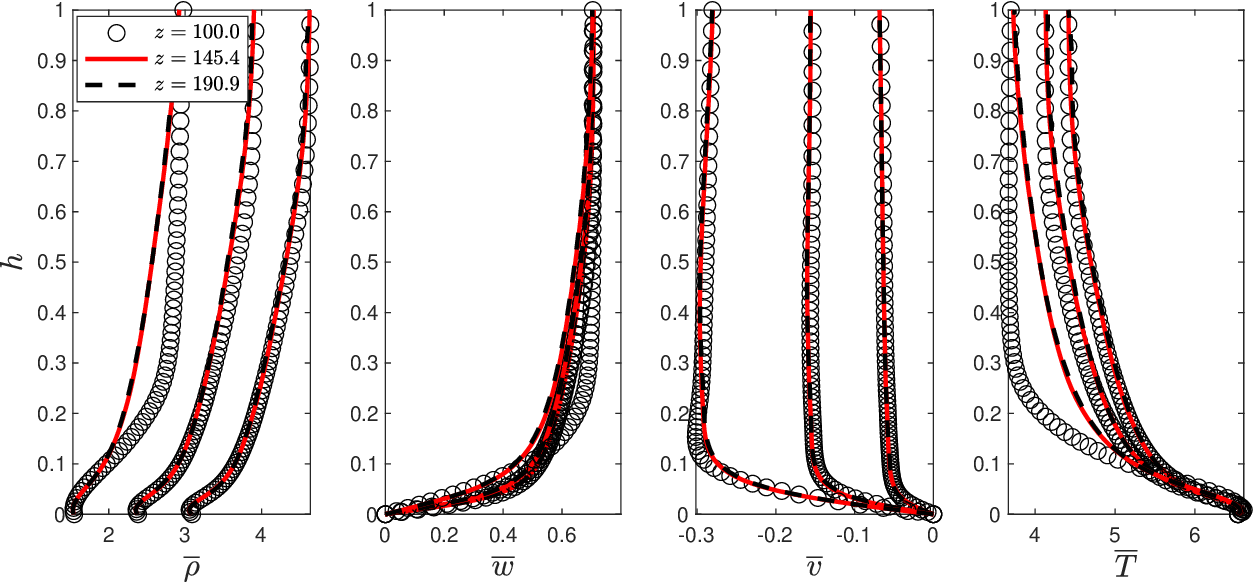}
  \end{overpic}
  \caption{The distributions for mean-flow variables with wall normal direction at the chordwise location $s_\xi = 8.71, s_\xi = 13.50$ and $s_\xi = 18.61$.}
  \label{Check_Mean_Flow_Cs}
\end{figure}

% Mean flow features, pressure gradients
The laminar and transitional boundary layer have been documented in the previous study. Here, we focus on the turbulent section. 
The basic features of turbulent boundary layer are reflected in their profiles along the wall normal directions. The profiles for major variables along the wall normal direction, at several locations, are shown in figure \ref{Check_Mean_Flow_AT} and \ref{Check_Mean_Flow_Cs}. 
It can be observed that as the attachment-line boundary layer develops from the upstream position $(z=136.3)$ to the downstream position $(z=163.6)$, the boundary layer does not thicken as expected for a well-known flat plate boundary layer. 
Instead, the thickness of the boundary layer remains almost constant. 
This indicates that the fully developed turbulent attachment-line boundary layer can also reach a corresponding asymptotic state, in the present cases. By broadening our perspective further, it becomes evident that this assumption remains valid even in regions where crossflow effects are significant, as demonstrated in figure \ref{Check_Mean_Flow_Cs}. All these phenomena indicate that, under the conditions of our study, the development of the turbulent boundary layer in the wall-normal $\eta-$direction as well as the spanwise $z-$direction are inhibited, and the boundary layer does not thicken in the spanwise direction.

\begin{figure}
  \centering
  \begin{tabular}{cc}
  \begin{overpic}[width=0.6\textwidth]{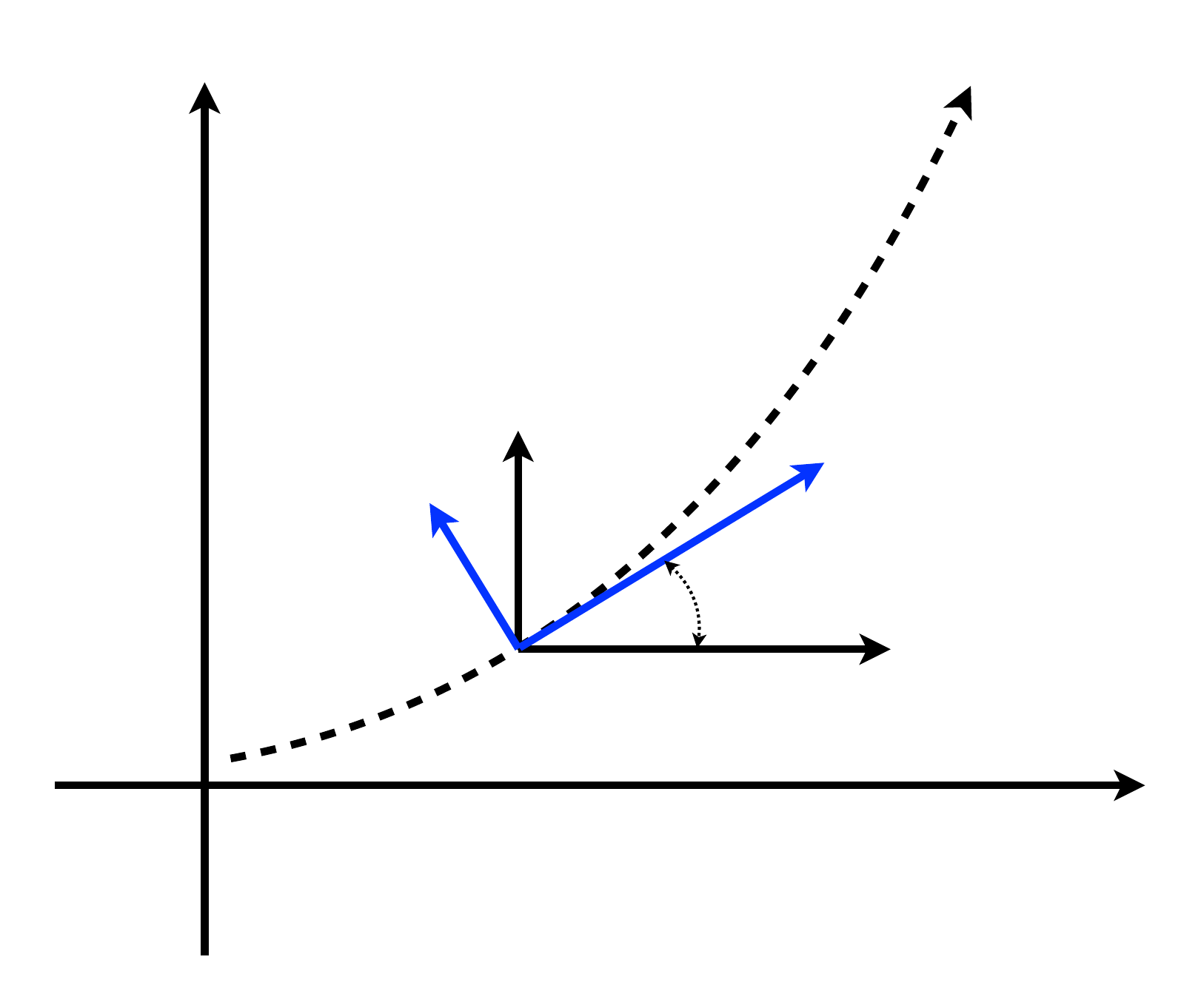}
  \put(3,76){$(a)$}
  \put(95,14){$z$}
  \put(20,75){$\xi$}
  \put(50,80){External streamline}
  \put(67,79){\vector(1,-2){7}}
  \put(75,26){$\overline{w}$}
  \put(70,45){${u}_s$}
  \put(60,34){$\theta_s$}
  \put(45,50){$\overline{u}_{\xi}$}
  \put(31,43){${u}_c$}
  \end{overpic} &
  \begin{overpic}[width=0.357\textwidth]{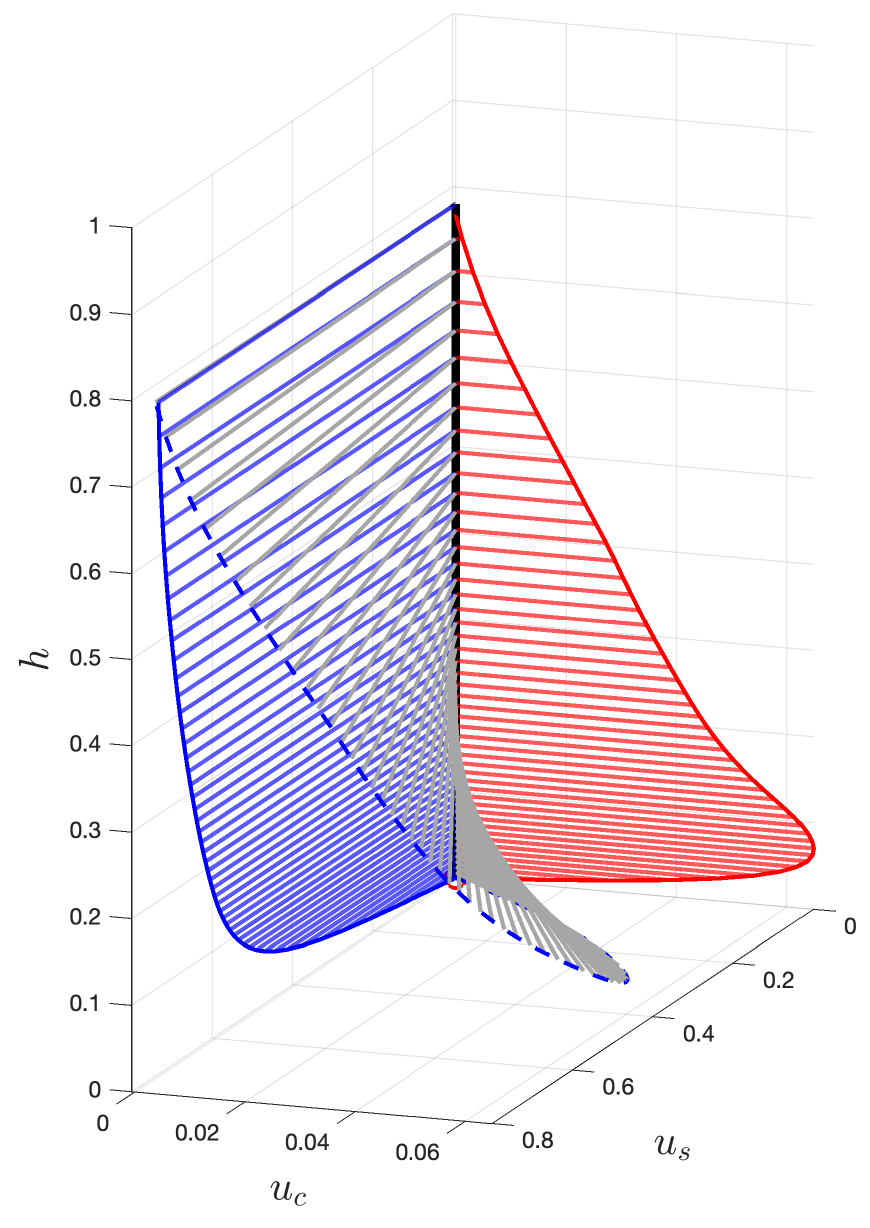}
  \put(3,90){$(b)$}
  \put(37,28.4){\vector(3,-1){18}}
  \put(54,24){$\overline{\tau}_w$}
  \put(37,28.4){\vector(-3,-2){18}}
  \put(12,12){External streamline direction}
  \end{overpic}
  \end{tabular}
  \caption{ $(a)$ Schematic of a local streamline coordinate system and $(b)$ Three dimensional velocity profiles of the present simulations at chordwise location $s_{\xi}=13.50$.}
  \label{Local_StreamLine}
\end{figure}

Away from the attachment-line, unlike traditional two-dimensional boundary layers, the presence of pressure gradients in three-dimensional boundary layers induces a crossflow component. The three-dimensional boundary layer are most conveniently described in a local streamline coordinate system as shown in figure \ref{Local_StreamLine}$(a)$, the main streamwise and the crossflow directions are defined based on the angle of external streamlines  $\theta_s = \arctan(\overline{u}_{\xi,e}/\overline{w}_e)$ at the edge of the turbulent boundary layer. The streamwise and the crossflow velocities are defined as
\begin{equation}\label{eqms}
\left.
\begin{aligned}
u_s &= \overline{w} \cos \theta_s + \overline{u}_{\xi} \sin \theta_s, \\
u_c &=-\overline{w} \sin \theta_s + \overline{u}_{\xi} \cos \theta_s.
\end{aligned}
\right\}.
\end{equation}

As shown in figure \ref{Local_StreamLine}$(b)$, when the wall surface is approached, the magnitude of the crossflow velocity $u_c$ (the red lines) increases from small value at the edge of boundary layer, reaches a maximum value, and then decreases back to zero at the wall surface to satisfy the no-slip condition. 
In contrast, the behaviour of the streamwise velocity $u_s$ (the blue lines) is qualitatively similar to that in two-dimensional boundary layers. The direction of the wall shear stress vector $\overline{\tau}_w$ determines the tangent to the limiting streamline at the wall. In general, the angle between the tangents to the external streamline and the limiting streamline at the wall surface is non-zero. As a result, when acrossing the hight of the boundary layer, the velocity vector rotates through an angle at the wall (the transparent grey lines). Similarly, the shear stress vector tends to rotate away from its direction at the wall as the distance from the surface increases.

\begin{figure}
  \centering
  \begin{tabular}{c}
  \begin{overpic}[width=\textwidth]{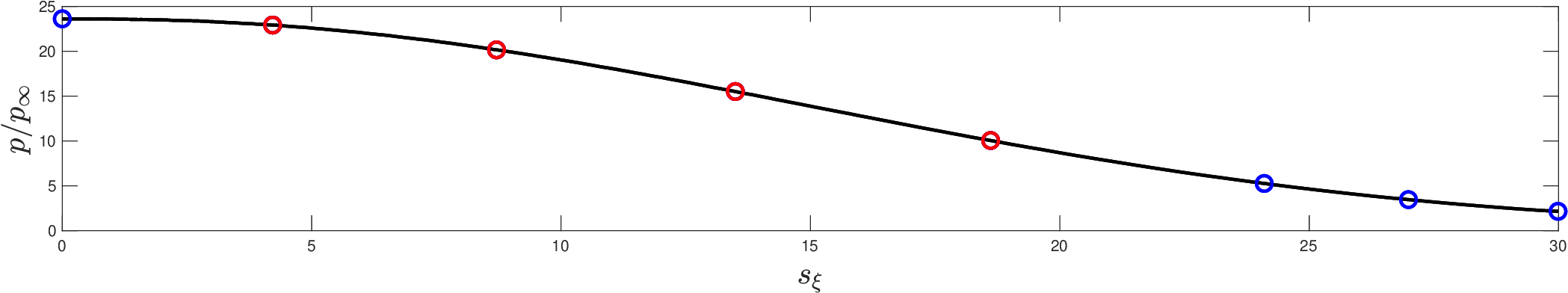}
   \put(72,15){$(a)$}
   \put(10,13){$s_{\xi} = 4.22$}
   \put(26,12){$s_{\xi} = 8.71$}
   \put(38,10){$s_{\xi} =13.50$}
   \put(55,7){$s_{\xi} =18.61$}
  \end{overpic} \\
  \begin{overpic}[width=\textwidth]{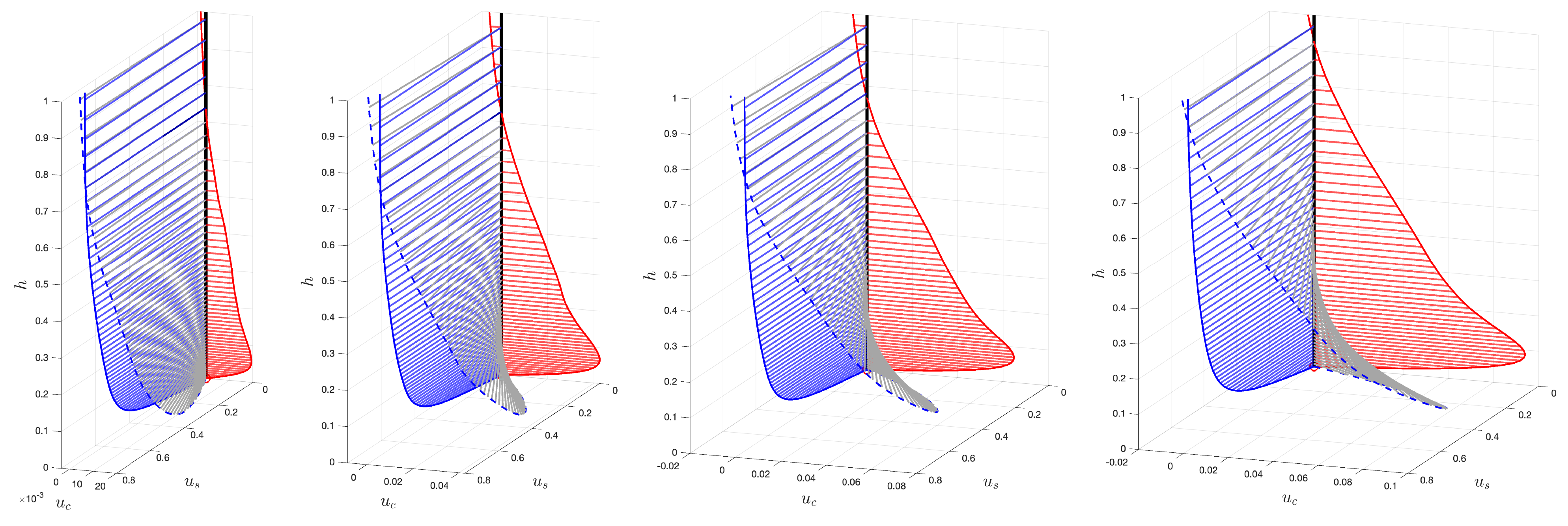}
  \put(3,30){$(b)$}
  \put(6,0){$s_{\xi}=4.22$}
  \put(21,30){$(c)$}
  \put(24,0){$s_{\xi}=8.71$}
  \put(44,30){$(d)$}
  \put(40,0){$s_{\xi}=13.50$}
  \put(72,30){$(e)$}
  \put(70,0){$s_{\xi}=18.61$}
  \end{overpic}
  \end{tabular}
  \caption{$(a)$ The pressure distributions along the surface. 
$(b-e)$ Three dimensional velocity profiles of the present simulations at several chordwise location, ranging from the attachment-line region to chordwise downstream. The specific locations corresponding to the three-dimensional velocity profiles are sequentially marked with red circles in $(a)$.}
  \label{Check_Mean_Flow_Cs_Change}
\end{figure}
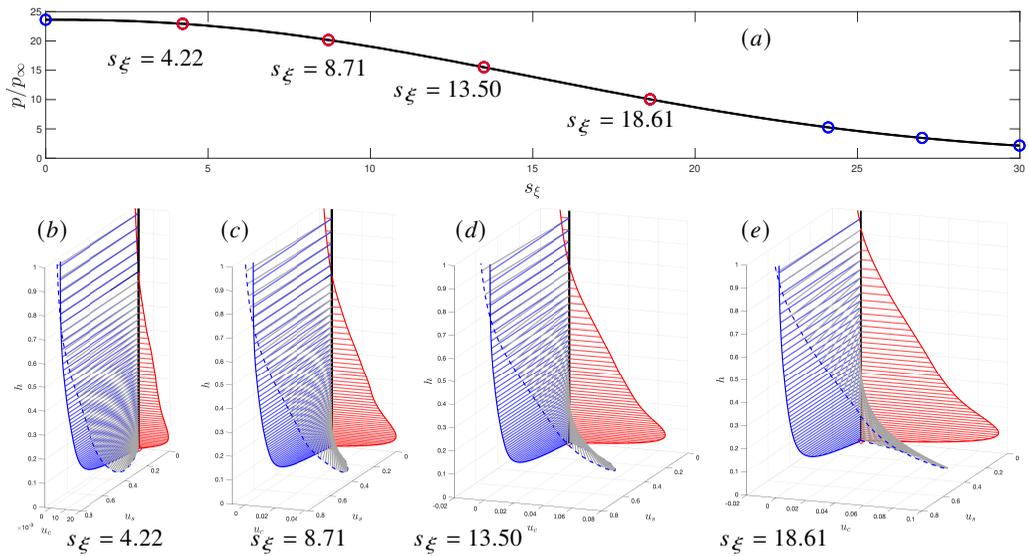

As shown in figure \ref{Check_Mean_Flow_Cs_Change}, when the flow develops from the attachment line to chordwise downstream, the pressure gradient increases (figure \ref{Check_Mean_Flow_Cs_Change}$(a)$), leading to stronger crossflow velocities (figure \ref{Check_Mean_Flow_Cs_Change}$(b-e)$). Consequently, the boundary layer becomes thicker, and the rotation of the velocity vectors within the boundary layer becomes more pronounced. 

\subsubsection{fluctuation field}
\begin{figure}
  \centering
  \begin{overpic}[width=\textwidth]{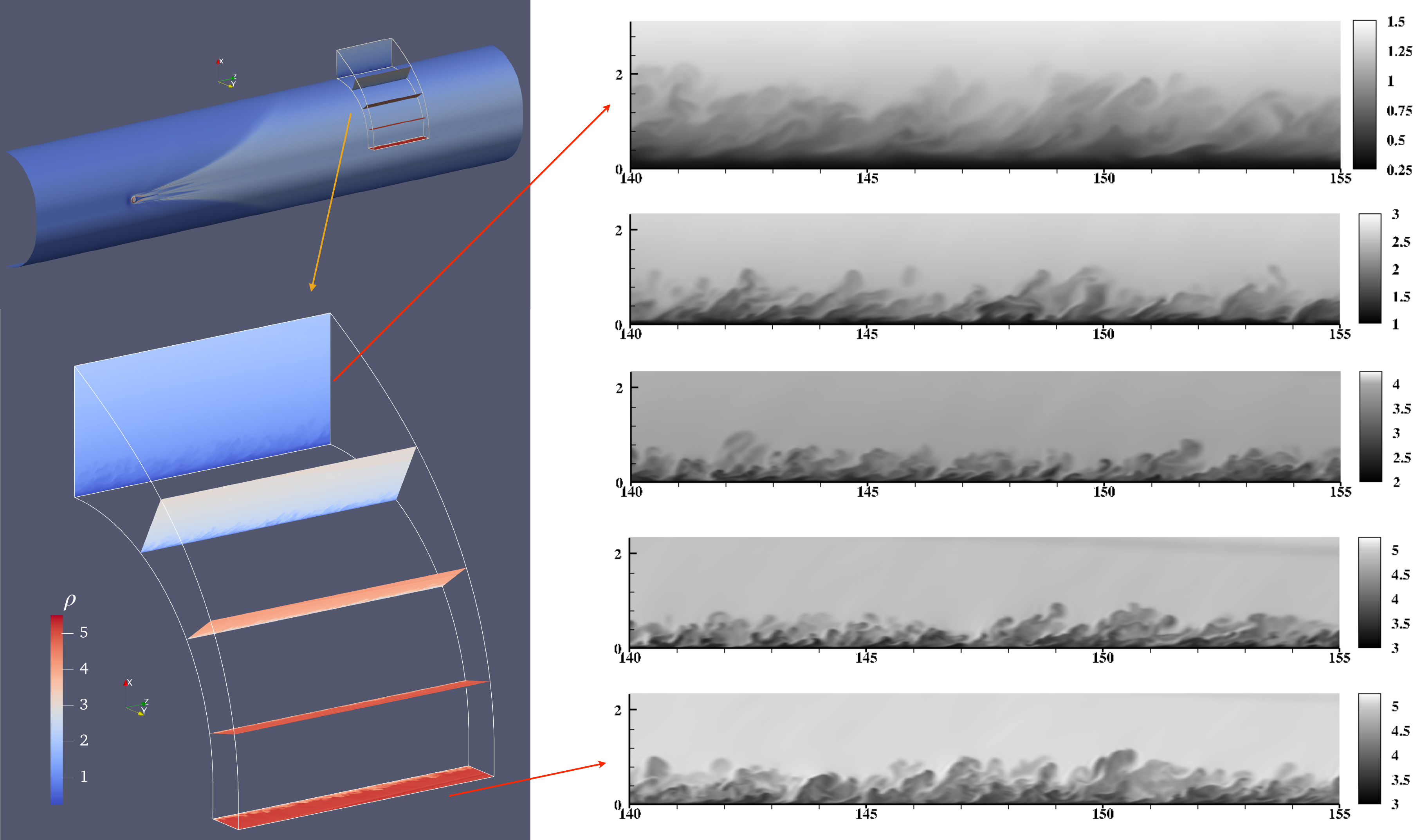}
  \put(1,57){$(a)$}
  \put(1,33){$(b)$}
  \put(38,57){$(c)$}
  \put(41,51.5){$h$}
  \put(41,39){$h$}
  \put(41,28){$h$}
  \put(41,16){$h$}
  \put(41,5){$h$}
  \put(69.5,0.5){$z$}
  \end{overpic}
  \caption{$(a)$ Schematics of the swept blunt leading edge with the average surface skin friction $\overline{\tau}_w$ and the selected region. $(b)$ the enlarged figure for the region selected in $(a)$. $(c)$ the instantaneous density of the selected cross-sections in $(a)$ and $(b)$.}
  \label{Cross_sections_Schematics}
\end{figure}

\begin{figure}
  \centering
  \begin{overpic}[width=\textwidth]{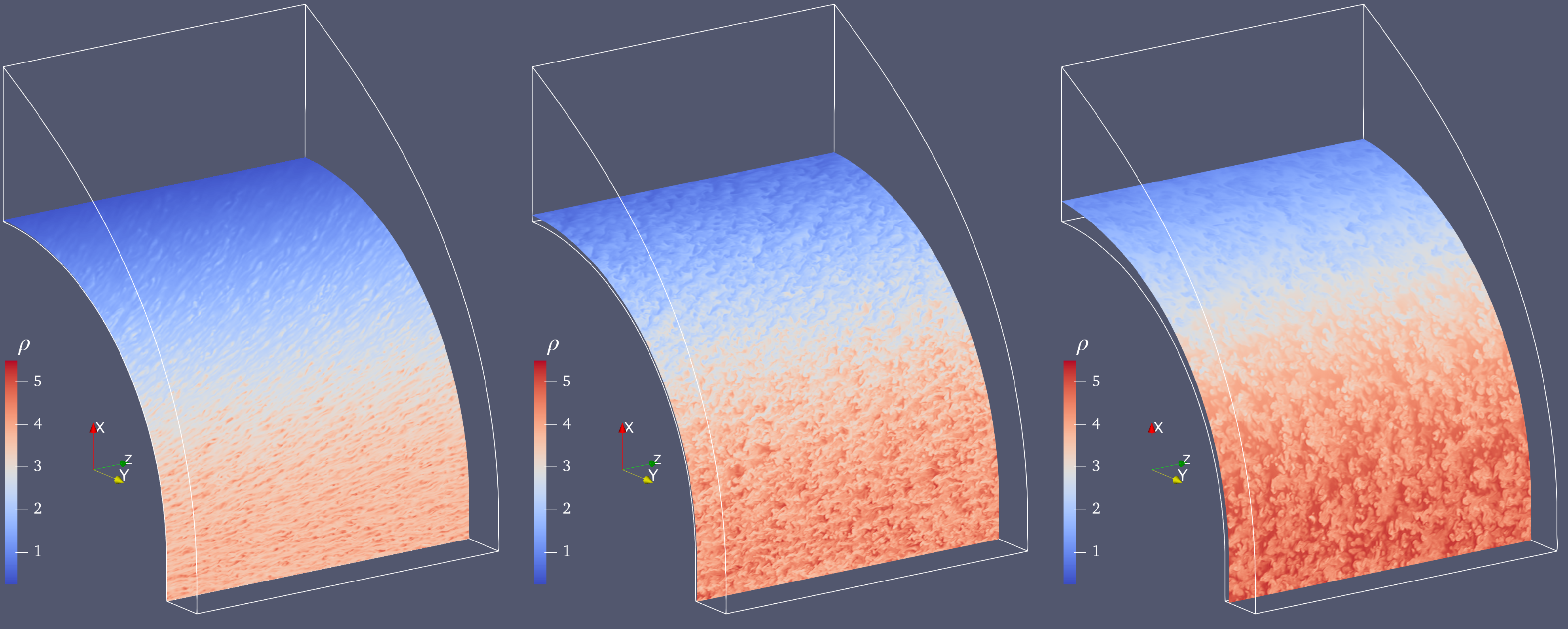}
  \put(5,25){$(a)$}
  \put(39,25){$(b)$}
  \put(73,25){$(c)$}
  \end{overpic}
  \caption{$(a-c)$ stand for instantaneous density $\rho$ distributions at spanwise cross sections at grid index $\eta = 33(h^+ = 15.5), 101(h/\delta_{99}=0.18)$ and $201(h/\delta_{99}=0.53)$, respectively. $h^+, h$ and $\delta_{99}$ are defined based on the variables at exact attachment-line plane. $\delta_{99}\approx 0.8$ at the present case.}
  \label{Normal_sections_Schematics}
\end{figure}

Then, we establish an intuitive visualization of the three-dimensional turbulent boundary layer flow by presenting the distribution of instantaneous or fluctuated variables contours at different sections.
The instantaneous density contours in several selected cross sections are shown in figure \ref{Cross_sections_Schematics}$(c)$, ranging from attachment line to chordwise exit. The exact locations of these slices are shown in figure \ref{Cross_sections_Schematics}$(a)$ and $(b)$.
The contour exhibite the typical structure, already found in flat-plate turbulent boundary layers \citep{Smith1995,Pirozzoli2011,Duan2014,Cogo2022}, dominated by large-scale patterns inclined at some angle. Therefore, simply observing the characteristics on a single cross-section makes it difficult to distinguish between two-dimensional and three-dimensional boundary layers.
As a supplement, some nearly wall-parallel slices for instantaneous density contours are shown in figure \ref{Normal_sections_Schematics}. From the attachment line to the chordwise exit, the density perturbations become sparser as the density amplitude decreases (changing from red to blue). With increasing distance from the wall (from Figure \ref{Normal_sections_Schematics}$(a)$ to Figure \ref{Normal_sections_Schematics}$(c)$), the characteristic structures of the perturbations grow larger. Near the wall, the disturbance structures exhibit a clear preferential orientation, and it can be faintly observed that the disturbance structures gradually shift from the spanwise direction to the chordwise direction.

\begin{figure}
  \centering
  \begin{overpic}[width=\textwidth]{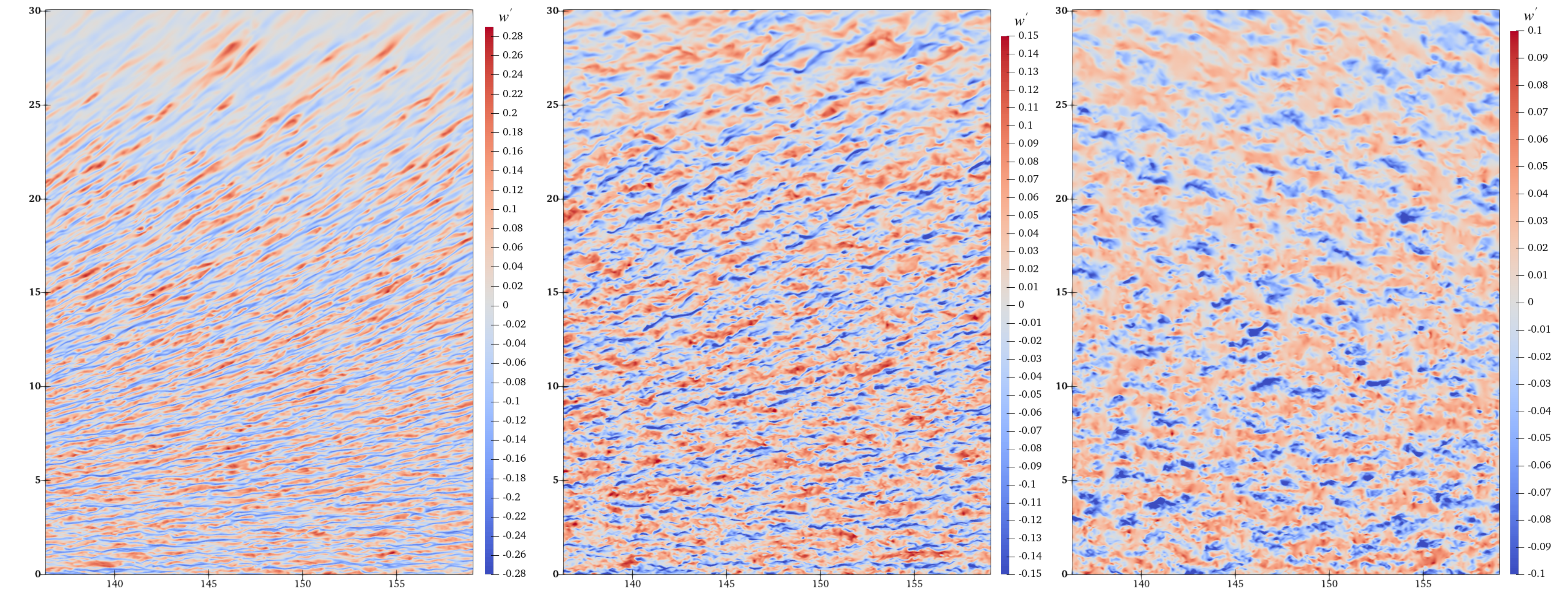}
  \put(-1,38){$(a)$}
  \put(32,38){$(b)$}
  \put(64,38){$(c)$}
  \put(-1.5,20){$s_{\xi}$}
  \put(15,-1){$z$}
  \put(48,-1){$z$}
  \put(82,-1){$z$}
  \end{overpic}
  \caption{$(a-c)$ stand for spanwise fluctuations $w^{\prime}$ at the cross sections at grid index $\eta = 33, 101$ and $201$, respectively.}
  \label{Normal_sections_Schematics_PertW}
\end{figure}

\begin{figure}
  \centering
  \begin{overpic}[width=\textwidth]{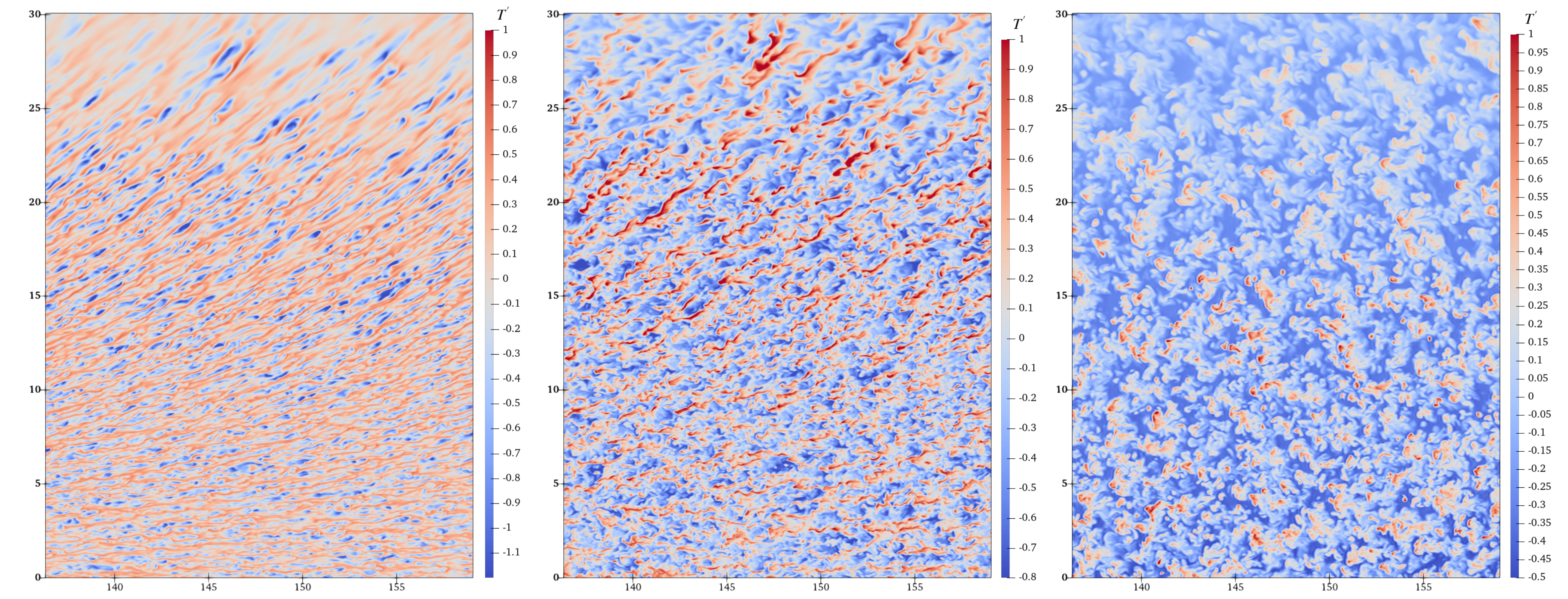}
  \put(0,38.5){$(a)$}
  \put(33,38.5){$(b)$}
  \put(65,38.5){$(c)$}
  \put(-1.5,19){$s_{\xi}$}
  \put(15,-1){$z$}
  \put(48,-1){$z$}
  \put(82,-1){$z$}
  \end{overpic}
  \caption{$(a-c)$ stand for temperature fluctuations $T^{\prime}$ at the cross sections at grid index $\eta = 33, 101$ and $201$, respectively.}
  \label{Normal_sections_Schematics_PertT}
\end{figure}

To obtain clearer images of the disturbances themselves, we subtract the mean flow to observe the fluctuations.
Figures \ref{Normal_sections_Schematics_PertW} and \ref{Normal_sections_Schematics_PertT} show contours
of the instantaneous spanwise velocity fluctuations $w^{\prime}$ and temperature fluctuations $T^{\prime}$ for the present cases in nearly wall-parallel slices taken at three different locations, standing for both the inner $(h^+ = 15.5)$ and outer region $(h/\delta_{99} = 0.18 / 0.53)$ at the selected stations.

As seen in figure \ref{Normal_sections_Schematics_PertW}$(a)$, the spanwise velocity field in the inner layer presents the typical streaky pattern. These streaky patterns are distributed along the spanwise $z-$direction in the attachment line region. As the chordwise distance $s_\xi$ increases, these structures tend to align with the direction of the external streamline. Therefore, all these patterns exhibit a certain radiating state, appearing as skirt-like structures extending from the attachment line along the chordwise direction. Similar visualizations of near-wall streaks in flat-plate boundary layer simulations were also reported by \citep{Duan2010,Pirozzoli2011,Duan2011,Cogo2022}. 
These findings suggest that the near-wall streak structures in the three-dimensional turbulent boundary layer resemble those in a two-dimensional flat-plate turbulent boundary layer, both aligning with the direction of the external streamline. The alternating distribution of high and low momentum zones forms the velocity streak structures depicted in this figure. This formation is associated with ejection and sweep events. These large-scale structures can extend several boundary layer thicknesses in the direction of the external streamline and contain a significant portion of the turbulent kinetic energy.
As the slice move away from the surface (figure \ref{Normal_sections_Schematics_PertW}$(b)$), a qualitatively similar pattern found in figure \ref{Normal_sections_Schematics_PertW}$(a)$ are also presented on a much larger scale. These structures can be identified as outer-layer streaks, which can be identified as large scale coherent structures or "superstructures" in outer region\citep{Ganapathisubramani2006,Hutchins2007}. From this perspective, the near-wall small-scale structures appear to be the footprints of the large-scale structures from the outer layer in the near-wall region. This observation is also consistent with what has been previously understood in flat-plate boundary layers. The scenario changes significantly as the nearly surface-parallel plane moves further away, the flow becomes extremely intermittent, as present in figure \ref{Normal_sections_Schematics_PertW}$(c)$.

The temperature field in the same layers (Figure \ref{Normal_sections_Schematics_PertT}$(a-c)$) also reveals a similar picture.
The figures $(a)$ and $(b)$ exhibit clear streak structure, which is qualitatively similar to that of spanwise velocity fluctuations $w^{\prime}$.
Upon closer inspection, one can observe a close correspondence between zones with positive temperature fluctuations (red regions) and low-speed streaks (blue regions), and vice versa. 
This is a typical manifestation of the well-known tendency for velocity and temperature fluctuations in shear flows to be negatively correlated. 
In this case, it is a consequence of the fact that outward wall-normal motions transport negative velocity fluctuations and positive temperature fluctuations from the inner, low-speed, high-temperature layers to the upper layers. Similar observations have been elucidated in previous studies in flat-plate boundary layer by \cite{Pirozzoli2011}. 

All these observations tell us that the typical near-wall disturbance structures in two-dimensional and three-dimensional turbulent boundary layers are very similar. However, due to the modulation effect of the mean flow, these flow structures exhibit a strong tendency to align with the direction of the external flow.

\subsection{Law-of-the-wall and velocity-temperature relationships}
\subsubsection{Law-of-the-wall}

All prior researches on mean velocity transformations in compressible turbulent boundary layers was mainly confined to statistically two-dimensional boundary layers, hence their performance in actual three-dimensional boundary layers remains largely unknown. Here, we will conduct a study of velocity transformations, in an attempt to further elucidate the impact of three-dimensional effects and pressure gradients on the effectiveness of these compressibility transformations.

As analyzed in the previous mean flow analysis, in a typical three-dimensional boundary layer, the velocity vector parallel to the wall can point in any direction, and this direction changes with the height away from the wall. 
Therefore, directly transforming the mean velocity parallel to the wall, as done in a two-dimensional flat-plate boundary layer, is not meaningful, since the direction of the velocity vector is not uniform. 
Thus, we need to decompose this velocity vector into two different directions to ensure the uniformity of the velocity vector direction. 
One can simply decompose and project the velocity vector onto the spanwise $z-$ direction and the chordwise $\xi-$direction. 
The reason to chose these two directions is that, in the current three-dimensional boundary layer, the pressure gradient exists only in the $x-y$ plane and not in the spanwise $z-$direction. By using this decomposition, one can decouple the pressure gradient effects from the flow in a specific direction.
Therefore, in each chordwise location, except the exact attachment line, two velocities $w$ and $u_\xi$ along the $z$ and $\xi$ directions, are used to assess various compressibility scaling relations in case of three dimensional flows. However, from a statistical perspective, in regions very close to the attachment line, the values of $u_\xi$ are very small over a substantial area, thus showing limited significance. Instead, we chose the mainstream velocity direction $u_s$ parallel to the external streamline as the second velocity vector to characterize the three-dimensional flow

The van Driest transformation\citep{Van1951} is widely used to transform the mean velocity profile in a compressible boundary layer to a profile from an incompressible boundary layer by taking the variations of mean density into account. 
In general three-dimensional boundary layer, the velocity vector parallel to the wall surface can be 
In the present three dimensional boundary layer, the similar form in the present study can be taken as:
\begin{equation}
w_{V_D}^+ = \int_0^{\overline{w}^+} \sqrt{\frac{\overline{\rho}}{\overline{\rho}_w}}d{\overline{w}^+},\quad
u_{s V_D}^+ = \int_0^{\overline{u_{s}}^+} \sqrt{\frac{\overline{\rho}}{\overline{\rho}_w}}d{\overline{u_{s}}^+}
\end{equation}   
This transoformation works well in collapse velocity profiles in normal adiabatic boundary layer flows. However, in diabatic boundary layers, the transformation fails and can not collapse into the incompressible law of the wall.
\citet{Trettel2016} proposed a new transformation based on the stress-balance conditions and can be written as 
\begin{equation}
\left.
\begin{aligned}
w_{T_L}^+ &= \int_0^{\overline{w}^+}\sqrt{\frac{\overline{\rho}}{\overline{\rho}_w}}
\left[1 + \frac{1}{2\overline{\rho}}\frac{\partial \overline{\rho}}{\partial h}h - \frac{1}{\overline{\mu}}\frac{\partial \overline{\mu}}{\partial h}h \right]d\overline{w}^+, \\
u_{s T_L}^+ &= \int_0^{\overline{u}_{s}^+}\sqrt{\frac{\overline{\rho}}{\overline{\rho}_w}}
\left[1 + \frac{1}{2\overline{\rho}}\frac{\partial \overline{\rho}}{\partial h}h - \frac{1}{\overline{\mu}}\frac{\partial \overline{\mu}}{\partial h}h \right]d\overline{u}_{s}^+, \\
h^* &= \frac{\overline{\rho}}{\overline{\mu}}\sqrt{\frac{\overline{\tau}_w}{\overline{\rho}}}h
\end{aligned}
\right\}
\end{equation}
The Trettle \& Larsson transformation is functioning well in collapsing the profiles for a non-adiabatic turbulent boundary layer to an equivalent incompressible profile, especially for channel flow. Their performances in boundary layers are not that good. To improve the overall collaspe of the mean velocity profile in boundary layers, \citet{Volpiani2020} developed a new transformation based on data-driven approaches, this transformation can be written as
\begin{equation}
\begin{aligned}
w_V^+ = \int_{0}^{\overline{w}^+}\sqrt{\frac{\overline{\rho}/\overline{\rho}_w}{\overline{\mu}/\overline{\mu}_w}}d\overline{w}^+, 
u_{s V}^+ &= \int_{0}^{\overline{u}_{s}^+}\sqrt{\frac{\overline{\rho}/\overline{\rho}_w}{\overline{\mu}/\overline{\mu}_w}}d\overline{u}_{s}^+, 
h_V^+ = \int_{0}^{\overline{h}^+}\sqrt{\frac{\overline{\rho}/\overline{\rho}_w}{(\overline{\mu}/\overline{\mu}_w)^3}}d h^+
\end{aligned}
\end{equation}
\citet{Griffin2021} further introduced a new transformation, which involves scaling the viscous stress through semi-local non-dimensionalization within the viscous sublayer, and addressing the Reynolds shear stress to preserve the approximate balance between turbulence production and dissipation in the log-law region. This transformation has been demonstrated to effectively unify the velocity profiles across a diverse array of flows, encompassing heated, cooled, and adiabatic boundary layers, as well as fully developed channel and pipe flows. The transformation is referred to the total-stress-based transformation, can be written as
\begin{equation}
\left.
\begin{aligned}
w_{T_S}^+ &= \int_0^{w^+}\frac{S_{eq,w}^+}{1 + S_{eq,w}^+ - S_{T_L,w}^+}dh^*,
S_{eq,w}^+ =\frac{\overline{\mu}_w}{\overline{\mu}}\frac{\partial w^+}{\partial h^*},
S_{T_L,w}^+ = \frac{\overline{\mu}}{\overline{\mu}_w}\frac{\partial w^+}{\partial h^+},\\
u_{s,T_S}^+ &= \int_0^{u^+_s}\frac{S_{eq,u_{s}}^+}{1 + S_{eq,u_{s}}^+ - S_{T_L,u_{s}}^+}dh^*, 
S_{eq,u_{s}}^+ = \frac{\overline{\mu}_w}{\overline{\mu}}\frac{\partial u_{s}^+}{\partial h^*},
S_{T_L,u_{s}}^+ =  \frac{\overline{\mu}}{\overline{\mu}_w}\frac{\partial u_{s}^+}{\partial h^+},\\
\end{aligned}
\right\}.
\end{equation}

\begin{figure}
  \centering
  \begin{tabular}{cc}
  \begin{overpic}[width=0.43\textwidth]{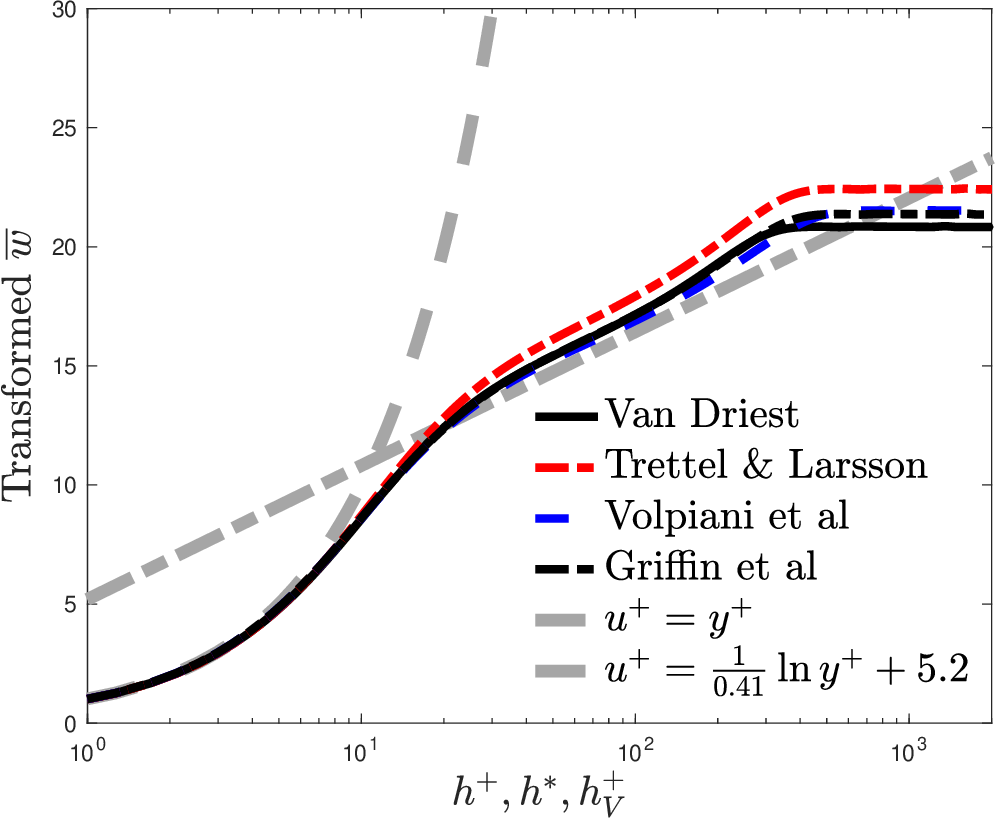}
  \put(10,75){$(a)$}
  \end{overpic} &
  \begin{overpic}[width=0.43\textwidth]{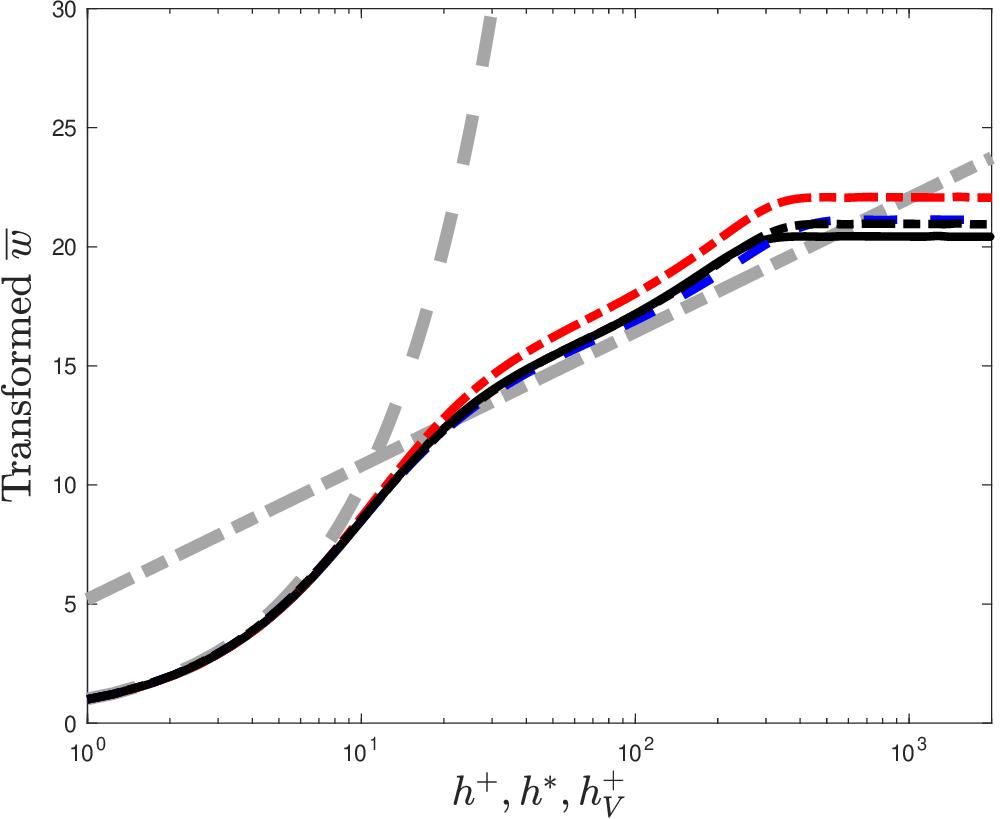}
  \put(10,75){$(b)$}
  \end{overpic} \\
  \begin{overpic}[width=0.43\textwidth]{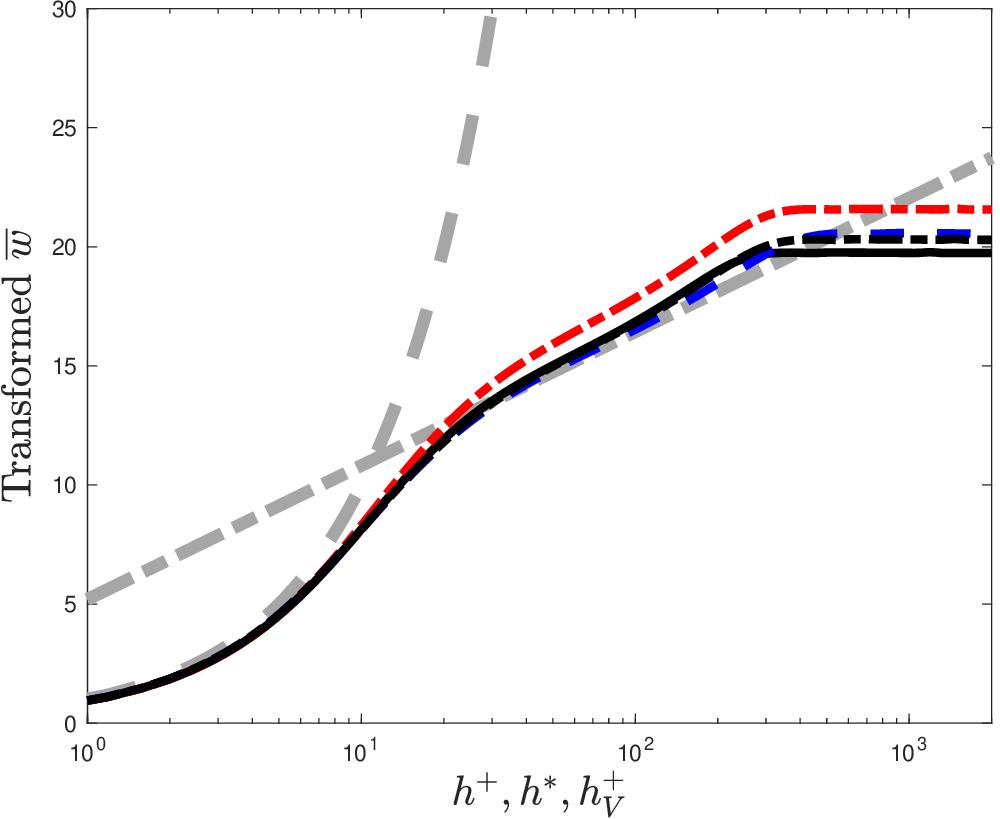}
  \put(10,75){$(c)$}
  \end{overpic} &
  \begin{overpic}[width=0.43\textwidth]{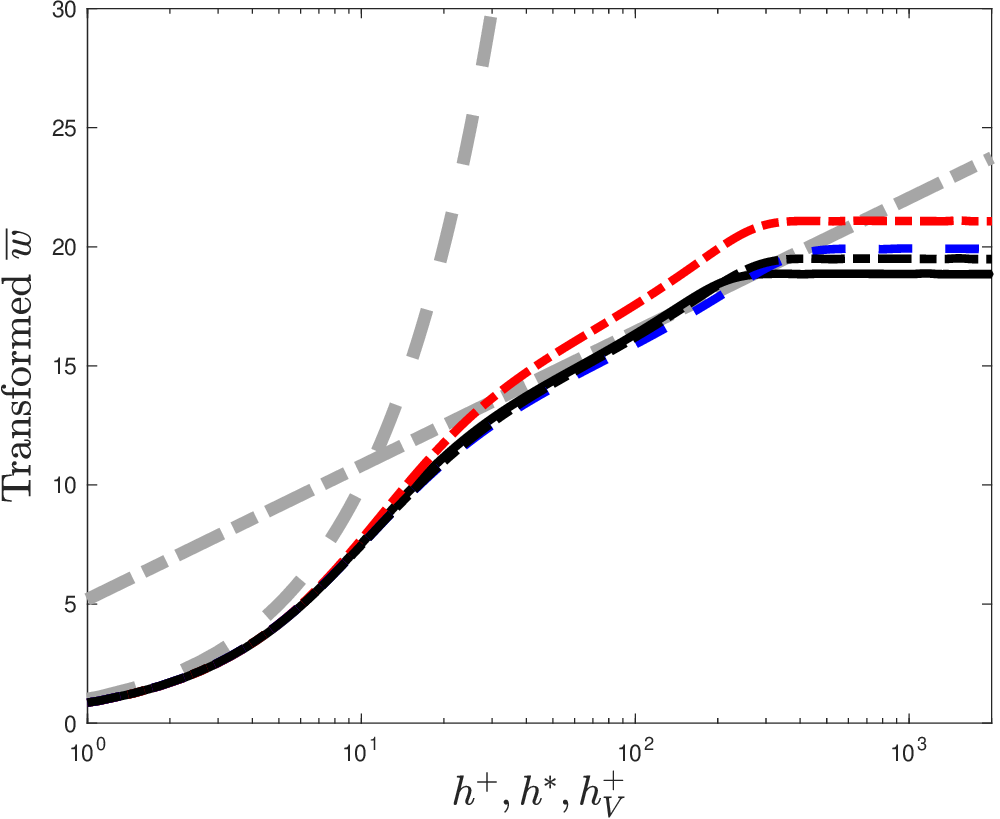}
  \put(10,75){$(d)$}
  \end{overpic} \\
  \begin{overpic}[width=0.43\textwidth]{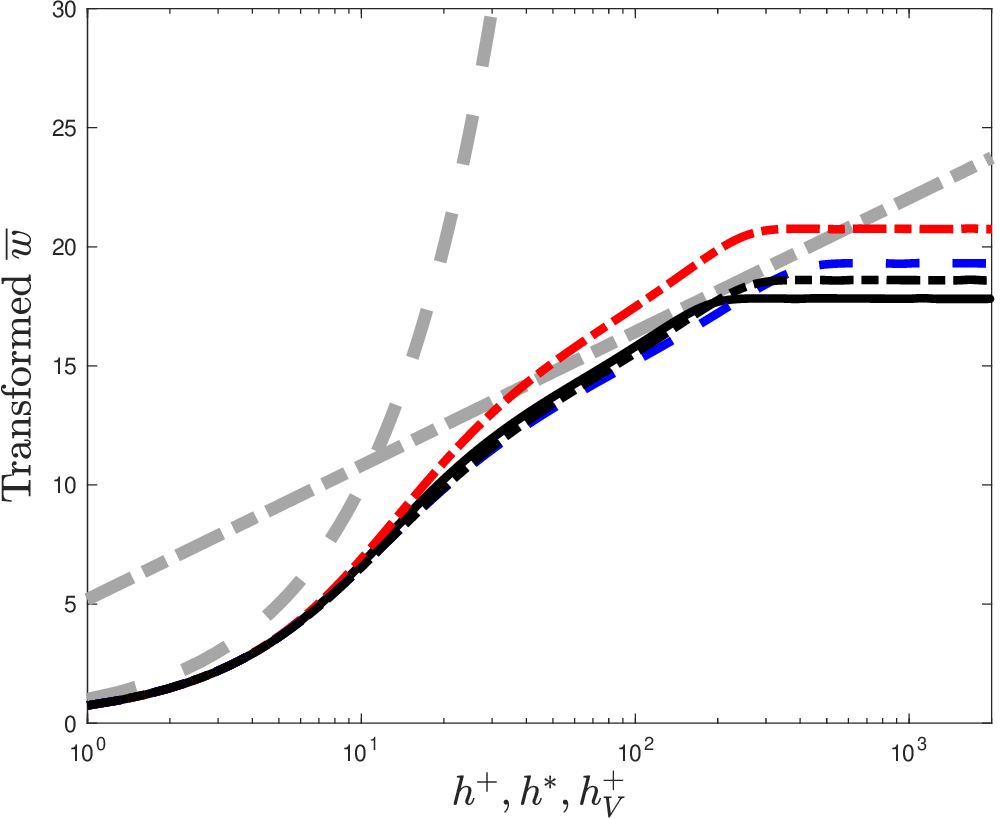}
  \put(10,75){$(e)$} 
  \end{overpic} &
  \begin{overpic}[width=0.43\textwidth]{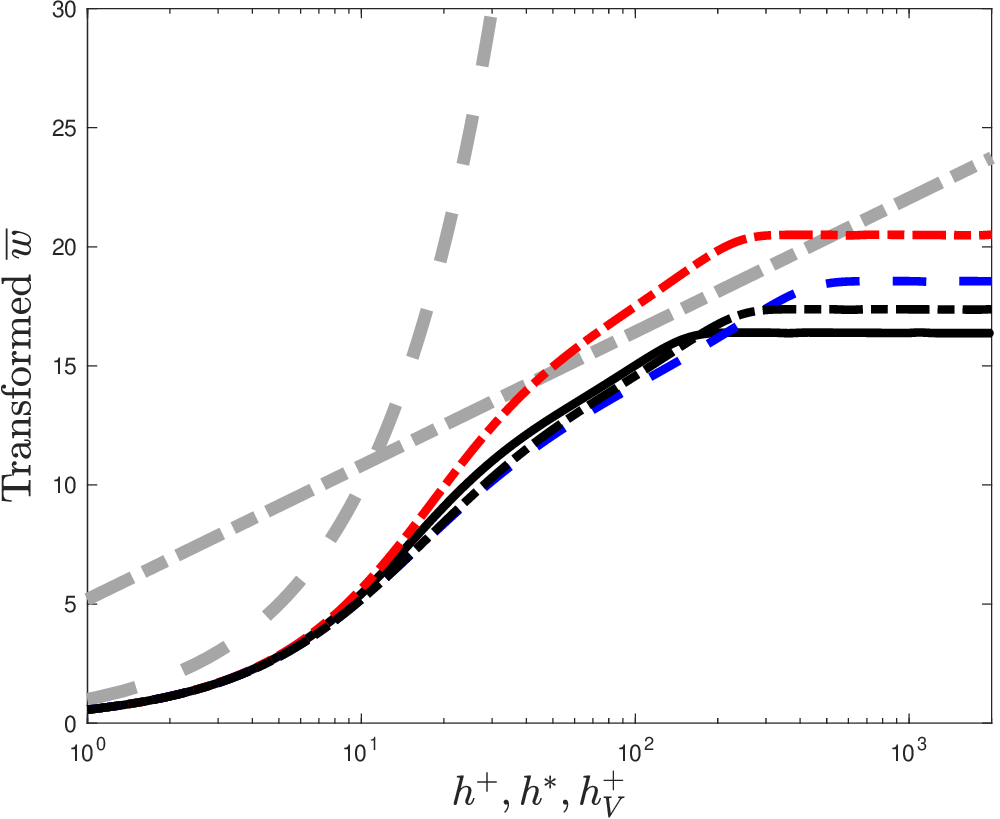}
  \put(10,75){$(f)$}
  \end{overpic} \\
  \begin{overpic}[width=0.43\textwidth]{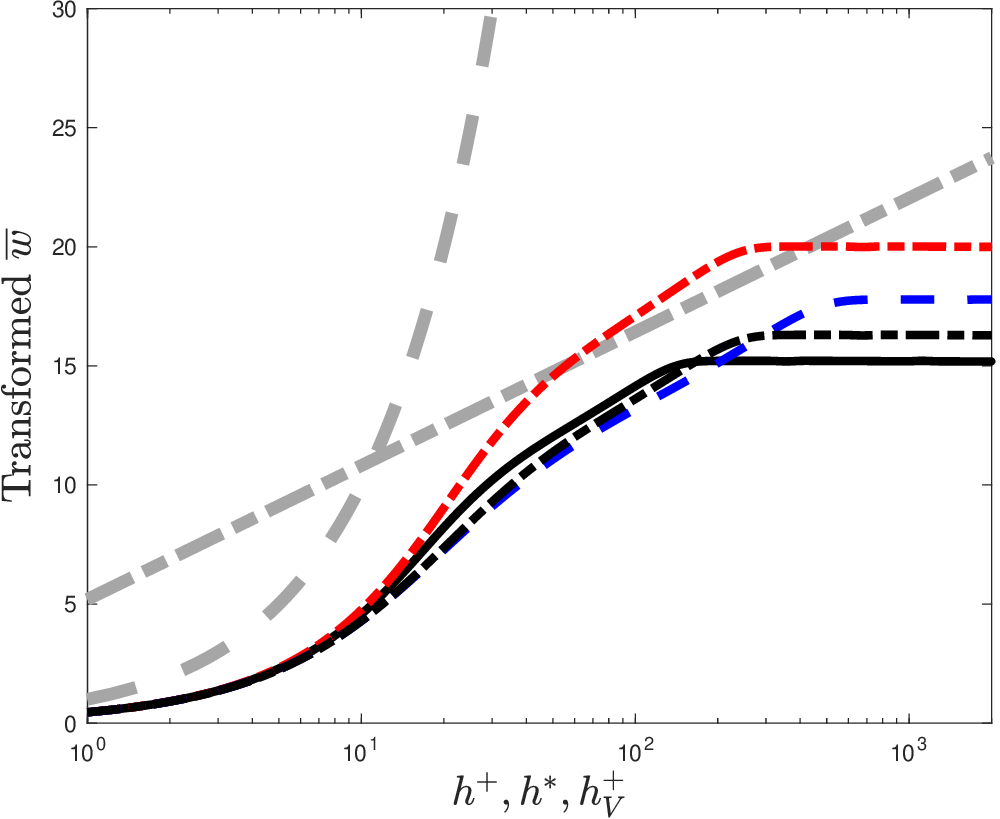}
  \put(10,75){$(g)$} 
  \end{overpic} &
  \begin{overpic}[width=0.43\textwidth]{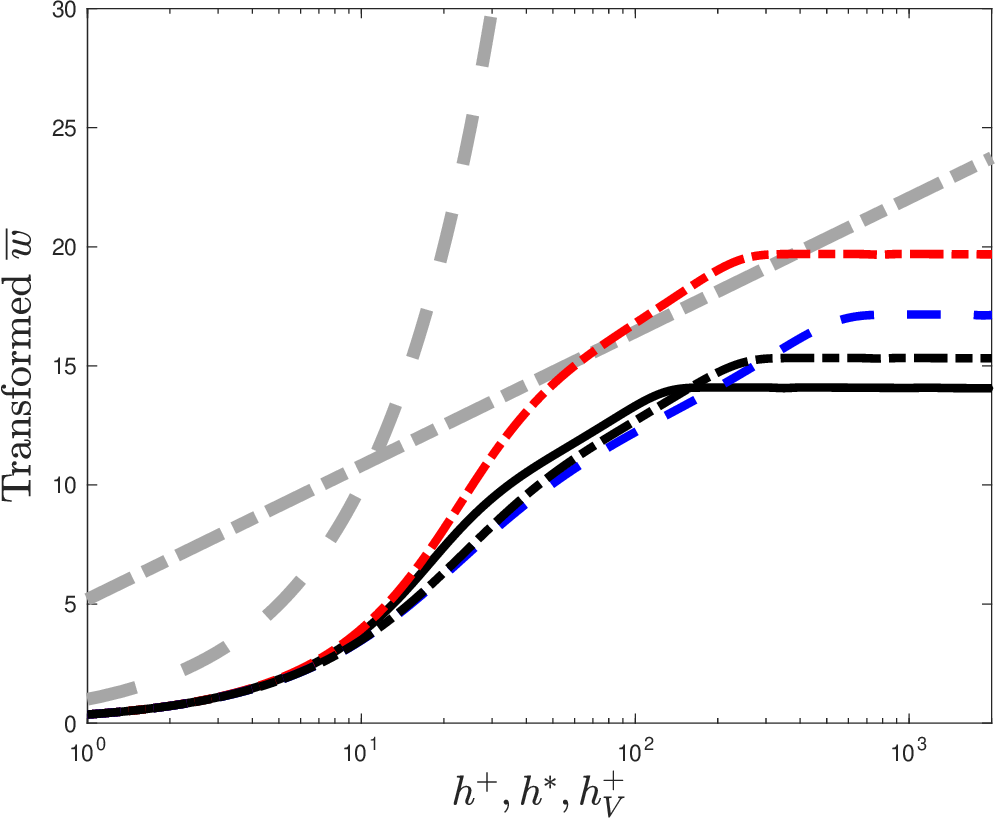}
  \put(10,75){$(h)$}
  \end{overpic}
  \end{tabular}
  \caption{Spanwise mean-velocity profiles, transformed according to several transformations, at different chordwise locations. $(a)-(h)$ stand for the location at chordwise $s_{\xi}=0.00,~4.22,~8.71,~13.50,~18.61,~24.10,~26.99$ and $30.00$, respectively.}
  \label{Mean_Profiles_Transformation}
\end{figure}

\begin{figure}
  \centering
  \begin{tabular}{cc}
  \begin{overpic}[width=0.43\textwidth]{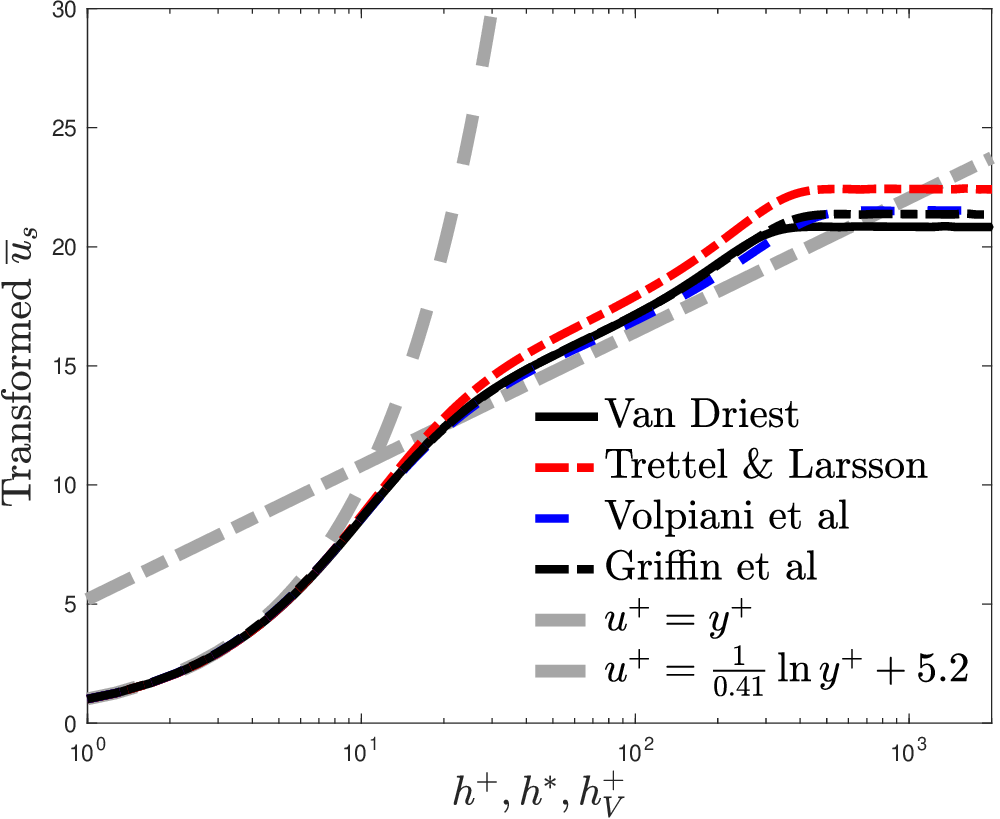}
  \put(10,75){$(a)$}
  \end{overpic} &
  \begin{overpic}[width=0.43\textwidth]{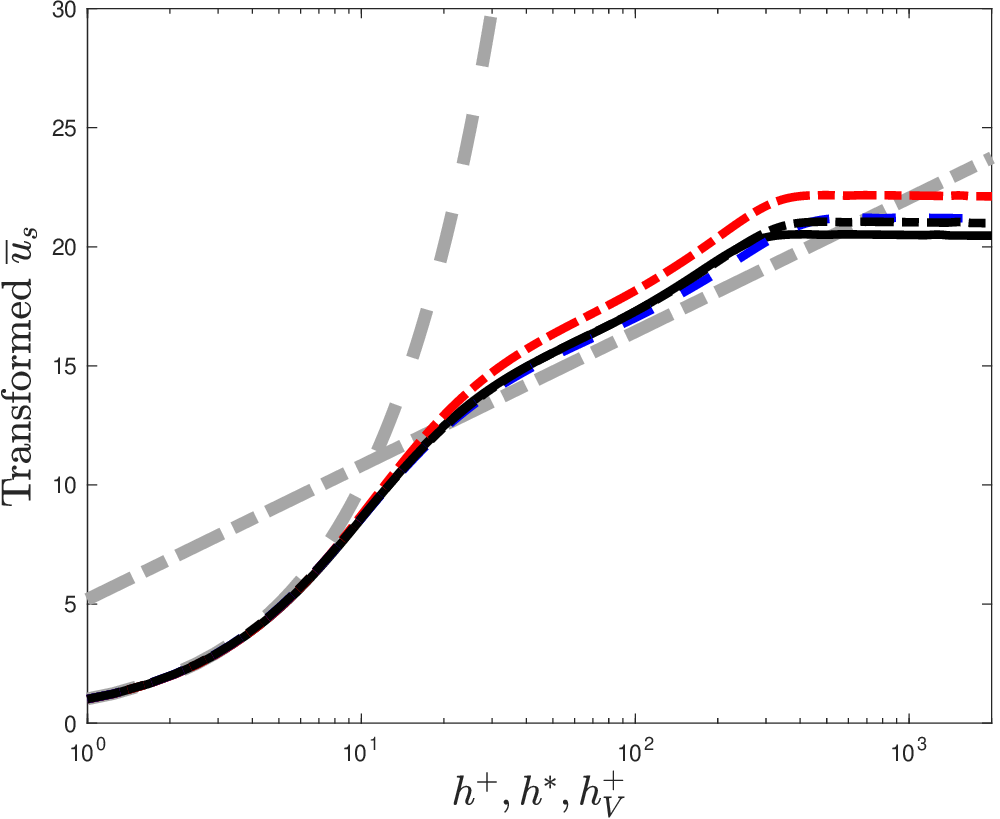}
  \put(10,75){$(b)$}
  \end{overpic} \\
  \begin{overpic}[width=0.43\textwidth]{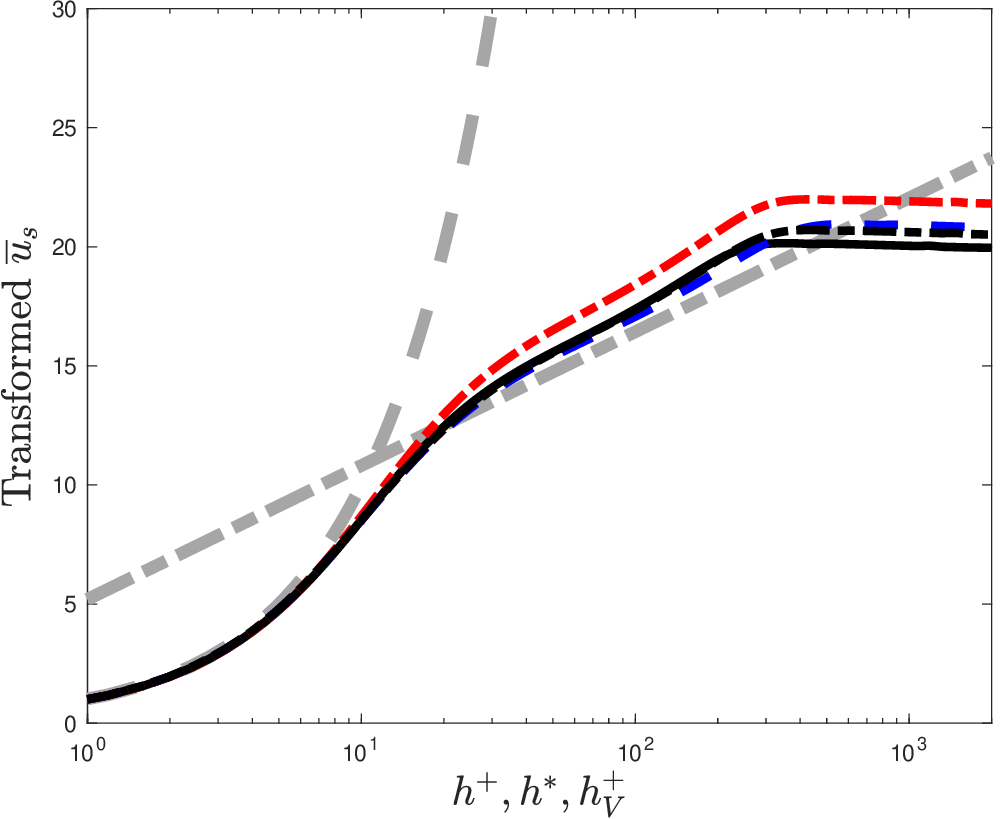}
  \put(10,75){$(c)$}
  \end{overpic} &
  \begin{overpic}[width=0.43\textwidth]{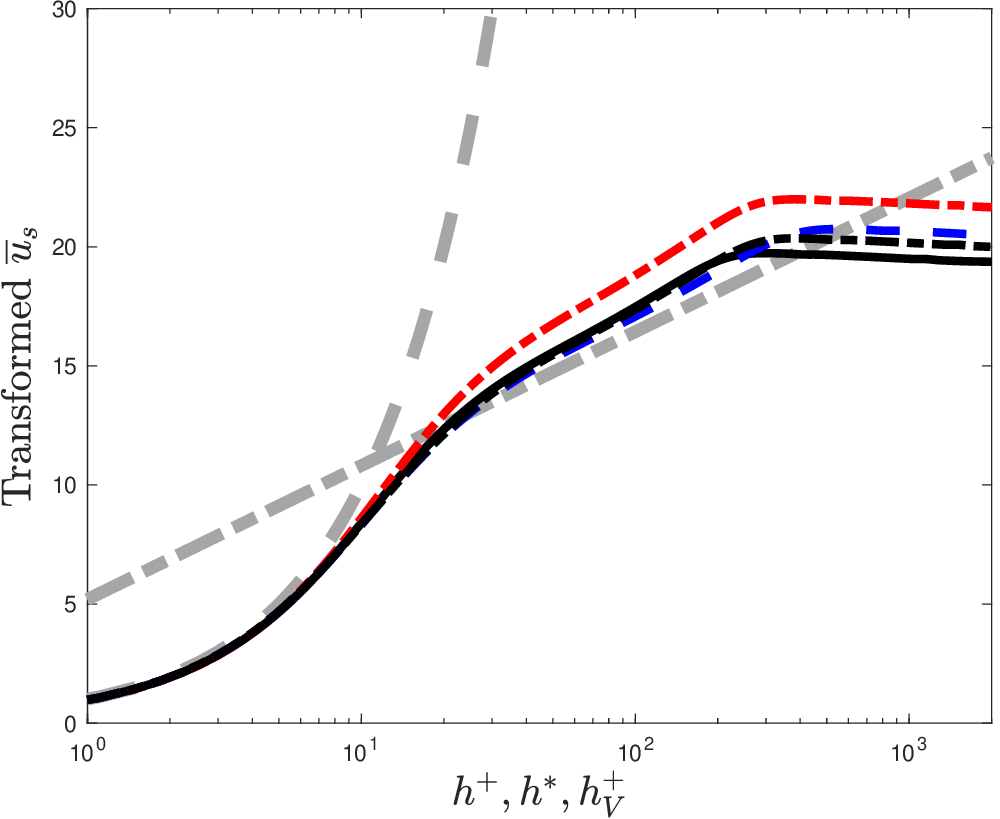}
  \put(10,75){$(d)$}
  \end{overpic} \\
  \begin{overpic}[width=0.43\textwidth]{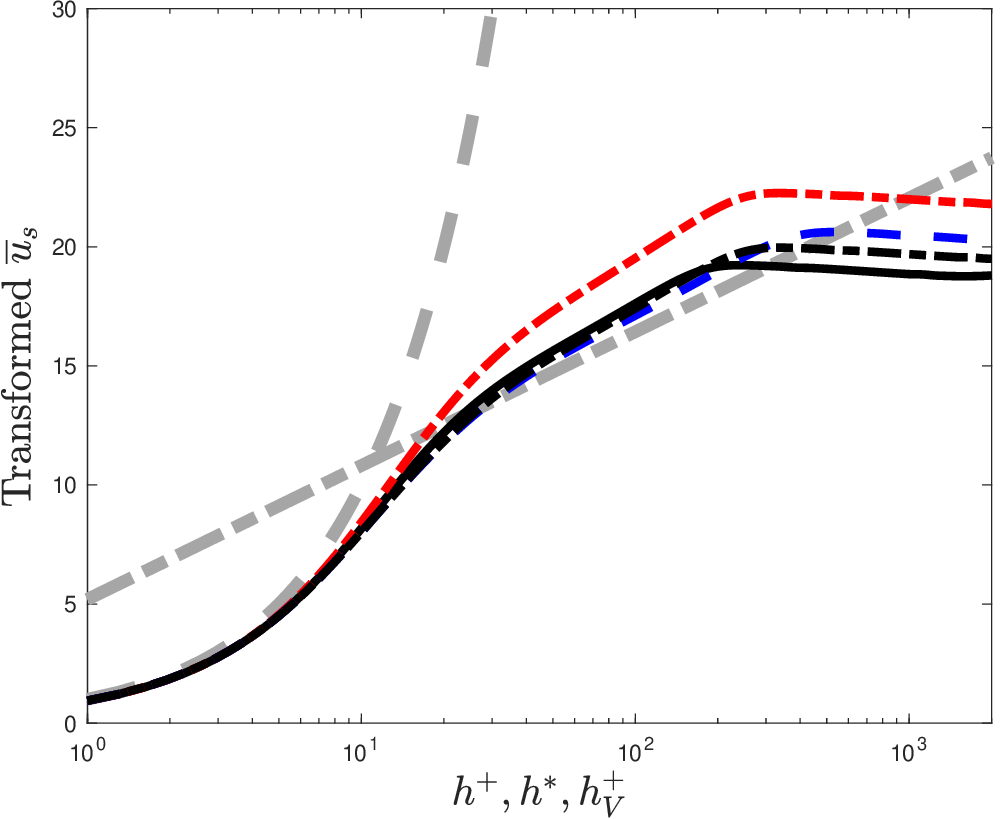}
  \put(10,75){$(e)$} 
  \end{overpic} &
  \begin{overpic}[width=0.43\textwidth]{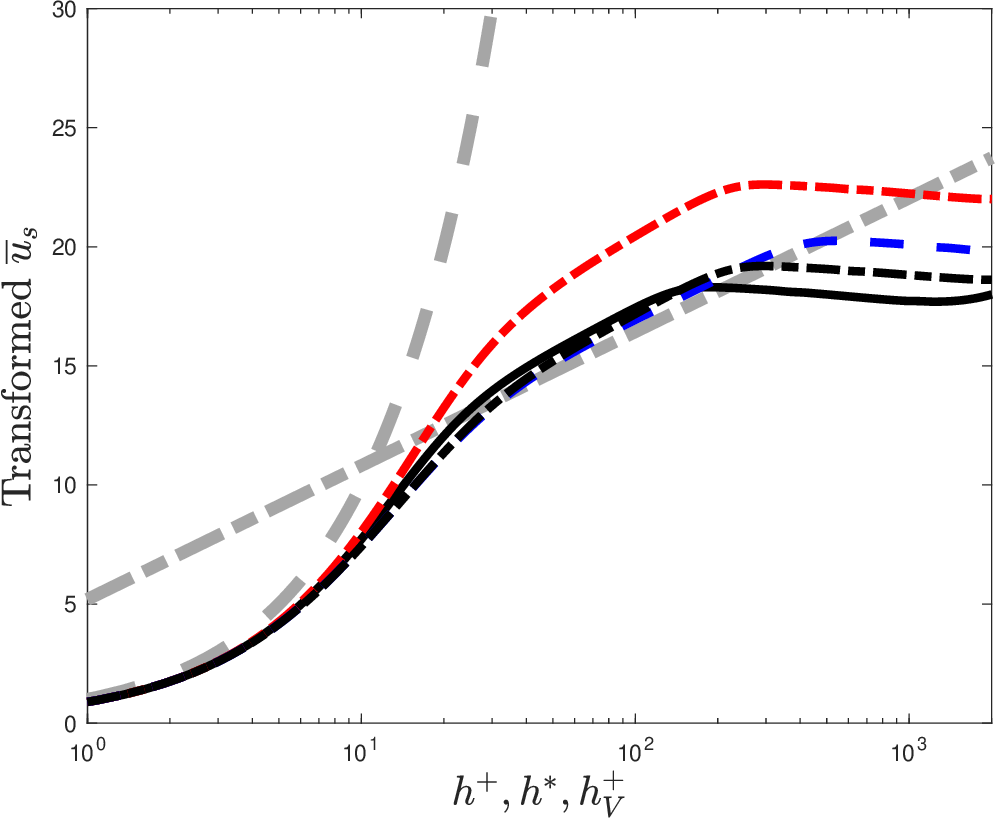}
  \put(10,75){$(f)$}
  \end{overpic} \\
  \begin{overpic}[width=0.43\textwidth]{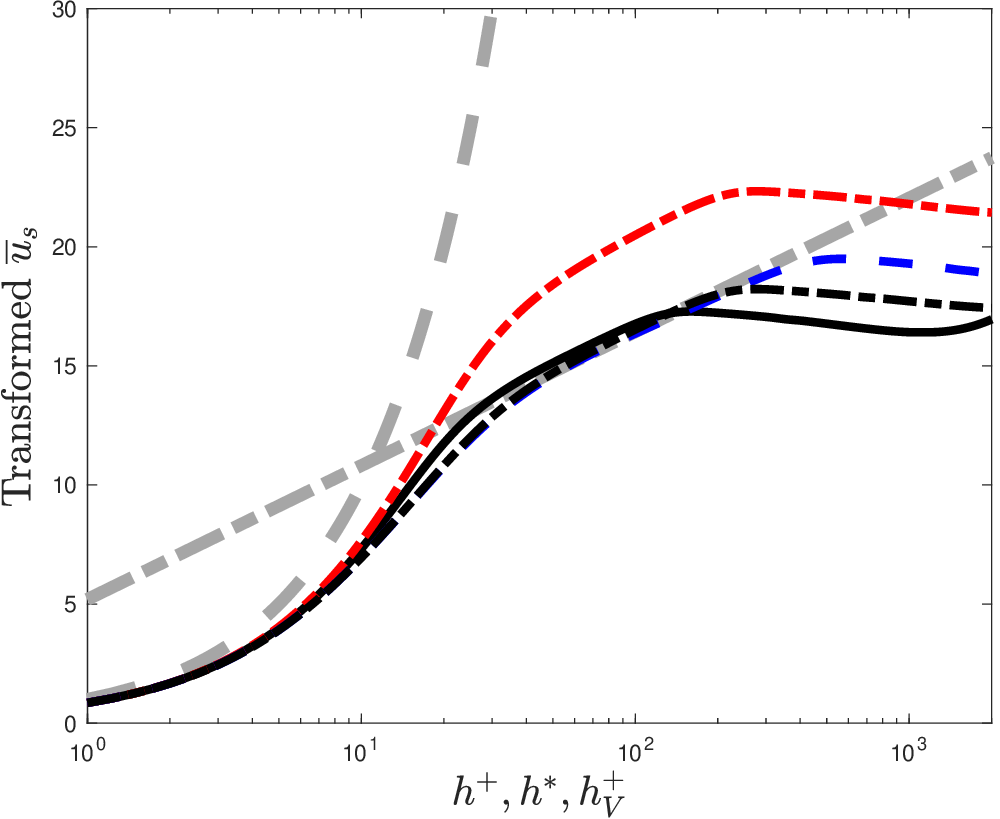}
  \put(10,75){$(g)$} 
  \end{overpic} &
  \begin{overpic}[width=0.43\textwidth]{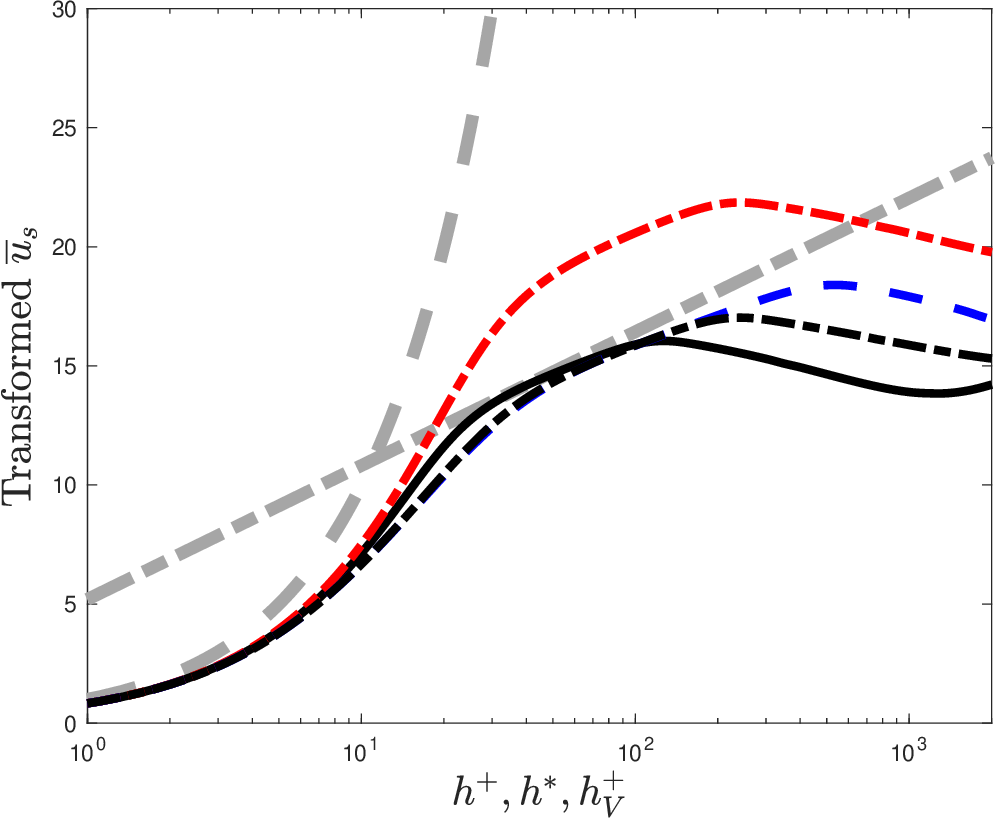}
  \put(10,75){$(h)$}
  \end{overpic}
  \end{tabular}
  \caption{Streamwise mean-velocity profiles, transformed according to several transformations, at different chordwise locations. $(a)-(h)$ stand for the location at chordwise $s_{\xi}=0.00,~4.22,~8.71,~13.50,~18.61,~24.10,~26.99$ and $30.00$, respectively.}
  \label{Mean_Profiles_Transformation_Us}
\end{figure}

% Mean Velocity Profiles Results
The transformed spanwise and streamwise velocity profiles at several chordwise locations are shown in figures \ref{Mean_Profiles_Transformation} and \ref{Mean_Profiles_Transformation_Us}, ranging from the exact attachment line to downstream chordwise direction. Firstly, we look at the profiles at the exact attachment line (Figure~\ref{Mean_Profiles_Transformation}$(a)$ and Figure~\ref{Mean_Profiles_Transformation_Us}$(a)$).
Excluding the Trettel \& Larsson transformation, the other three transformations yielded satisfactory results. Based on previous research, the Trettel \& Larsson transformation is expected to perform well in fully developed, non-spatially varying boundary layers such as channel flow; however, in the conditions investigated in this study, it inaccurately predicted the logarithmic region. 
It is well-known that the Van Driest transformation generally performs poorly in compressible, heat-exchanging flat-plate turbulent boundary layer flows. Nevertheless, under the conditions of this study, the Van Driest transformation produced acceptable results. As the position gradually moves away from the attachment line to the point of maximum transverse pressure gradient, different velocity transformations continue to exhibit similar patterns to those observed at the attachment line. 
This indicates that in the present fully developed three-dimensional turbulent compressible boundary layer, even in the presence of transverse pressure gradients, the mainstream velocity component still largely conforms to the velocity transformations  previously established for flat-plate boundary layers. Meanwhile, it is important to note that, under the conditions studied, the transverse velocity components gradually increase from zero along the chordwise direction starting from the attachment line. Therefore, the chordwise velocities at the locations shown in figures $(b-d)$ are not very large. As the transverse velocity increases, shown in figures $(e-h)$, the slopes and intercepts of various transformations for spanwise velocity profiles begin to deviate from the classical relationships. Meanwhile, the transformed streamwise velocity profiles still exhibite a clear log-law region, as shown in figure \ref{Mean_Profiles_Transformation_Us}. This indicate that along the streamwise directions, the streamwise mean velocities still exhibit some feature similar to the two-dimensional turbulent boundary layer. When the transverse velocity increases further to a magnitude comparable to the spanwise velocity, the regularity of the entire transformation no longer holds for spanwise velocity. However, the streamwise velocities still show a common behaviour and regions similar to the log-law of the two-dimensional flat-plate boundary layer can also be identified. It should be noted that the results presented here are merely illustrative. Whether a general law-of-the-wall, similar to that in two-dimensional flat-plate boundary layers, exists in three-dimensional boundary layers remains an open question that requires further in-depth investigation.

% Mainstream profiles

\subsubsection{velocity-temperature relationship}
The temperature profiles also play a vital role in compressible boundary, therefore, the widly used relationship between the mean velocity and mean temperature is also shown to assess the scaling relations in three-dimensional boundary layers. The mean temperature-velocity relation model given by \citet{Zhang2014}, which is also known as the generalized reynolds analogy,  is defined as 
\begin{equation}
\left.
\begin{aligned}
\frac{\overline{T}}{T_{e}} &= \frac{T_w}{T_{e}} + \frac{T_{rg} - T_w}{T_{e}}\frac{\overline{u}_p}{\overline{u}_{p,e}}
+ \frac{T_e - T_{rg}}{T_e}\left(\frac{\overline{u}_p}{\overline{u}_{p,e}}\right)^2, \\
T_{rg} &= T_e + \frac{r_g \overline{u}_{p,e}^2}{2 C_p},
r_g = \frac{2 C_p(T_w - \overline{T}_e)}{\overline{u}_{p,e}^2} - 2 Pr \frac{\theta_{tw}}{\overline{u}_{p,e} \overline{\tau}_w},
\end{aligned}
\right\},
\end{equation}
where the subscript $e$ stands for the variables at the edge of boundary layer and the velocity $\overline{u}_p$ represents the velocity parallel to the local surface. It should be noted that, in this study, the mean temperature $T_e$ and velocity $\overline{u}_{p,e}$ outside the boundary layer change with the chordwise position $s_{\xi}$. The relationship between the mean temperature and mean velocity at different chordwise locations is illustrated in the figure \ref{Mean_Profiles_Transformation}. Figure $(a)$ stands for the mean temperature profiles against mean velocity at the exact attachment line $s_{\xi} = 0$ and the chordwise exit $s_{\xi}=30$.
It reveals that the relation proposed by \cite{Zhang2014} yields an acceptable results. In fact, the accurate prediction results at the exact attachment line are as expected, since it statistically returns to a zero-pressure-gradient flat-plate boundary layer. However, achieving good predictive results at the chordwise exit was surprising. This can be attributed to two potential reasons: in the current flow study, the governing spanwise momentum equation for the spanwise velocity $w$ is actually weakly coupled with the other equations, as the boundary layer seems to be homogeneous in spanwise direction. Secondly, at the exit location, although there is a strong crossflow velocity, there is no pressure gradient along either the chordwise or spanwise directions. As a result, the energy generated by the pressure work accumulated upstream is entirely absorbed into the mean flow temperature and velocity.

% Mean Temperature Profiles Results
\begin{figure}
  \centering
  \begin{tabular}{cc}
  \begin{overpic}[width=0.48\textwidth]{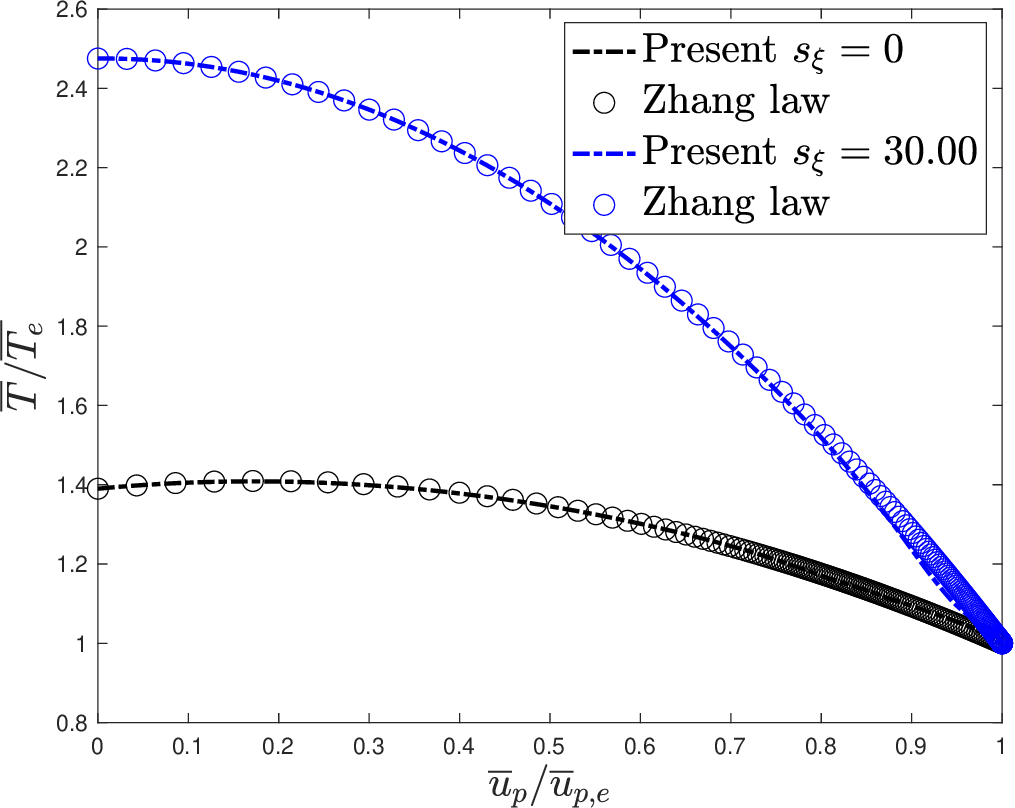}
  \put(12,15){$(a)$}
  \end{overpic} &
  \begin{overpic}[width=0.48\textwidth]{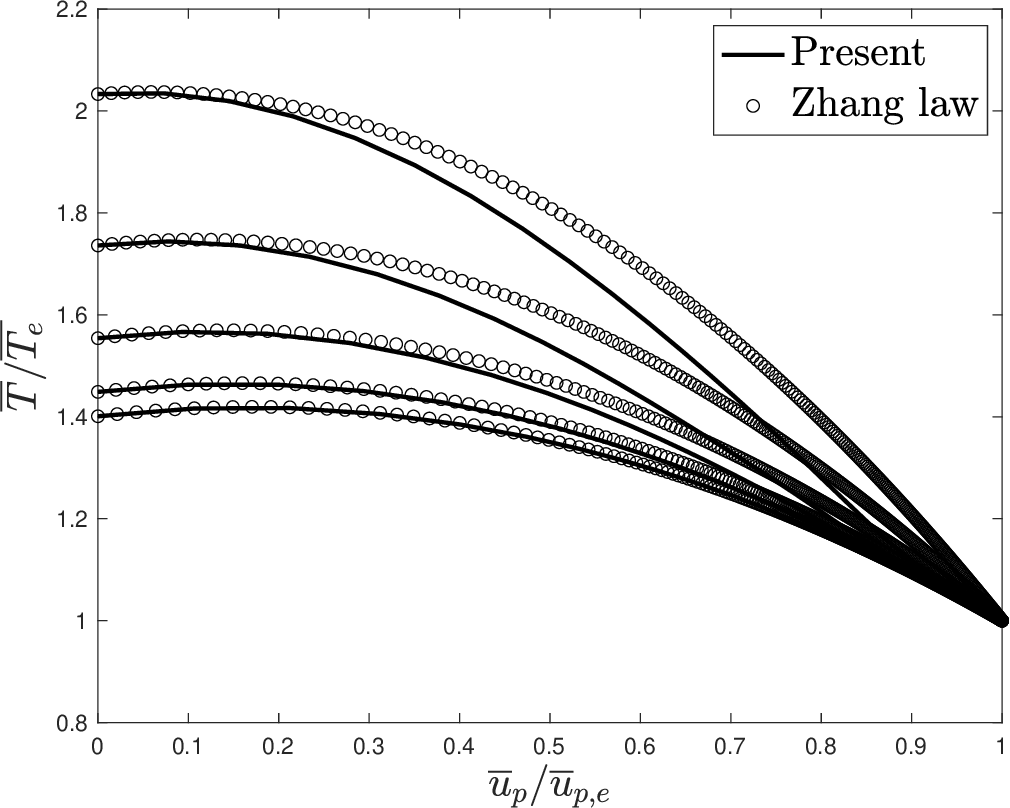}
  \put(12,15){$(b)$}
  \put(40,30){\vector(1,1){30}}
  \put(72,60){$s_{\xi}$ increase}
  \end{overpic}
  \end{tabular}
  \caption{Panels (a) and (b) show the mean temperature profile against mean velocity at several chordwise locations, compared with relation of \citet{Zhang2014}.}
  \label{Mean_Profiles_TV}
\end{figure}

Then we look at the profiles at the region with both crossflow and pressure gradient effect (figure \ref{Mean_Profiles_TV}$(b)$). The figure shows the results from small $s_{\xi}$ to further chordwise downstream, as pointed by the vector. It can be observed that when the pressure gradient and velocity in the transverse direction are relatively small, the temperature-velocity relationship proposed by \cite{Zhang2014} still aligns well with the computational results. As the transverse pressure gradient and transverse velocity increase, the original temperature-velocity relationship begins to deviate in the outer boundary layer region. As the flow further develops downstream, the transverse pressure gradient starts to decrease (figure \ref{Check_Mean_Flow_Cs_Change}$(a)$), although the transverse velocity continues to increase. The deviation between the original temperature-velocity relationship and the computational results in the outer boundary layer region gradually diminishes, eventually returning to the trend indicated by the blue line in figure \ref{Mean_Profiles_TV}$(a)$ at the chordwise exit. These phenomena indicate that the traditional temperature-velocity relationship is significantly influenced by the transverse pressure gradient in three-dimensional boundary layers. 
Attention must also be given to the observation that, in the chordwise direction downstream from the attachment line, the overall turbulence intensity exhibits a decreasing trend. 
Additionally, the chordwise velocity in certain regions remains at relatively low levels compared to the spanwise velocity. 
Therefore, this flow condition might thus differ from typical three-dimensional turbulent boundary layers.

% Why they act as this?
%The spanwise time-averaged momental equation along $z-$direction in the present simulation as well as the normal streamwise momental equation in statistical two-dimensional channel can be written as
%\begin{equation}
%\begin{aligned}
%\bar{\rho} \tilde{u} \frac{\partial \tilde{w}}{\partial x}+\bar{\rho} \tilde{v} \frac{\partial \tilde{w}}{\partial y}&=\left[\frac{\partial \bar{\tau}_{xz}}{\partial x}+\frac{\partial \bar{\tau}_{yz}}{\partial y}\right]-\frac{\partial \bar{\rho} \widetilde{u^{\prime\prime} w^{\prime \prime}}}{\partial x}-\frac{\partial \bar{\rho} \widetilde{v^{\prime \prime} w^{\prime \prime}}}{\partial y}, \\
%\bar{\rho} \tilde{v} \frac{\partial \tilde{u}}{\partial y}&=\frac{\partial \bar{\tau}_{x y}}{\partial y} - \frac{\partial \bar{\rho} \widetilde{u^{\prime \prime} v^{\prime \prime}}}{\partial y}
%\end{aligned}
%\end{equation}
%and the terms that marked red can be 

\subsection{Reynolds stresses}
\begin{figure}
  \centering
  \begin{tabular}{cc}
  \begin{overpic}[width=0.48\textwidth]{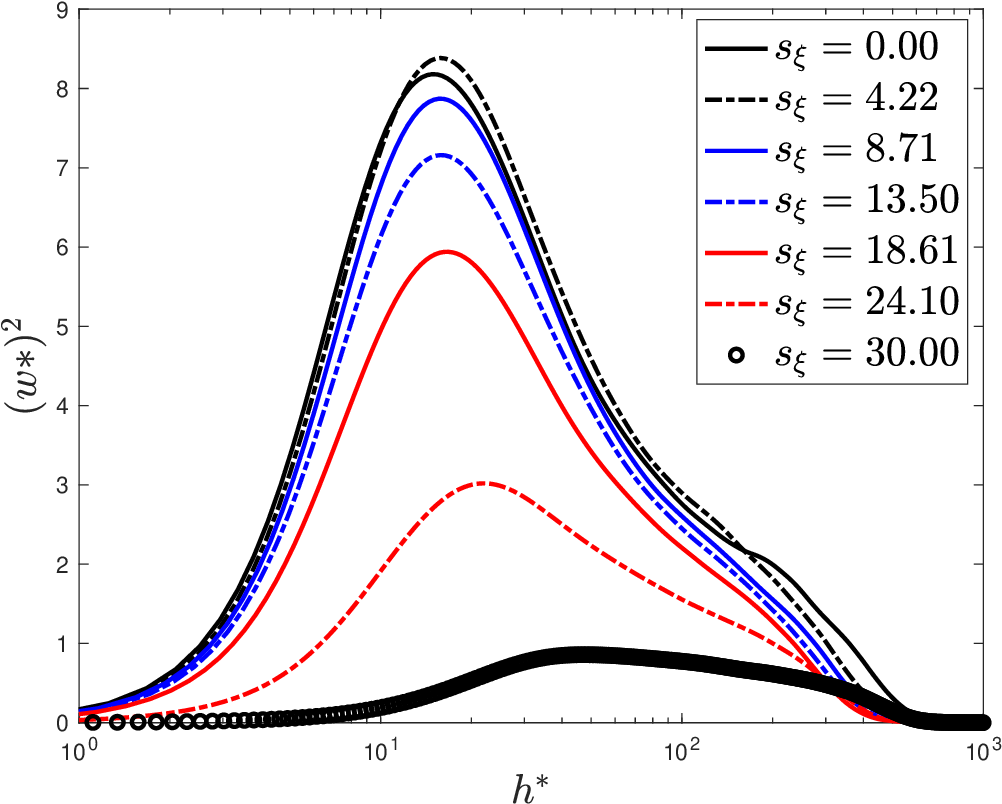}
  \put(-1,75){$(a)$}
  \end{overpic} &
  \begin{overpic}[width=0.48\textwidth]{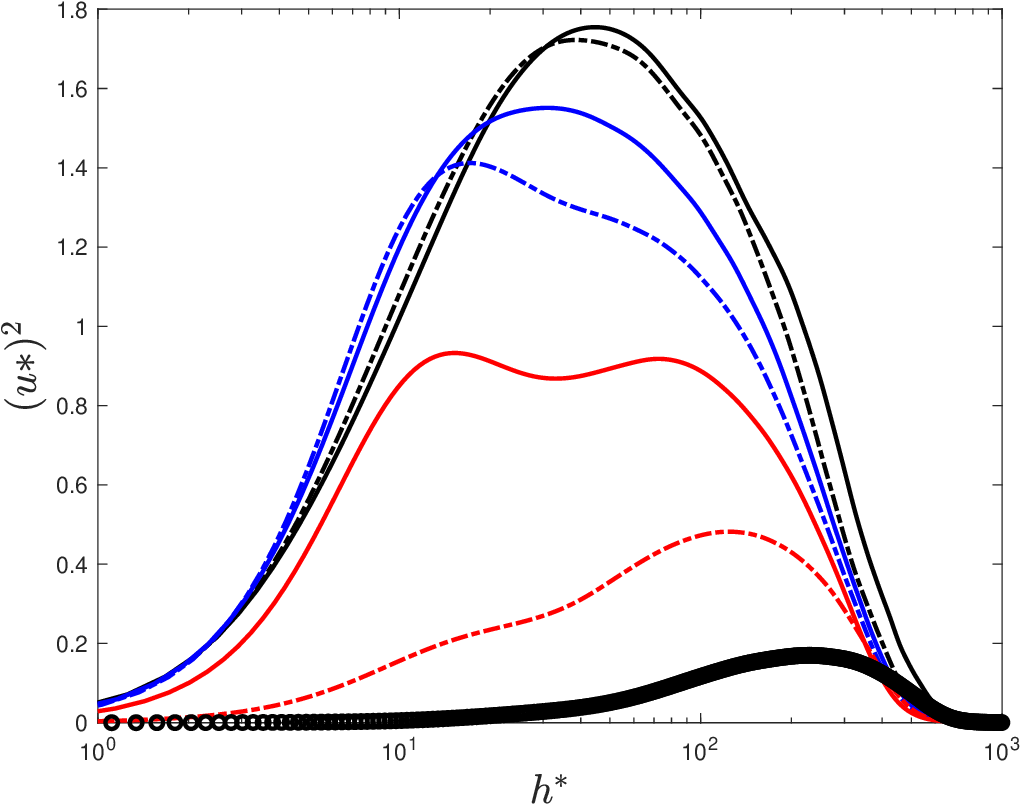}
  \put(-1,75){$(b)$}
  \end{overpic}  \\
  \begin{overpic}[width=0.48\textwidth]{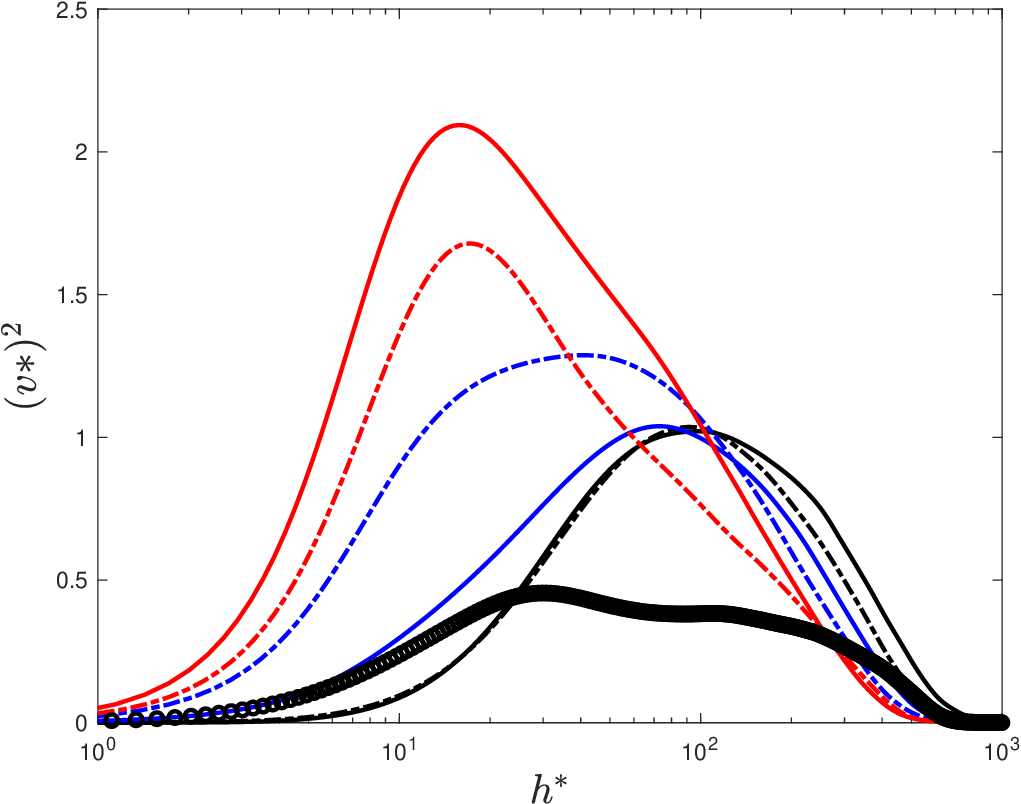}
  \put(-1,75){$(c)$}
  \end{overpic} & 
  \begin{overpic}[width=0.48\textwidth]{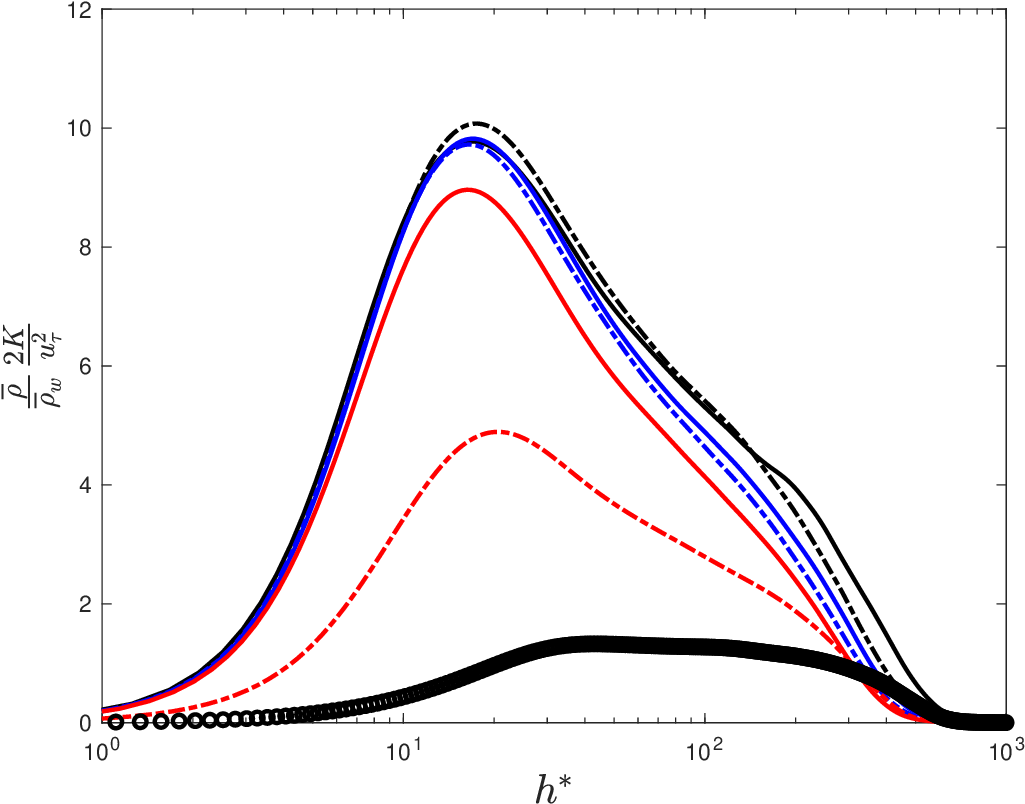}
  \put(-1,75){$(d)$}
  \end{overpic} \\
  \begin{overpic}[width=0.48\textwidth]{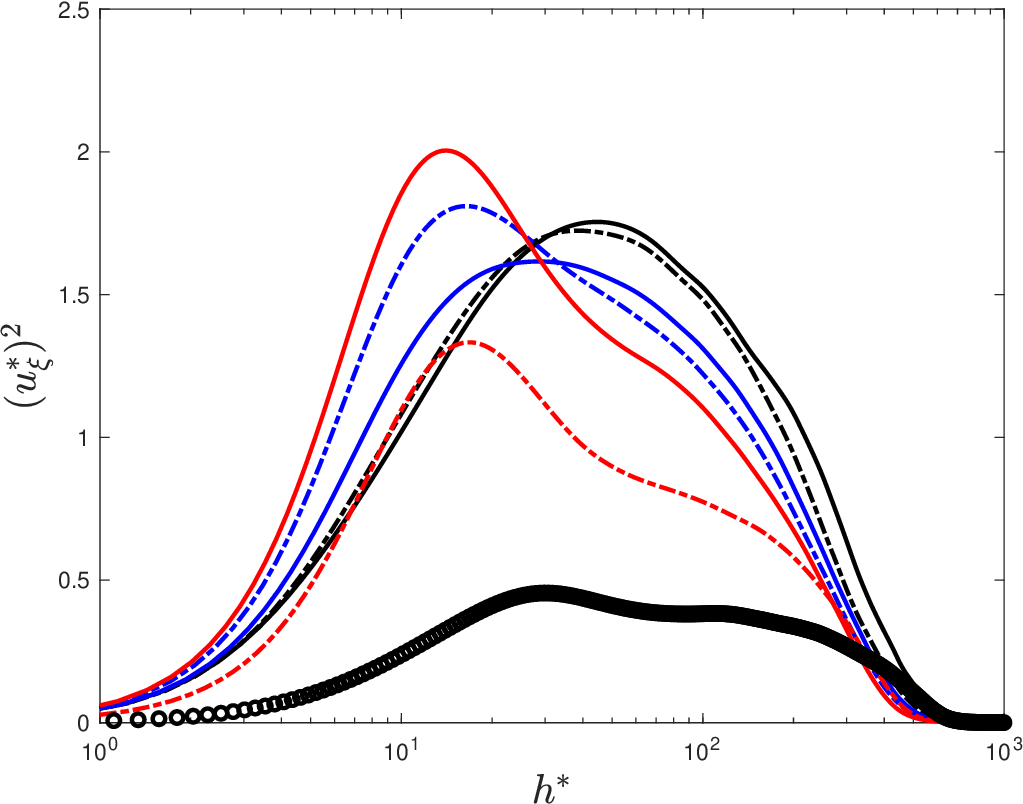}
  \put(-1,75){$(e)$}
  \end{overpic} & 
  \begin{overpic}[width=0.48\textwidth]{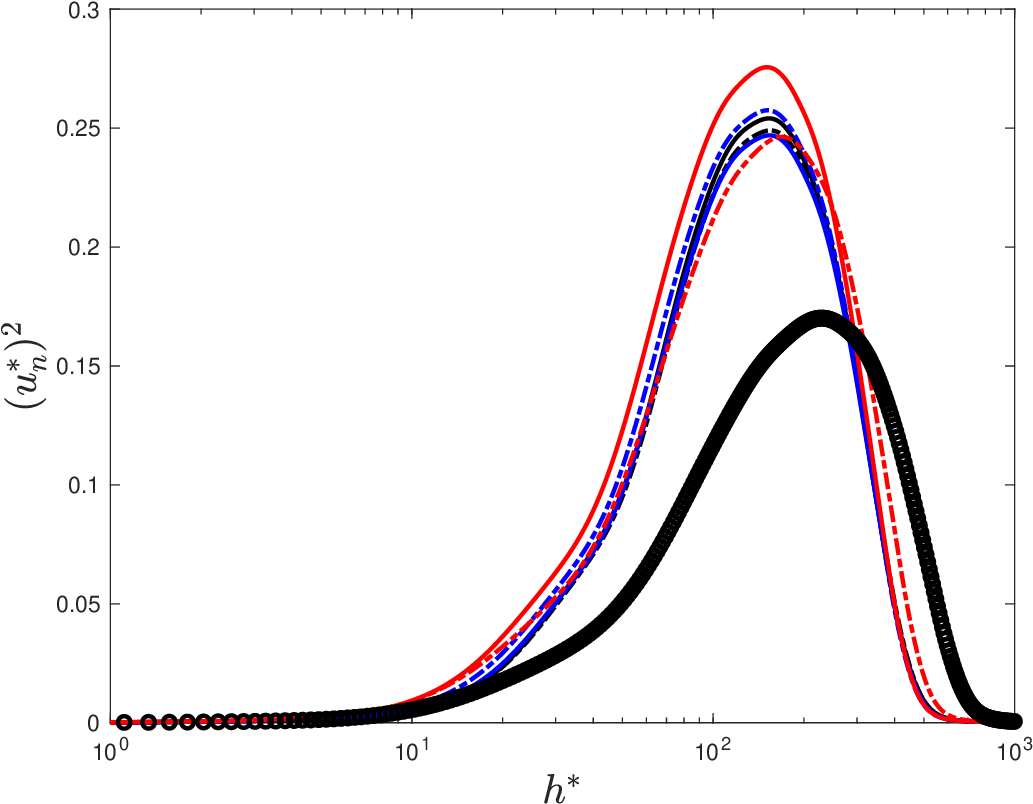}
  \put(-1,75){$(f)$}  
  \end{overpic}
  \end{tabular}
  \caption{Comparison of the scaled Reynolds stress and kinetic energy at different chordwise locations $s_{\xi}$. $(a-f)$ stand for the $(w*)^2, (u*)^2, (v*)^2$, scaled turbulent kinetic energy $2\overline{\rho}K/(\rho_w u_{\tau}^2)$, $(u_{\xi}^*)^2$ and $(u_{n}^*)^2$, respectively.}
  \label{Reynold_Stress_Variations}
\end{figure}

% Scaled Reynolds stresses
In this part, we further look at the features of Reynolds stresses of the present simulations. The scaled Reynolds stresses at the exact attachment line are shown in previous and are found in good agreements with the results from incompressible and compressible flat-plate boundary layer. As the transverse pressure gradient begin to increase, the scaled Reynolds stress are shown in figure \ref{Reynold_Stress_Variations}. Two more scaled Reynolds stresses are defined based on the direction of surface coordinate as follows
\begin{equation}
\left.
\begin{aligned}
\left(u_{\xi}^*\right)^2&=\frac{\overline{\rho}}{\overline{\rho}_w}\frac{\widetilde{u_{\xi}^{\prime\prime}u_{\xi}^{\prime\prime}}}{\overline{u}_\tau^2}, \widetilde{u_{\xi}^{\prime\prime}u_{\xi}^{\prime\prime}}
= \bm{t}_x^2  \widetilde{u^{\prime\prime}u^{\prime\prime}}
+ \bm{t}_y^2  \widetilde{v^{\prime\prime}v^{\prime\prime}}
+2 \bm{t}_x \bm{t}_y \widetilde{u^{\prime\prime}v^{\prime\prime}}, \\
\left(u_{n}^*\right)^2&=\frac{\overline{\rho}}{\overline{\rho}_w}\frac{\widetilde{u_{n}^{\prime\prime}u_{n}^{\prime\prime}}}{\overline{u}_\tau^2}, 
\widetilde{u_{n}^{\prime\prime}u_{n}^{\prime\prime}}
= \bm{n}_x^2  \widetilde{u^{\prime\prime}u^{\prime\prime}}
+ \bm{n}_y^2  \widetilde{v^{\prime\prime}v^{\prime\prime}}
+2 \bm{n}_x \bm{n}_y \widetilde{u^{\prime\prime}v^{\prime\prime}}
\end{aligned}
\right\},
\end{equation}
where $\bm{t}=(\bm{t}_x, \bm{t}_y)$ and $\bm{n} = (\bm{n}_x, \bm{n}_y)$ stands for the unit surface tangent and normal vectors, in the $x-y$ plane. From these figures, the evolution of Reynolds stress with chordwise location is clearly presented. When the pressure gradient first affects the turbulence in the attachment line region, there is a slight increase in turbulence intensity, reflected in the slight rise in spanwise Reynolds stress and turbulent kinetic energy (as shown in figures \ref{Reynold_Stress_Variations}$(a)$ and $(d)$). Subsequently, under the influence of the chordwise favorable pressure gradient, the fluid not only experiences shear in the spanwise direction but also begins to be affected in the chordwise direction. This results in the peak Reynolds stress in the $x-y$ plane shifting towards the wall (figures \ref{Reynold_Stress_Variations}$(b,c)$). With further acceleration of the fluid in the chordwise direction, chordwise fluctuations become significantly enhanced (figure \ref{Reynold_Stress_Variations}$(e)$). This indicates that the pressure gradient can, to some extent, induce a redistribution of turbulent energy. However, as the fluid continues to accelerate and develops downstream, fluctuations in all directions show a notable decreasing trend. A noticeable phenomenon is that when the fluctuation intensity shows a significant decrease, the primary Reynolds stress along the chordwise direction exhibits a distribution pattern very similar to that of the primary Reynolds stress in a flat-plate boundary layer (figure \ref{Reynold_Stress_Variations}$(e)$). Additionally, the fluctuations in the wall-normal direction do not seem to be noticeably affected by the pressure gradient, apart from the last station (figure \ref{Reynold_Stress_Variations}$(f)$). The variations in general reflect the combined effects of shear caused by the three-dimensional nature of the flow field and the favorable pressure gradient along the chordwise direction.

% anisotropy tensor 
\subsection{Anisotropy tensor}
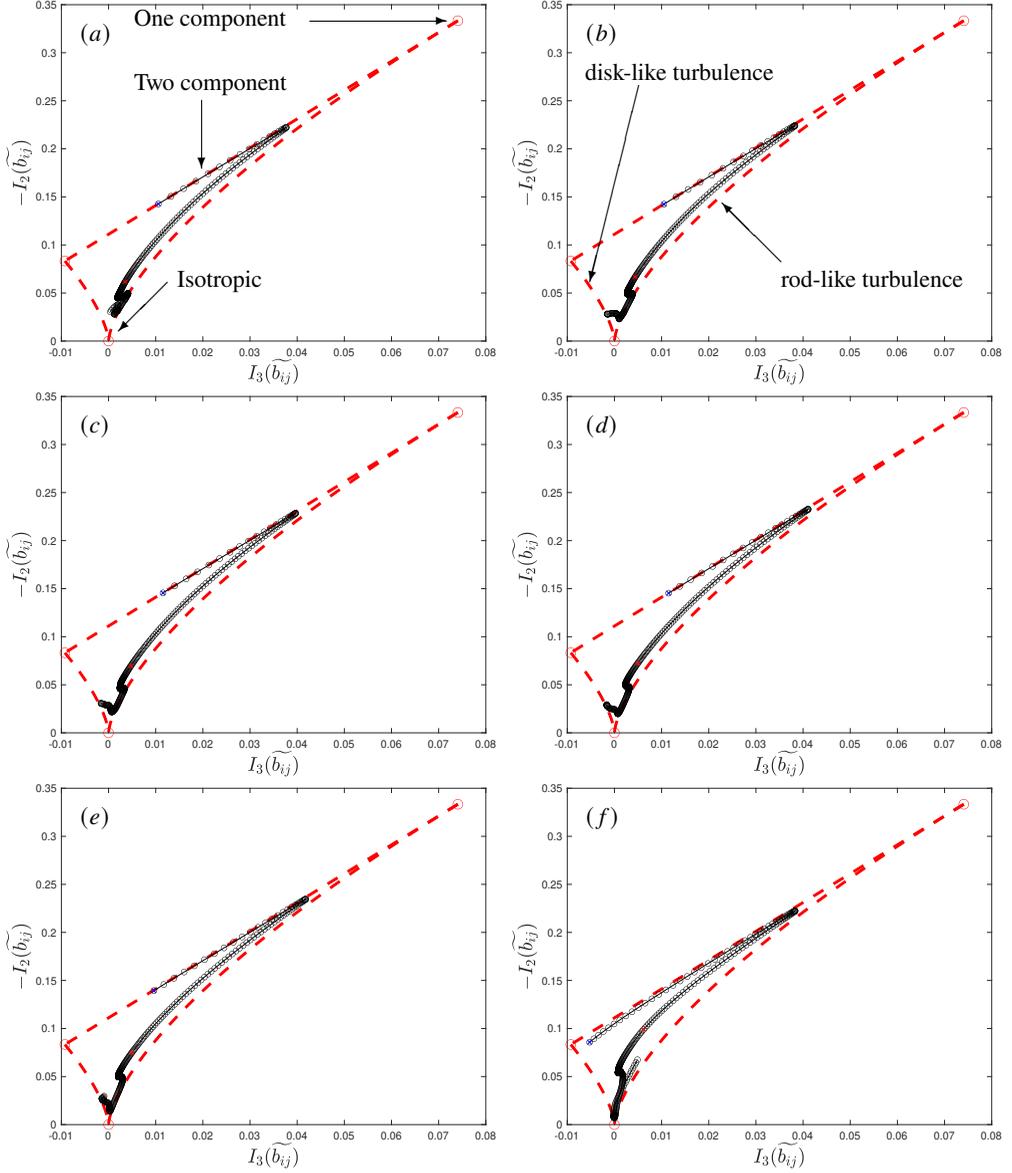
\begin{figure}
  \centering
  \begin{tabular}{cc}
  \begin{overpic}[width=0.48\textwidth]{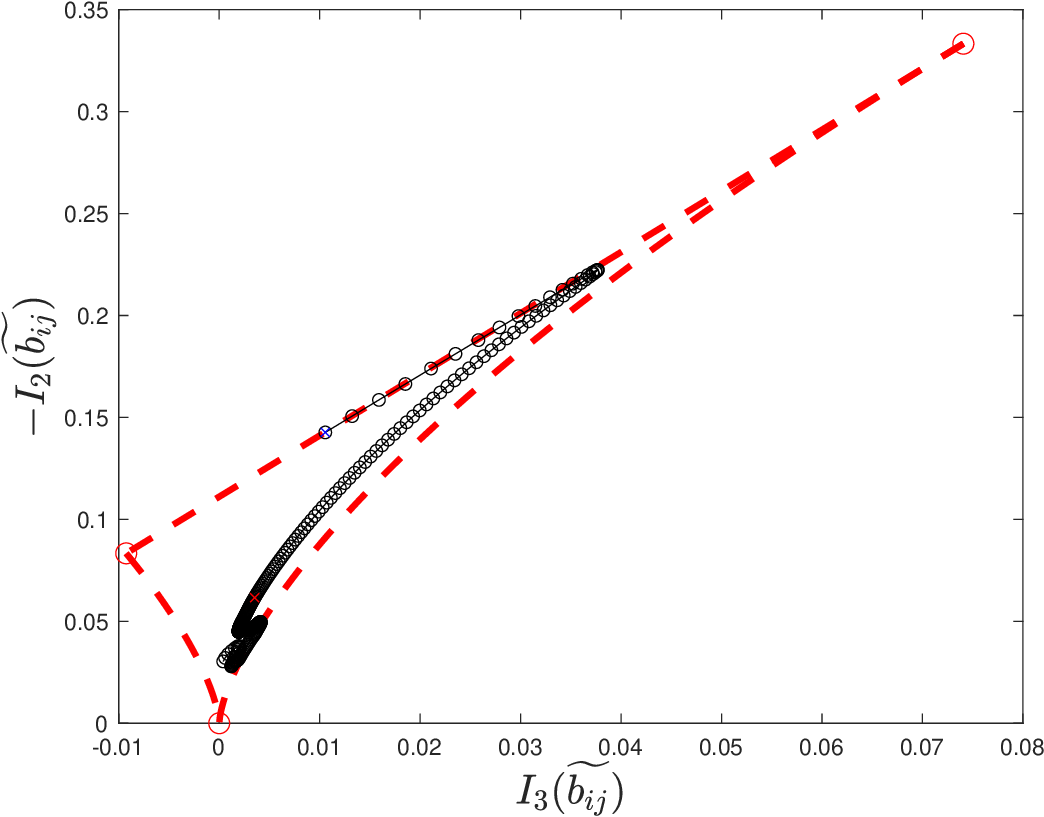}
  \put(15,70){$(a)$}
  \put(25,73){ One component}
  \put(25,60){ Two component}
  \put(62,74){\vector(1,0){28}}
  \put(40,58){\vector(0,-1){14}}
  \put(35,20){Isotropic}
  \put(33,21){\vector(-1,-1){10}}
  \end{overpic} &
  \begin{overpic}[width=0.48\textwidth]{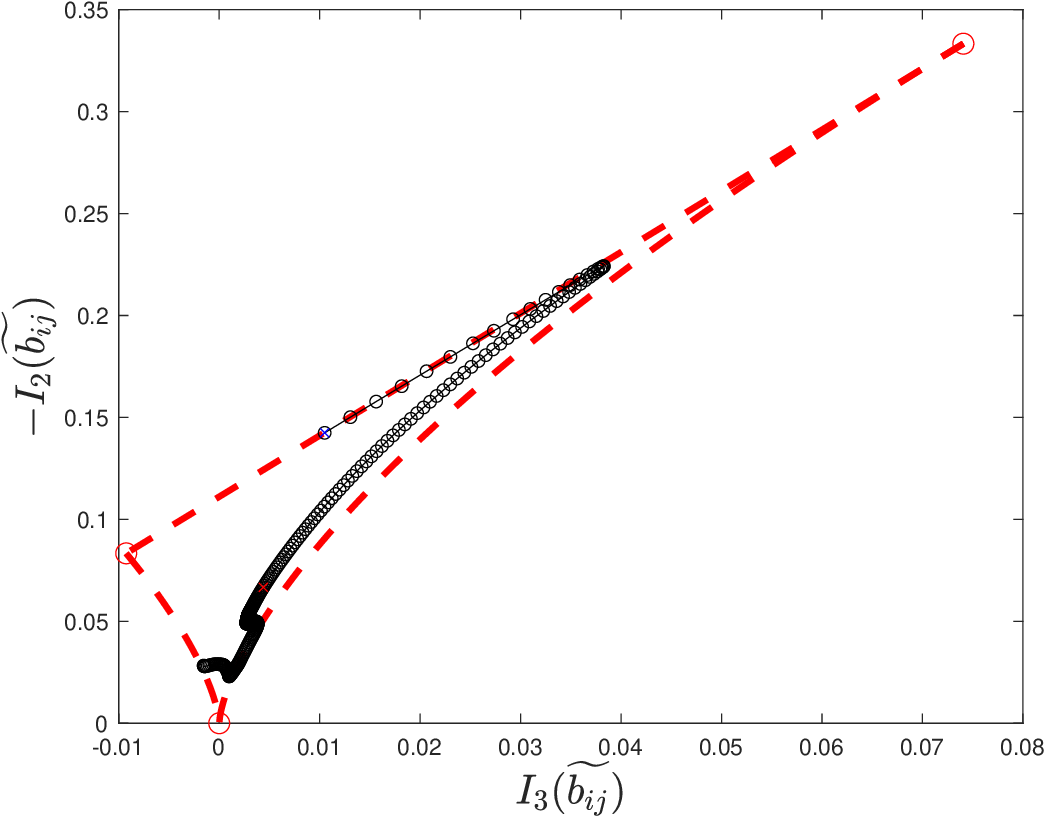}
  \put(15,70){$(b)$}
  \put(55,20){rod-like turbulence}
  \put(15,62){disk-like turbulence}
  \put(55,25){\vector(-1,1){12}}
  \put(26,61){\vector(-1,-4){10}}
  \end{overpic}  \\
  \begin{overpic}[width=0.48\textwidth]{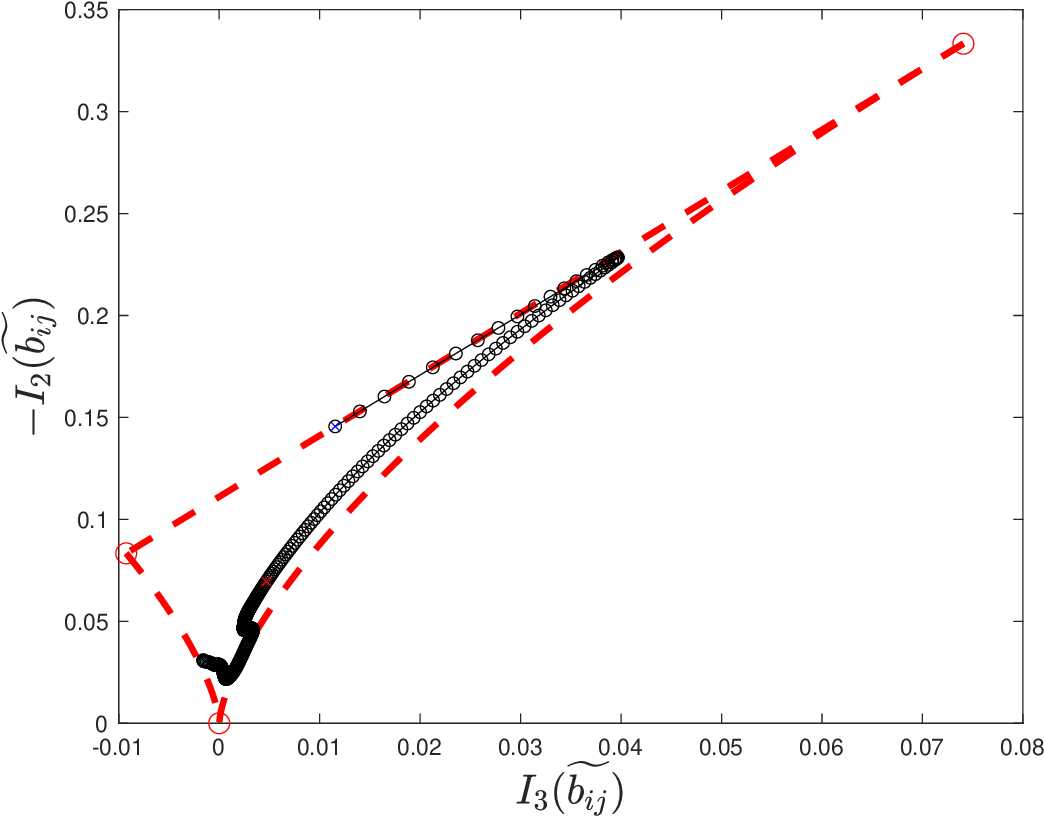}
  \put(15,70){$(c)$}
  \end{overpic} & 
  \begin{overpic}[width=0.48\textwidth]{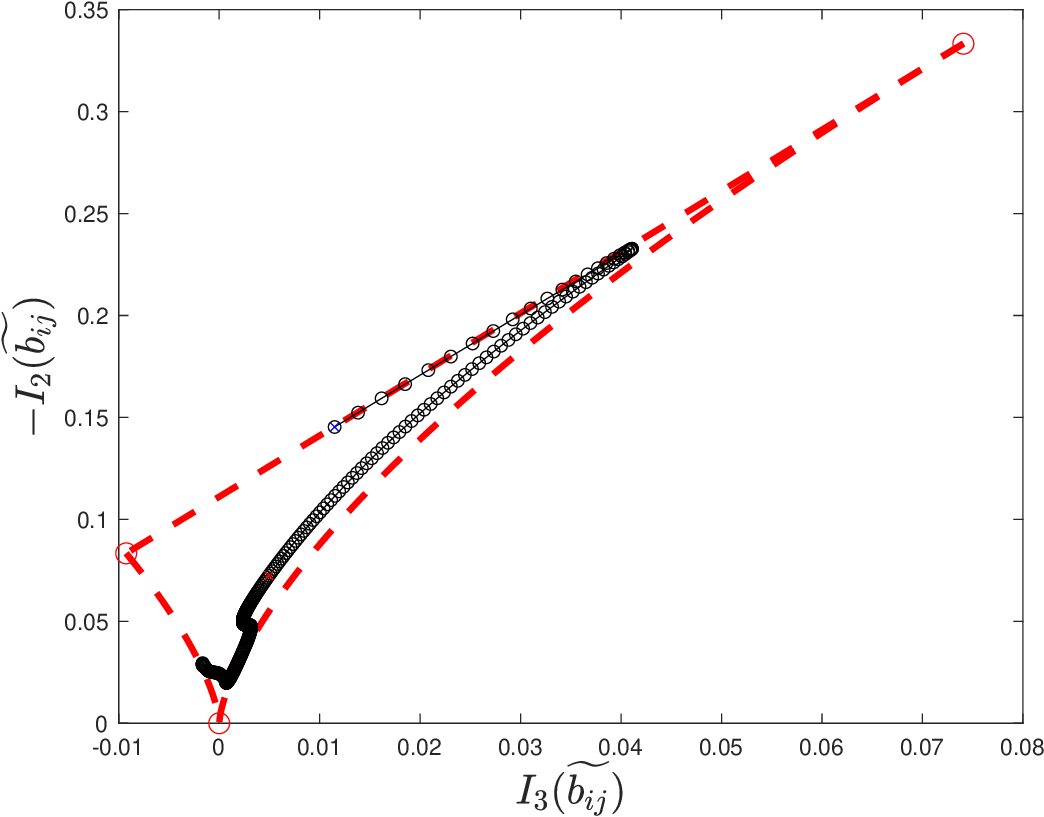}
  \put(15,70){$(d)$}
  \end{overpic} \\
  \begin{overpic}[width=0.48\textwidth]{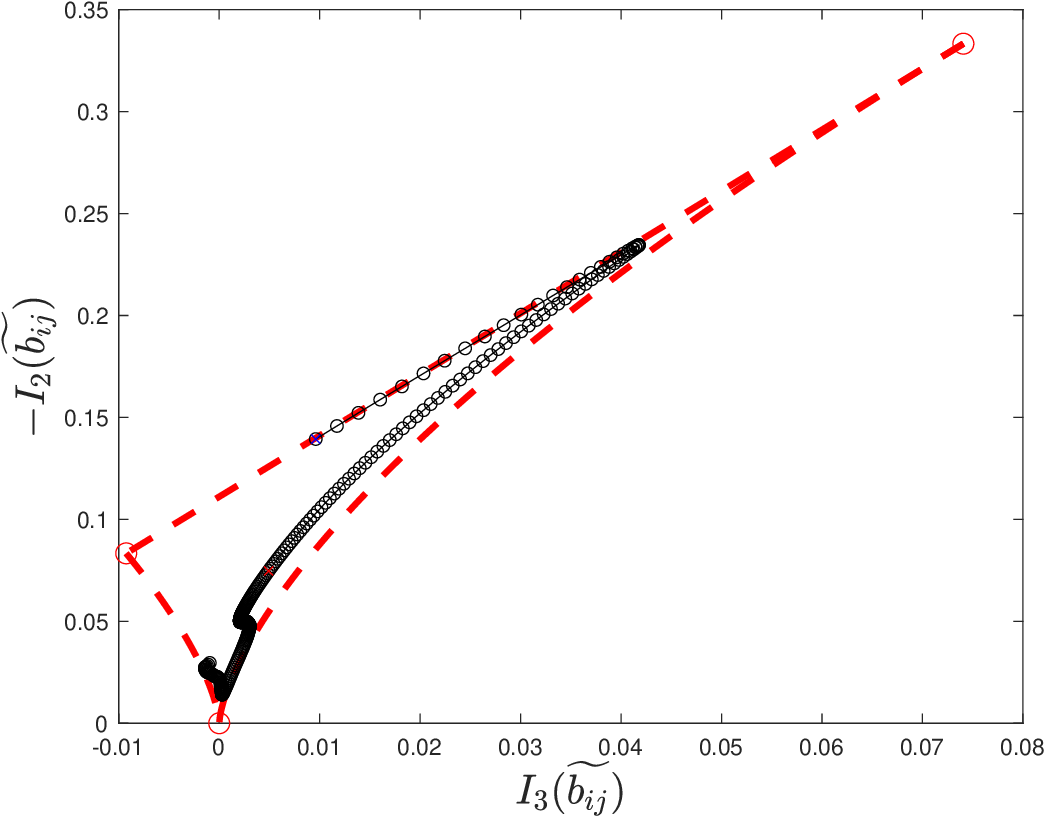}
  \put(15,70){$(e)$}
  \end{overpic} & 
  \begin{overpic}[width=0.48\textwidth]{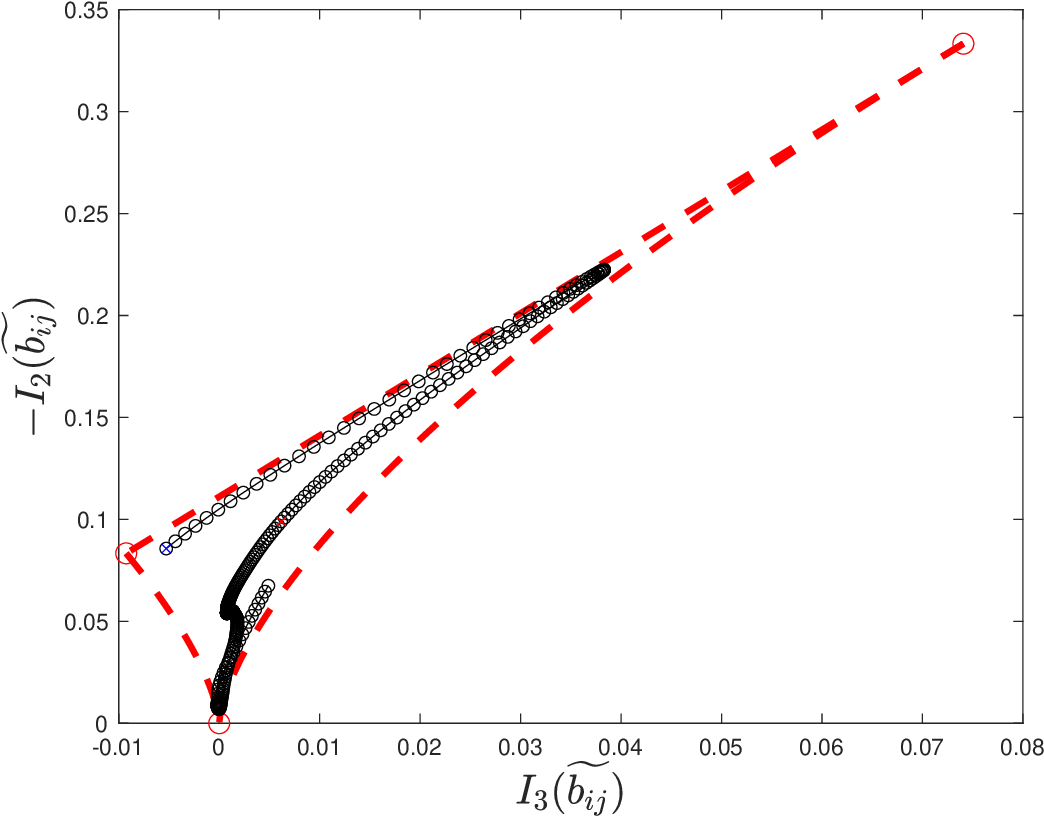}
  \put(15,70){$(f)$}  
  \end{overpic}
  \end{tabular}
  \caption{Invariant mapping of Reynolds stress anisotropy in three-dimensioanl boundary layer at different chordwise locations $s_{\xi}$. $(a-f)$ stand for chordwise locations at $s_{\xi} = 0.00, 4.22, 8.71, 13.50, 18.61$ and $30.00$, respectively. }
  \label{Reynold_Stress_Anisotropy_Variations}
\end{figure}

Here, in this section, we try to understand the state of turbulence by using anisotropy tensor of Reynolds stress.
The anisotropy tensor $\widetilde{b}_{ij}$, which is defined as 
\begin{equation}
\widetilde{b}_{ij} = \frac{\overline{\rho u_i^{\prime\prime}u_j^{\prime\prime}}}{\overline{\rho u_i^{\prime\prime}u_i^{\prime\prime}}} - \frac{\delta_{ij}}{3} = \frac{\widetilde{u_i^{\prime\prime}u_j^{\prime\prime}}}{\widetilde{u_i^{\prime\prime}u_i^{\prime\prime}}} - \frac{\delta_{ij}}{3}
\end{equation}
is used to understand the state of turbulence at different chordwise location. The invariants of this tensor can be written as\citep{Simonsen2005}
\begin{equation}
I_1 = \widetilde{b}_{ii} = 0, \quad
II_2 = -\frac{1}{2}\widetilde{b}_{ij}\widetilde{b}_{ji}, \quad
III_3 = det(\widetilde{b}_{ij}).
\end{equation}
The state of the anisotropy stress tensor can be shown in an anisotropy invariant map, as shown in figure \ref{Reynold_Stress_Anisotropy_Variations}. The blue cross stands for the location of the first grid away from the wall surface and the red cross represents the 101th grid away from the surface. All the points, along the wall-normal directions, inside the boundary layer are shown in the figures. 
The points close to the wall, as $h\to 0$, are also marked, which are very close to the two-component turbulence, as the wall significantly restricts turbulence fluctuations in the direction normal to the wall in the near-wall region. 
As the position moves away from the surface, the reynolds stress anisotropy tensor exhibits a tendency to one-component turbulence. 
Subsequently, as the position moves further away from the wall, the wall's constraint on the turbulence progressively weakens. 
The Reynolds stress begins to exhibit characteristics of rod-like turbulence and, with increasing distance from the wall, tends to become more isotropic.
As the position moves away from the attachment-line boundary layer (from figures \ref{Reynold_Stress_Anisotropy_Variations} $(b)$ to $(e)$), the distribution characteristics of the anisotropic Reynolds stress tensors remain consistent with those observed at the attachment line. 
This indicates that the anisotropic characteristics of the Reynolds stress tensors are not significantly affected by the three-dimensional effects, which could provide guidance for subsequent modeling of three-dimensional turbulent boundary layers.

\subsection{Turbulent kinetic energy budgets}
Further investigation of the turbulence in this three dimensional boundary layer is conducted by analysing the turbulent kinetic energy equation. The transport equation of the turbulent kinetic energy can be expressed as
\begin{equation}
\frac{\partial \overline{\rho} K}{\partial t} = \mathbf{C} + \mathbf{P} + \mathbf{\Pi} + \boldsymbol{\epsilon} + \mathbf{M} + \mathbf{D},
\end{equation}
where
\begin{equation}
\left.
\begin{aligned}
\mathbf{C} &= -\frac{\partial \overline{\rho} K \widetilde{u}_i}{\partial x_i}, 
\mathbf{P} = - \overline{\rho}\widetilde{u_i^{\prime\prime}u_k^{\prime\prime}}\frac{\partial \widetilde{u_i}}{\partial x_k}, 
\mathbf{\Pi} = \overline{p^{\prime}\frac{\partial u_i^{\prime}}{\partial x_i}}, \\
\boldsymbol{\epsilon} &= -\overline{\tau_{ik}^{\prime}\frac{\partial u_i^{\prime}}{\partial x_k}},
\mathbf{M} = \overline{u_i^{\prime\prime}}\left(\frac{\partial \overline{\sigma}_{ik}}{\partial x_k} - \frac{\partial \overline{p}}{\partial x_i}\right), \\
\mathbf{D} &= -\frac{\partial}{\partial x_k}\left[\frac{\overline{\rho}\widetilde{u_i^{\prime\prime}u_i^{\prime\prime}u_k^{\prime\prime}}}{2}
+\overline{p^{\prime}u_i^{\prime}}\delta_{ik} - \overline{\tau_{ik}^{\prime}u_i^{\prime}} \right],
\end{aligned}
\right\},
\end{equation}
stands for the convection term($\mathbf{C}$), production term($\mathbf{P}$), pressure-dilatation term($\mathbf{\Pi}$), dissipation rate term($\boldsymbol{\epsilon}$), mass flux term($\mathbf{M}$) and transport-diffusion term($\mathbf{D}$), respectively. 

\begin{figure}
  \centering
  \begin{tabular}{cc}
  \begin{overpic}[width=0.4\textwidth]{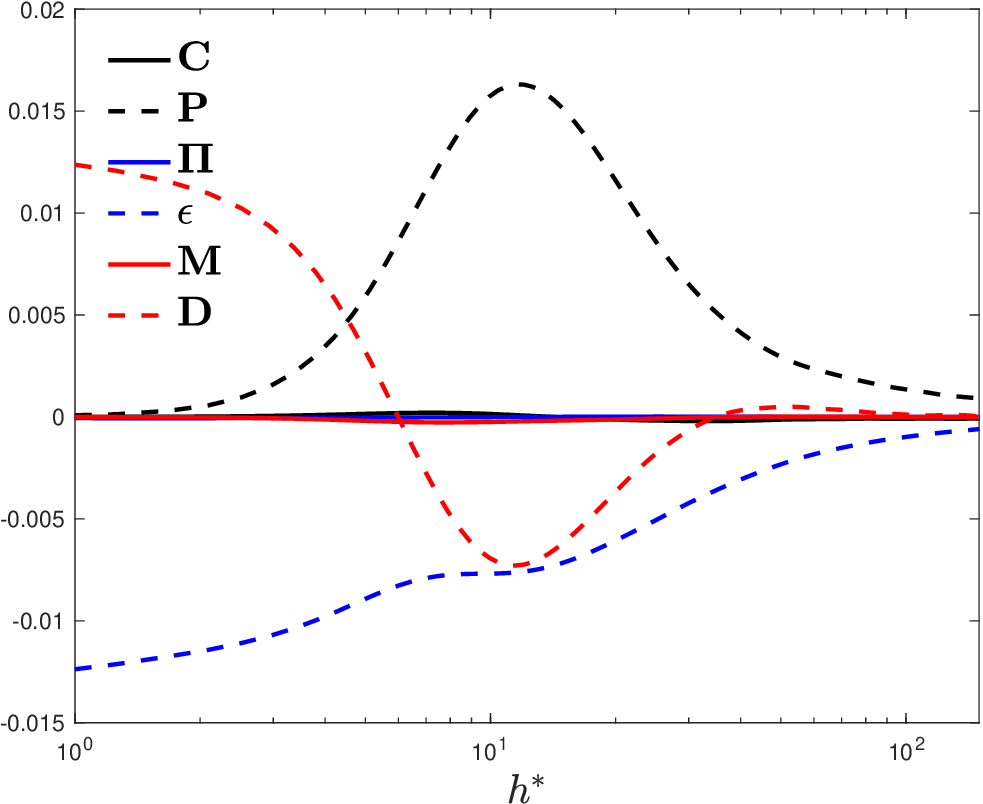}
  \put(65,70){$(a)$}
  \end{overpic} &
  \begin{overpic}[width=0.4\textwidth]{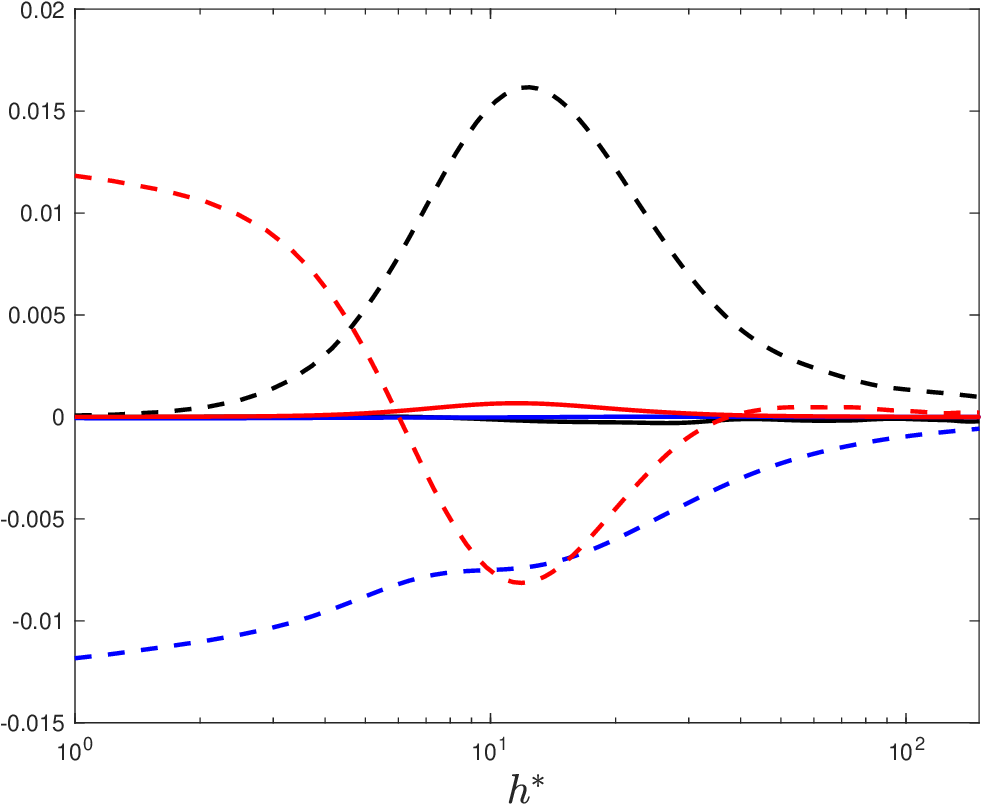}
  \put(15,70){$(b)$}
  \end{overpic}  \\
  \begin{overpic}[width=0.4\textwidth]{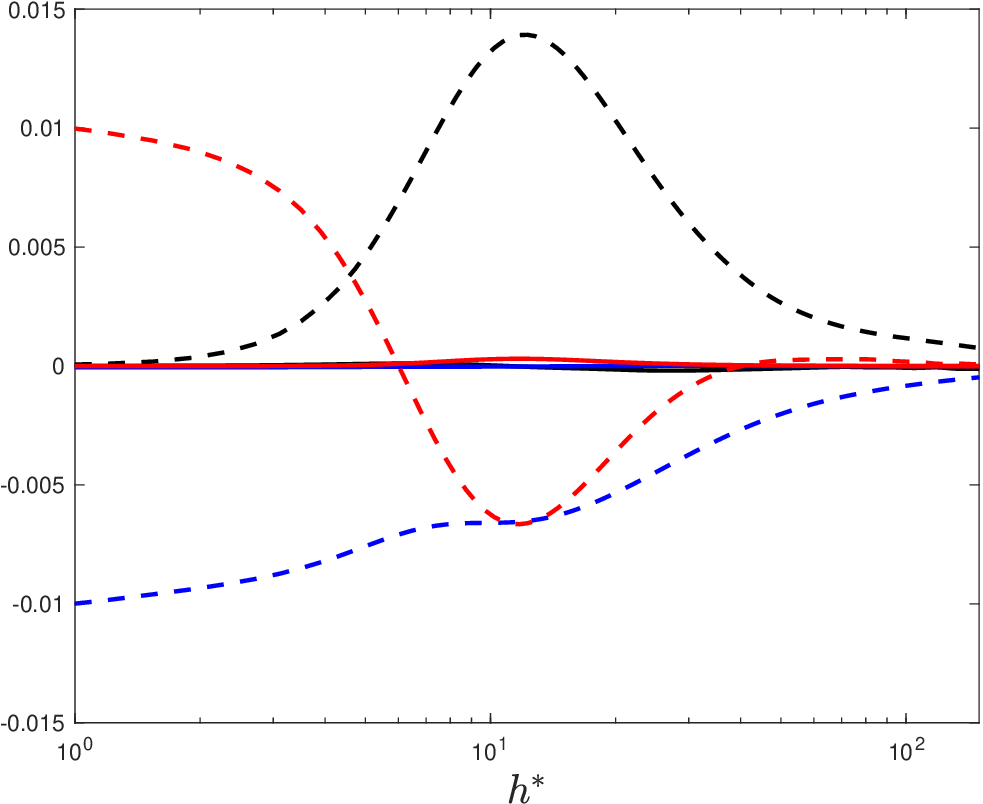}
  \put(15,70){$(c)$}
  \end{overpic} & 
  \begin{overpic}[width=0.4\textwidth]{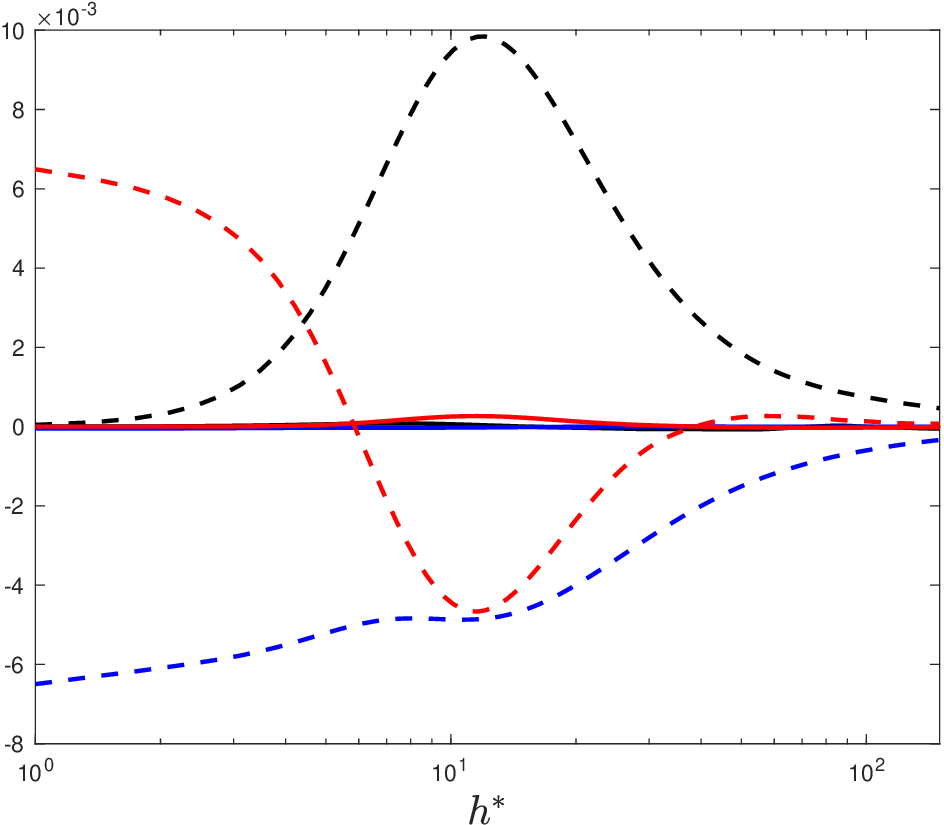}
  \put(15,70){$(d)$}
  \end{overpic} \\
  \begin{overpic}[width=0.4\textwidth]{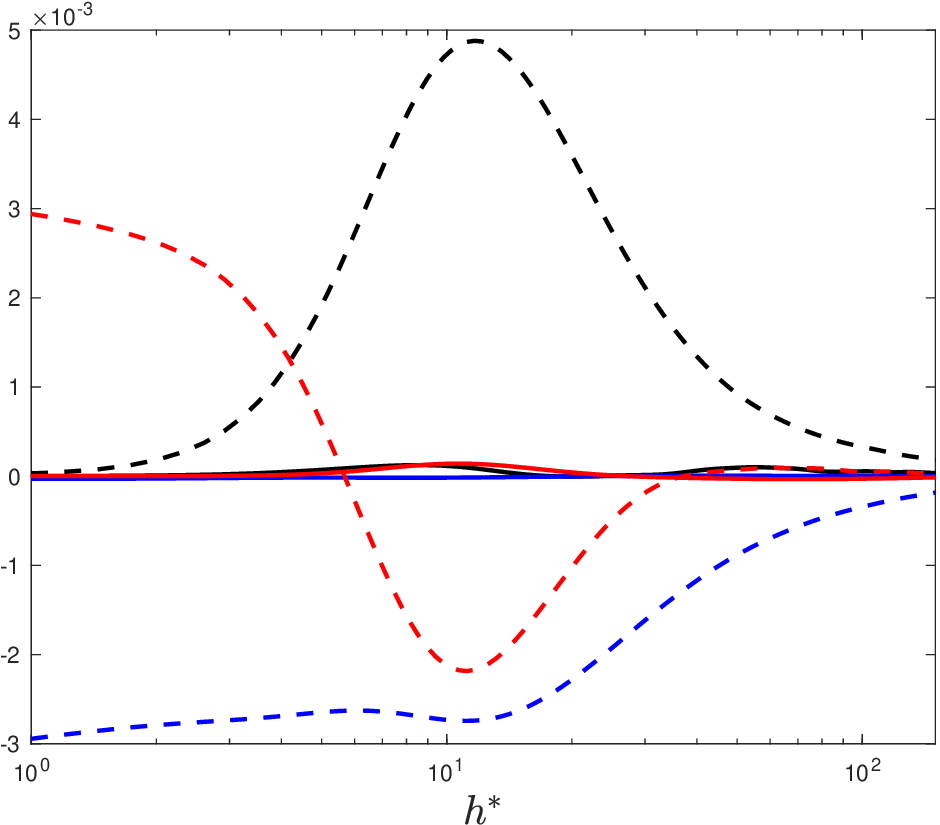}
  \put(15,70){$(e)$}
  \end{overpic} & 
  \begin{overpic}[width=0.4\textwidth]{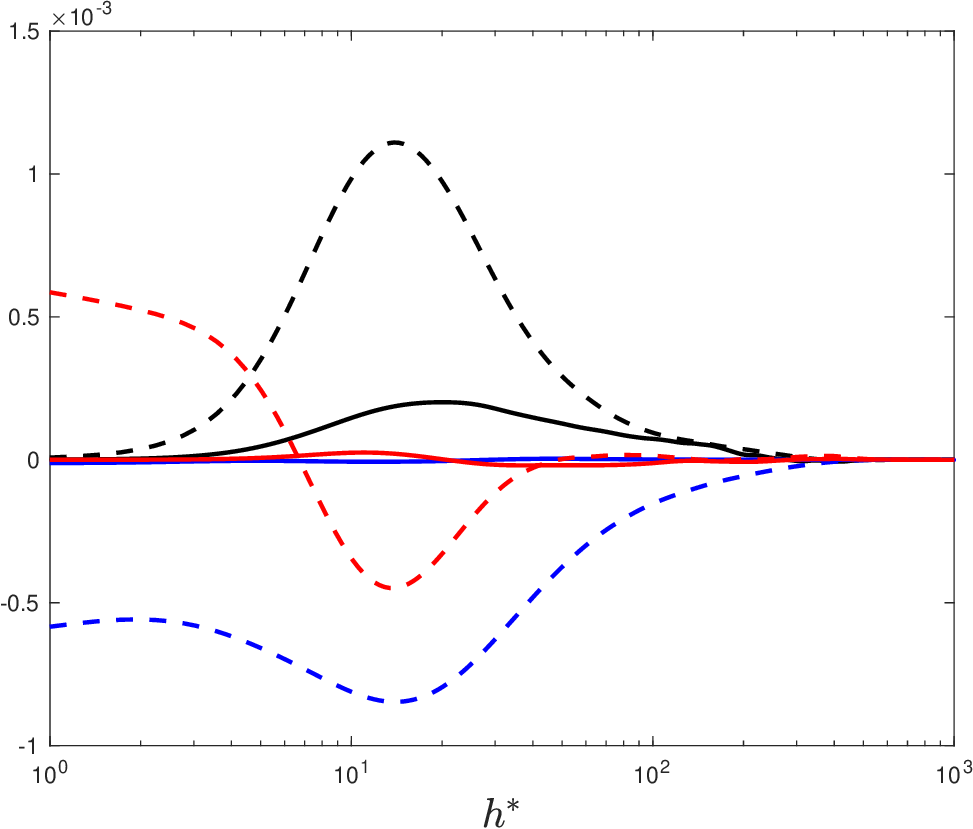}
  \put(15,70){$(f)$}
  \end{overpic} 
  \end{tabular}
  \caption{The energy budget distributiones along the wall normal direction in three-dimensioanl boundary layer at different chordwise locations $s_{\xi}$. $(a-f)$ stand for chordwise locations at $s_{\xi} = 0.00, 4.22, 8.71, 13.50, 18.61$ and $24.10$, respectively. }
  \label{Turbulent_budget}
\end{figure}

Figure \ref{Turbulent_budget} shows the basic results of
energy budget distributiones along the wall normal direction at several chordwise position. 
From the distribution shown in the figures \ref{Turbulent_budget}$(a-e)$, it is evident that the dominant terms are the production term $\mathbf{P}$, dissipation term $\boldsymbol{\epsilon}$, and the transport-diffusion term $\mathbf{D}$, consistent with the analysis in traditional two-dimensional boundary layers. This indicates that even though three-dimensional effects significantly alter the flow characteristics (introducing crossflow, pressure gradient effects, etc.), their impact on the turbulence energy balance mechanism is minimal. However, as shown in figure \ref{Turbulent_budget}$(f)$, the convection term (the black line) become obvious. Although the proportion of the convection term in the energy budget analysis increased, this does not imply an absolute intensification of the convection. In fact, under the influence of the chordwise pressure gradient, the corresponding turbulence is suppressed, leading to a reduction in the production, dissipation and transport-diffusion terms, which in turn makes the convection term more pronounced. This implies that the core modeling terms for turbulence in three-dimensional boundary layer flows remain consistent with those in two-dimensional boundary layers.

\subsection{Analysis of Shear Direction Discrepancies}
In the previous discussion, we briefly introduced some basic characteristics of the computed three-dimensional turbulent boundary layer. In this section, we will delve deeper into the shear effects within the three-dimensional boundary layer. Traditionally, in two-dimensional turbulent boundary layers (assuming the mainstream and wall-normal velocities are $u_{\xi}$ and $u_{n}$), the focus of shear effects is primarily on $\widetilde{u_{\xi}^{\prime\prime}u_{n}^{\prime\prime}}$ and the corresponding meanflow shear $du_{\xi}/dh$. However, in three-dimensional cases, the shear effects should actually encompass both $\widetilde{u_{\xi}^{\prime\prime}u_{n}^{\prime\prime}}$ and $\widetilde{w^{\prime\prime}u_{n}^{\prime\prime}}$, as well as the corresponding $d u_{\xi}/dh$ and $d w / dh$. From the previous analyses, we observed that the direction of the mean velocity within the boundary layer changes with the distance from the wall. Consequently, the direction of the resultant shear Reynolds stress also varies accordingly. This naturally prompts the question: does the direction of the shear Reynolds stress still align with the direction of the resultant mean flow shear under these conditions?

\begin{figure}
  \centering
  \begin{tabular}{cc}
  \begin{overpic}[width=0.48\textwidth]{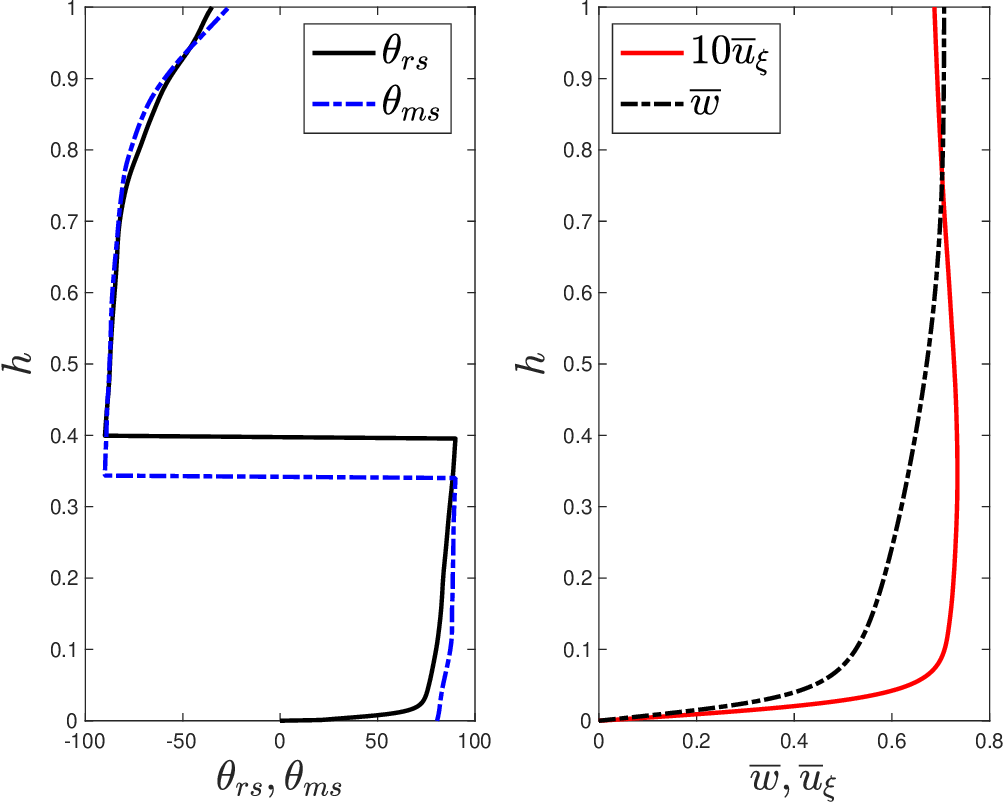}
  \put(35,60){$(a_1)$}
  \put(65,60){$(a_2)$}
  \end{overpic} &
  \begin{overpic}[width=0.48\textwidth]{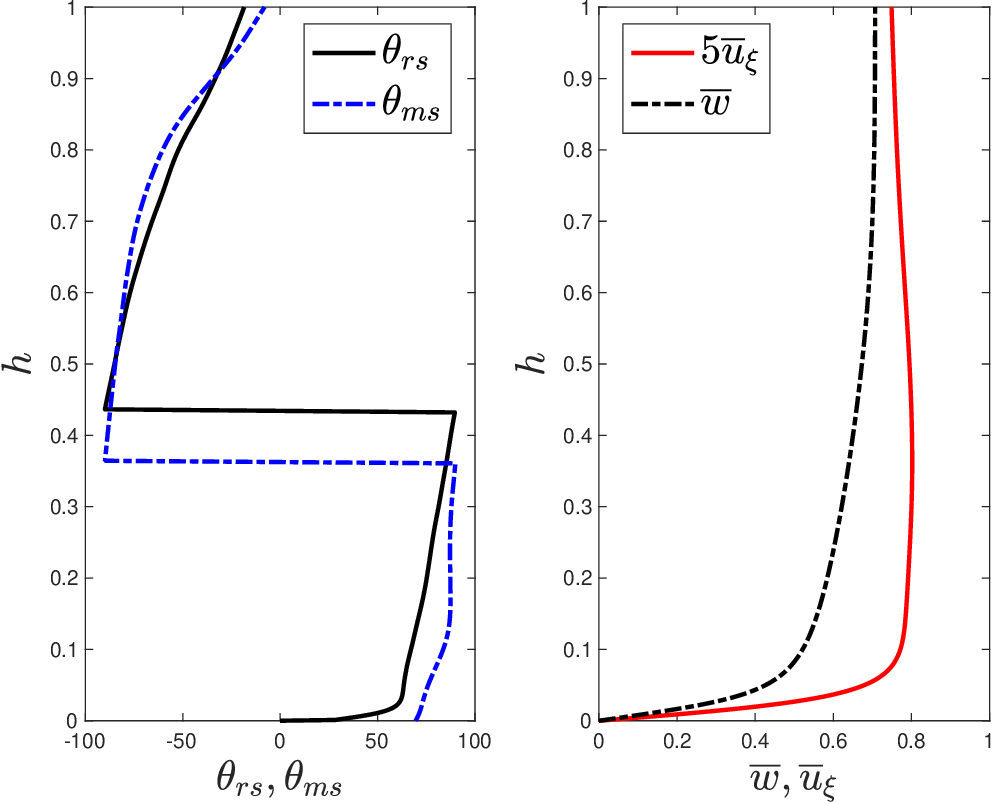}
  \put(35,60){$(b_1)$}
  \put(65,60){$(b_2)$}
  \end{overpic}  \\
  \begin{overpic}[width=0.48\textwidth]{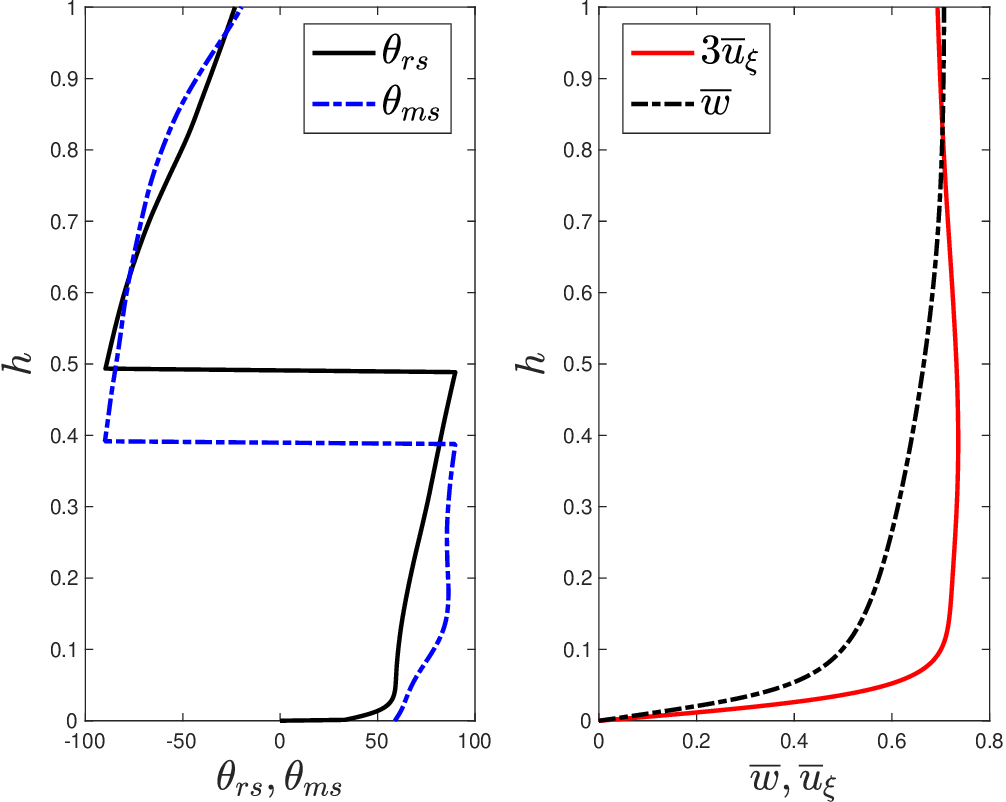}
  \put(35,60){$(c_1)$}
  \put(65,60){$(c_2)$}
  \end{overpic} & 
  \begin{overpic}[width=0.48\textwidth]{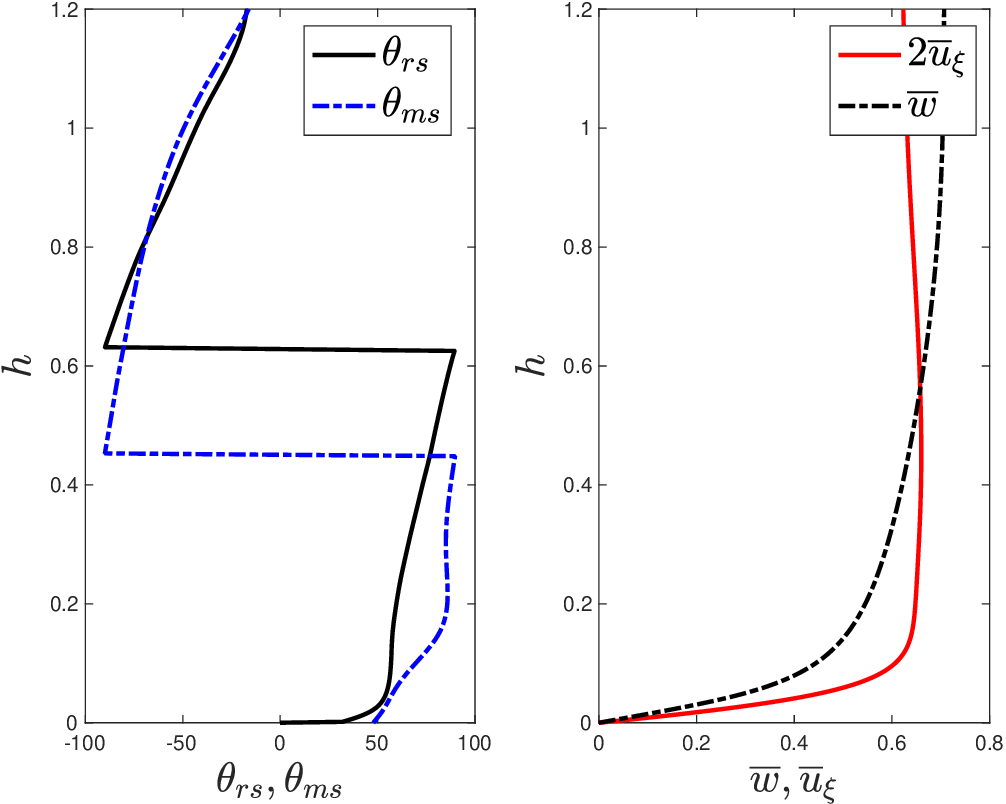}
  \put(35,60){$(d_1)$}
  \put(65,60){$(d_2)$}
  \end{overpic} 
  \end{tabular}
  \caption{$(a_1 - d_1)$ Variation of the directional angles ($\theta_{rs}$ and $\theta_{ms}$ in degree) of shear Reynolds stress and mean flow shear with distance from the wall, at several chordwise locations. $(a_2 - d_2)$ The corresponding mean velocity profiles at relative spanwise locations. $(a-d)$ stand for chordwise locations at $s_{\xi} = 4.22, 8.71, 13.50$ and $18.61$, respectively. Some velocity component is enlarged for clarity. }
  \label{Shear_discrepancies}
\end{figure}
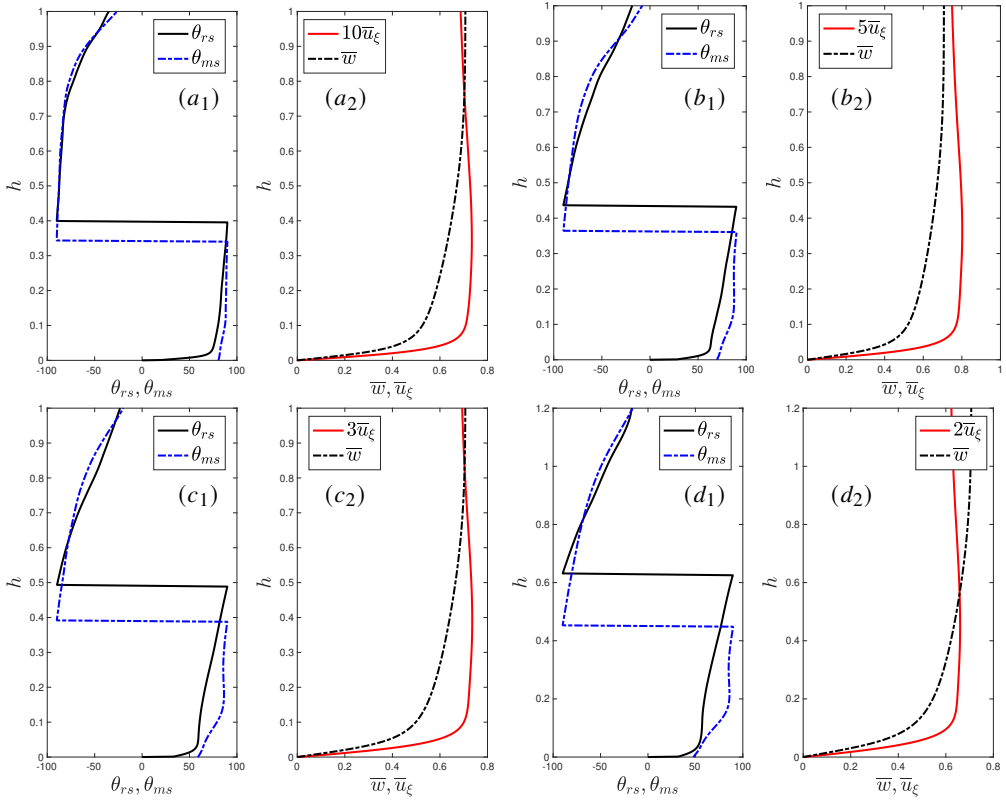

In general, the Reynolds shear stresses can be expressed as
\begin{equation}
\left.
\begin{aligned}
\widetilde{u_{\xi}^{\prime\prime}u_{n}^{\prime\prime}} &= \bm{t}_x \bm{n}_x \widetilde{u^{\prime\prime}u^{\prime\prime}} + \left(\bm{t}_x\bm{n}_y + \bm{t}_y\bm{n}_x\right)\widetilde{u^{\prime\prime}v^{\prime\prime}}
+ \bm{t}_y \bm{n}_y \widetilde{v^{\prime\prime}v^{\prime\prime}}, \\
\widetilde{w^{\prime\prime}u_{n}^{\prime\prime}} &= \bm{n}_x \widetilde{u^{\prime\prime}w^{\prime\prime}} + \bm{n}_y\widetilde{v^{\prime\prime}w^{\prime\prime}}
\end{aligned}
\right\},
\end{equation}
and we define the angle $\theta_{rs} = \arctan{\left(\widetilde{w^{\prime\prime}u_{n}^{\prime\prime}} / \widetilde{u_{\xi}^{\prime\prime}u_{n}^{\prime\prime}}\right)}$ between the resultant shear Reynolds stress and the chordwise $\xi-$direction, and the angle $\theta_{ms} =  \arctan{\left(\overline{\partial{w}/\partial h}/\overline{\partial u_{\xi}/\partial h}\right)}$ between the mean shear direction and the chordwise direction. The variation of the directional angles of shear Reynolds stress and mean-flow shear at different chordwise locations are shown in figure \ref{Shear_discrepancies} and some mean-flow velocity components are also shown for better understanding the behaviours. 

At different chordwise stations, the distribution of the angles $(\theta_{rs}, \theta_{ms})$ is generally similar and can be divided into three regions along the wall-normal direction. First is the near-wall region, where both $\theta_{rs}$ and $\theta_{ms}$ exhibit positive angles. 
There is a significant discrepancy between the two at the wall due to the shear Reynolds stress being zero there, while the mean shear gradient has a substantial value. 
As the distance from the wall increases, the two angles rapidly converge. 
In regions with negligible transverse pressure gradients and velocity, 
$\theta_{rs}$ and $\theta_{ms}$ approximate equality, indicating a linear relationship between the shear Reynolds stress and the mean flow shear gradient (figure \ref{Shear_discrepancies}$(a_1)$).
As the transverse velocity increases, the difference between the two angles in this region also grows, suggesting a nonlinear relationship between the Reynolds shear stress and the mean flow shear when the transverse components ($\overline{u}_{\xi}$ in figures \ref{Shear_discrepancies}$(a_2 - d_2)$) become sufficiently large (figures \ref{Shear_discrepancies}$(b_1 - d_1)$). 
When the wall-normal distance further increases, a sudden change in the angles is observed. 
This is attributable to the pressure gradient causing the velocity profile to overshoot, resulting in a reversal of the mean flow gradient from positive to negative. 
A similar phenomenon occurs in the shear Reynolds stress, but its change lags behind the mean flow shear, with the extent of this lag increasing with higher transverse velocities.
After these angular transitions, both $\theta_{rs}$ and $\theta_{ms}$ quickly return to similar levels, indicating that in the outer region of the boundary layer, at a certain distance from the wall, the shear Reynolds stress and mean flow shear again appear to exhibit a linear relationship. This observation indicates that, in the case of three-dimensional flows, the relationship between the shear Reynolds stress and the mean flow shear is inherently nonlinear. Consequently, a simple linear approximation would introduce certain inaccuracies.

\section{Conclusions}\label{sec5}
Overall, this study represents the second part of our series on hypersonic swept blunt-body flows. In this segment, we have examined the characteristics of three-dimensional turbulent boundary layers in the presence of transverse flow and pressure gradients. The main findings of this study are as follows:

\begin{itemize}
\item In the flow over a swept blunt body discussed in this paper, even without the assumption of infinite sweep, if the boundary layer reaches a fully developed turbulent state, the subsequent flow state also satisfies the infinite sweep assumption—the boundary layer is homogeneous along the spanwise direction.

\item We examined the law-of-the-wall and the temperature-velocity relationship, which have been widely applied in two-dimensional turbulent boundary layers, within the context of three-dimensional boundary layers. Our findings indicate that, when the transverse velocity is not excessively high, the spanwise velocity still adheres to some classical velocity transformation relationships. Excluding the Trettel \& Larrson transformation, existing transformation forms generally yield good predictive results, with the transformations proposed by \citet{Griffin2021} and \citet{Volpiani2020} aligning best with incompressible scaling. However, as the transverse velocity further increases, the spanwise velocity no longer conforms to traditional scaling laws. Instead, the streamwise velocity $u_s$, in the present condition, conforms to traditional scaling laws and classical transformations can collapse the data to the log-law of the incompressible ones. We further examined the applicability of the temperature-velocity relationship proposed by \citet{Zhang2014} under such three-dimensional conditions. The study found that, in the absence of pressure gradients or when the transverse velocity is relatively low, the traditional temperature-velocity relationship still holds well. However, when the pressure gradient and transverse velocity are significant, the predictive accuracy of the temperature-velocity relationship markedly diminishes. The potential reason for this deviation is that in boundary layers with pressure gradients, the velocity profile along the direction of the pressure gradient exhibits an overshoot, which significantly alters the corresponding relationship between temperature and velocity, leading to an overestimation in the predictive results.

\item Analysis of fluctuating velocities, temperature, Reynolds stresses, Reynolds stress anisotropy tensors, and turbulent kinetic energy in three-dimensional boundary layers indicates that the fundamental turbulent characteristics remain qualitatively consistent with those in two-dimensional turbulent boundary layers. Typical near-wall streak structures still appear in three-dimensional turbulent boundary layers and correspond with the energetic structures in the outer region. However, due to the three-dimensional effects, turbulent energy is redistributed within the three-dimensional turbulent boundary layer. This redistribution causes the turbulent energy to align more with the three-dimensional mean flow direction, resulting in the streak structures tending to orient along the direction of the external streamlines.

\item By analyzing and comparing the directions of the shear Reynolds stress and the mean flow shear, the study found that in the near-wall region, when the transverse flow velocity is relatively low, the two align well. However, when the transverse velocity is high, there is a significant deviation between their directions. Additionally, with the introduction of a pressure gradient in the transverse direction, the velocity profile in the corresponding direction exhibits an overshoot, which alters the mean flow shear direction. The shear Reynolds stress also undergoes a similar change, but with a certain lag, and this lag increases with higher transverse velocity. Finally, in the outer region, some distance away from the wall, the directions of the shear Reynolds stress and mean flow shear are essentially consistent. This indicates that in three-dimensional boundary layers with transverse flow, the relationship between shear Reynolds stress and mean flow shear is partitioned: it is nonlinear in the inner region and can be approximated to return to a linear relationship in the outer region.
\end{itemize}

Finally, we would like to say that the leading edge of the swept configuration is an ideal model for studying three-dimensional turbulent boundary layers. It effectively bridges the traditional, unified concepts of two-dimensional turbulent boundary layers with the unique effects inherent to three-dimensional turbulent boundary layers.

\backsection[Supplementary data]{}

\backsection[Acknowledgements]{We acknowledge Yancheng MetaStone Tech. Co. for prociding us with all the computational resources required by this work. 
Useful discussions with Professor Chunxiao Xu, Professor Qibing Li of Tsinghua University, Professor Jie Ren of Beijing Institute of Technology, Dr. Jian Liu of China Aerodynamics Research and Development Center and Dr. Yi Huang of China Academy of Launch Vehicle Technology are gratefully acknowledged.}

\backsection[Funding]{This work received support from the NSFC Grants 12202242, 12172195 and 12388101. The authors are also grateful for the support from the National Key Research and Development Plan of China through project no. 2019YFA0405201 and the National Key Project GJXM92579 and the Grants 20231001 in Supercomputing Center in Yancheng.}

\backsection[Declaration of interests]{The authors report no conflict of interest.}

\backsection[Data availability statement]{The full data set of the simulations is of the order of thousands of gigabytes. By contacting the authors, a smaller subset can be made available.}

\backsection[Author ORCIDs]{

Youcheng Xi, https://orcid.org/0000-0002-6484-0231;

Bowen Yan, https://orcid.org/0009-0002-0655-9414;

Guangwen Yang, https://orcid.org/0000-0002-8673-8254

Song Fu, https://orcid.org/0000-0003-2052-7435}

\backsection[Author contributions]{
Youcheng Xi: Funding acquisition, Conceptualization, Data curation, Formal analysis, Coding, Investigation, Methodology, Validation, Writing original draft. 
Bowen Yan \& Guangwen YANG: Software and optimization on high performance computing, Computational Resources.
Song Fu: Funding acquisition, Supervision, Resources, Writing-review \& editing.}

\appendix

%\section{}\label{appA}
 
\bibliographystyle{jfm}
\bibliography{jfm}

\end{document}